\documentclass[%
 reprint,
 amsmath,amssymb,
 aps,
]{revtex4-1}

\usepackage{graphicx}
\usepackage{dcolumn}
\usepackage{bm}
\usepackage{enumerate}   
\usepackage{xcolor}
\usepackage{xspace}
\usepackage{comment}
\usepackage{multirow}
\usepackage{placeins}
\usepackage{hyperref}

\renewcommand{\vec}[1]{\mathbf{#1}}

\newcommand{\trento}{T\raisebox{-.5ex}{R}ENTo\xspace}

\newcommand{\twosevensixnn}{\ensuremath{\sqrt{s_{\mathrm{NN}}}=2.76\,\mathrm{TeV}}\xspace}
\newcommand{\fivenn}       {\ensuremath{\sqrt{s_{\mathrm{NN}}}=5.02\,\mathrm{TeV}}\xspace}

\newcommand{\meanpt}           {\ensuremath{\langle p_{\mathrm{T}}\rangle}\xspace}
\newcommand{\snn}          {\ensuremath{\sqrt{s_{\mathrm{NN}}}}\xspace}

\begin{document}

\title{Enhancing Bayesian parameter estimation by adapting to multiple energy scales in RHIC and LHC heavy-ion collisions}

\author{~M.~Virta$^{1,2}$}
\author{~J.E.~Parkkila$^{3}$}%
\author{~D.J.~Kim$^{1,2}$}%
\email{maxim.virta@cern.ch}

\affiliation{$^{1}$University of Jyväskylä, Department of Physics, P.O. Box 35, FI-40014 University of Jyväskylä, Finland}
\affiliation{$^{2}$Helsinki Institute of Physics, P.O.Box 64, FI-00014 University of
Helsinki, Finland}
\affiliation{$^{3}$Warsaw University of Technology, Warsaw, Poland}

\date{\today}

\begin{abstract}
\noindent Improved constraints on current model parameters in a heavy-ion collision model are established using the latest measurements from three distinct collision systems. Various observables are utilized from Au--Au collisions at $\sqrt{s_\mathrm{NN}}=200$~GeV and Pb--Pb collisions at $\sqrt{s_\mathrm{NN}}=5.02$~TeV and $\sqrt{s_\mathrm{NN}}=2.76$~TeV. Additionally, the calibration of centrality is now carried out separately for all parametrizations. The inclusion of an Au–Au collision system with an order of magnitude lower beam energy, along with separate centrality calibration, suggests a preference for smaller values of nucleon width, minimum volume per nucleon, and free-streaming time. The results with the acquired \textit{maximum a posteriori} parameters show improved agreement with the data for the second-order flow coefficient, identified particle yields, and mean transverse momenta. This work contributes to a more comprehensive understanding of heavy-ion collision dynamics and sets the stage for future improvements in theoretical modeling and experimental measurements.

\end{abstract}

\maketitle
\section{Introduction}\label{sec:introduction}
\noindent Experiments utilizing ultra-relativistic heavy-ion collisions (HIC) are crucial for improving understanding of Quantum Chromodynamics (QCD) in many-body systems \cite{Borsanyi:2013bia,HotQCD:2014kol,Braun-Munzinger:2015hba,Busza:2018rrf}. The high-energy collisions at facilities such as the Relativistic Heavy Ion Collider (RHIC) and Large Hadron Collider (LHC) enable the release of confined quarks and gluons within nuclei, causing a medium known as quark-gluon plasma (QGP) to be formed~\cite{Adams:2005dq,Adcox:2004mh,Arsene:2004fa,Back:2004je, ALICE:2022wpn, Akiba:2015jwa}. Multi-stage heavy-ion collision models, which include stages such as initial, pre-equilibrium, QGP, and hadron gas, have been instrumental in describing heavy-ion physics \cite{Bernhard2019}. Explicitly, the evolution of the QGP has been successfully expressed with the viscous relativistic hydrodynamics. Models applying hydrodynamics have been developed to the second-order in viscous corrections described by the first-order transport coefficients, the shear and bulk viscosity to entropy density ratios ($\eta/s$ and $\zeta/s$, respectively). Due to the low values of viscosity over entropy density — as suggested by the experimental measurements~\cite{Bernhard2019} --- the QGP produced in HICs is seen as the most perfect fluid in nature~\cite{Kovtun:2004de,Bazavov:2019lgz,Meyer:2007ic,Astrakhantsev:2017nrs,Meyer:2007dy,Astrakhantsev:2018oue}. However, studying it is challenging, as it exists in a strongly coupled regime where perturbative techniques are not applicable~\cite{Shuryak:2008eq}. In addition, the current non-perturbative techniques are limited to certain scenarios, making accurate experimental measurements vital for enriching the understanding of the QCD~\cite{Busza:2018rrf,Heinz:2013th}.

In these multi-stage models, the number of free parameters, including temperature-dependent ones, typically ranges from 10 to 20. For most of these parameters, the theory doesn't restrict them to exact values, and the values must be extracted by fitting to experimental observations such as particle yields, anisotropy in final particle distribution in momentum space, particle mean transverse momentum, etc \cite{Aamodt:2010cz, Abelev:2013vea, ALICE:2011ab}. This complex extraction process has made significant strides through the application of Bayesian analysis~\cite{Bernhard:2015hxa,Bernhard:2016bar,Bernhard:2016tnd,Bernhard:2018hnz,Bernhard2019,Auvinen:2020mpc,Nijs:2020ors,Nijs:2020roc,JETSCAPE:2020mzn}. Certain experimental observables, like symmetric cumulants, provide detailed insights into a system's collective evolution due to their sensitivity to the medium properties. 

Previous Bayesian analyses have been able to narrow down the possible values of several model parameters, especially those characterising the transport coefficients~\cite{Parkkila:2021tqq, Parkkila:2021yha, Bernhard2019}. Over a few years, the analyses have improved by including more observables that are independent of each other and more sensitive to different model parameters. To further improve the laid out constraints, in this paper, measurements from Au--Au collisions at RHIC and Pb--Pb collisions at the LHC across three different collision energies are included. Calibrating with systems varying up to an order of magnitude in energy, will allow one to probe the model's ability to adjust to different energy scales simultaneously. As such, this paper aims to provide a more accurate and general estimate for the viscosity over entropy density coefficients and quantify the sensitivity of anisotropic flow estimations to model parameters.

The paper is composed as follows. First, details of the analysis procedure are provided in Sec.~\ref{sec:bayesian}, and the experimental observables used in the analysis are described in Sec.~\ref{sec:obs}, along with a discussion on the suitability of the observables. Then, in Sec.~\ref{sec:results}, the results are presented, i.e., the posterior distributions and the final predictions on the observables. Finally, the observable sensitivity and remaining caveats are examined in Sec.~\ref{sec:discussion}, and the summary is given in Sec.~\ref{sec:summary}.

\section{Bayesian analysis}\label{sec:bayesian}
The Bayesian analysis has been shown to be a potent tool for extracting optimal model parameters from experimental measurements. Here, a brief overview of the procedure is given, and for more comprehensive details, readers are referred to Ref.~\cite{Bernhard:2018hnz}. In the following, the sets of model parameters and output observables are represented by vectors $\vec{x}$ and $\vec{y}$, respectively. The initial information about the parameter values is encoded into \textit{prior} distribution $P(\vec{x})$. Without presuming anything of the parameter values, the prior distributions are set as uniform distributions with intervals defined in Tab.~\ref{tab:design}. The distributions are then updated by calibrating with experimental measurements based on Bayes' theorem. The reconditioned, \textit{posterior}, distributions are, thus, obtained as $P(\vec{x}|\vec{y})\propto P(\vec{y}|\vec{x}) P(\vec{x})$. The probability $P(\vec{y}|\vec{x})$, i.e., the likelihood, is acquired by comparing outputs of the parameter space $\vec{x}$ with the experimental measurements $\vec{y}$. The extraction is carried out by employing a Markov Chain Monte Carlo (MCMC) model~\cite{foreman2013emcee} which probes the parameter phase space. Due to the computational cost of heavy-ion collision models, calculations are performed on 500 parameter design points distributed using a Latin hypercube scheme~\cite{TangHypercube,MORRIS1995381}. Based on these design points, a Gaussian process (GP)~\cite{10.5555/1162254} is trained to emulate the model in a continuous parameter phase space. With the emulator, results for parameter values between the calculated points can be evaluated without explicitly computing every possible combination. At each designed point, $3\times 10^6$ events are generated for the 5.02~TeV collision energy, $5\times 10^6$ for the 2.76 TeV, and $4\times 10^6$ for the 200~GeV including the ten samples of the hydrodynamic hypersurface.

\subsection{Hydrodynamic model and parameters}\label{subsec:hydro}
In the current investigation, the multi-stage model configuration is primarily same as that used in Refs.~\cite{Bernhard2019,Parkkila:2021tqq}, denoted as {\tt \trento{}+VISH(2+1)+UrQMD}. First, the description of the initial conditions are given by {\tt \trento{}} model~\cite{Moreland:2014oya}, which evaluates the initial entropy density as a generalized mean of the energy densities of the colliding nuclei. Then, at the pre-equilibrium stage, the initial stage is bridged to the QGP stage with free-streaming. The evolution of the system in the deconfined phase is simulated with a 2+1 causal hydrodynamic model, {\tt VISH2+1}~\cite{Shen:2014vra,Song:2007ux}. Finally, a particlization model transitions the partonic degrees of freedom to hadrons \cite{Pratt:2010jt,Bernhard:2018hnz} and the evolution in the hadron gas continues with the {\tt UrQMD} model~\cite{Bass:1998ca,Bleicher:1999xi}.

A hydrodynamic modelling relies on the energy and momentum conservation laws of the fluid dynamics. The conservation is expressed in terms of
\begin{equation}
	\label{eq:hydro1}
	\partial_\mu T^{\mu\nu}(x)=0,
\end{equation}
where $T^{\mu\nu}(x)$ is the energy-momentum tensor. In the case of viscous hydrodynamics, the energy-momentum tensor becomes
\begin{equation}
	\label{eq:hydro2}
	T^{\mu\nu}=\epsilon u^\mu u^\nu-(P+\Pi)\Delta^{\mu\nu}+\pi^{\mu\nu},
\end{equation}
where $\epsilon$ is the energy density, $P$ is the local pressure given by the equation of state, and $\Delta^{\mu\nu}=g^{\mu\nu}-u^\mu u^\nu$ is a projector onto the transverse four-velocity. The shear and bulk viscosities are represented by $\pi^{\mu\nu}$ and $\Pi$, respectively. This model has several free parameters, which include initial conditions, $\eta/s(T)$, and $\zeta/s(T)$. The model is characterized by a total of 16 parameters (see Tab.~\ref{tab:design}), controlling the model's main features. These parameters allow for the simultaneous characterization of the initial state and medium response. However, the parameter ranges for overall normalization vary according to the collision beam energy.

Compared to Refs.~\cite{Parkkila:2021tqq,Parkkila:2021yha}, the setup used in this analysis differs only in the way the centrality, which describes the collision geometry and event activity, is defined. It is calibrated individually for each parametrization by sorting the resulting events into centrality bins.

\subsection{Parameter ranges}\label{subsec:paramest_overview}
The full list of parameters used in the analysis are shown in Tab.~\ref{tab:design}. The framework attempts to find the best-fit values for the parameters, which are obtained from the posterior distribution. However, initial distributions for the parameters' values, called prior distributions, are required to extract initial information from the parameter values. Setting up pre-defined distributions inevitably allows for some form of bias to be formed. To reduce the bias, the prior distributions used for the parameters in this analysis are uniform distributions and cover a certain range of values. The ranges are shown on the final column in Tab.~\ref{tab:design}, and are taken to be same as those used in Ref.~\cite{Parkkila:2021tqq}. The range of an additional parameter, the {\tt \trento} normalization constant for Au--Au collisions at $\snn = 200$~GeV, is set from 3 to 10.
\begin{table*}[tbh!]
  \caption{
    \label{tab:design}
    Input parameter ranges for the initial condition and hydrodynamic models.
  }
  \begin{tabular}{|llc|}
    \hline
    Parameter         & Description                        & Prior range    \\
    \hline
    Norm(5.02~TeV)            & Overall normalization              & 16.542 -- 25  \\
    Norm(2.76~TeV)             & Overall normalization              & 11.152 -- 18.96  \\
    Norm(200~GeV)             & Overall normalization              & 3 -- 10  \\
    $p$               & Entropy deposition parameter       & 0.0042 -- 0.0098   \\
    $\sigma_k$ & Std. dev. of nucleon multiplicity fluctuations & 0.5508 -- 1.2852 \\
    $w$ & Nucleon width & 0.6692 -- 1.2428 \\
    $d_{\min}^3$ & Minimum volume per nucleon & $0.889^3$ -- $1.524^3$   \\
    $\tau_\mathrm{fs}$ & Free-streaming time & 0.03 -- 1.5  \\
    \hline
    $T_c$ & Temperature of const. $\eta/s(T)$, $T < T_c$ & 0.135 -- 0.165 \\
    $\eta/s(T_c)$ & Minimum $\eta/s(T)$  & 0 -- 0.2  \\
    $(\eta/s)_\mathrm{slope}$ & Slope of $\eta/s(T)$ above $T_c$ & 0 -- 4 \\
    $(\eta/s)_\mathrm{crv}$ & Curvature of $\eta/s(T)$ above $T_c$ & $-1.3$ -- $1$  \\
    $(\zeta/s)_\mathrm{peak}$  & Temperature of $\zeta/s(T)$ maximum & 0.15 -- 0.2  \\
    $(\zeta/s)_{\max}$ & Maximum $\zeta/s(T)$ & 0 -- 0.1  \\
    $(\zeta/s)_\mathrm{width}$ & Width of $\zeta/s(T)$ peak & 0 -- 0.1  \\
    $T_\mathrm{switch}$ & Switching / particlization temperature & 0.135 -- 0.165  \\
    \hline
  \end{tabular}
\end{table*}

\section{Experimental Observables}\label{sec:obs}
\begin{table*}[tbh!]
  \caption{
    \label{tab:observables}
    Experimental data included in Bayesian analysis. The observables in the brackets are excluded from the main analysis.
  }
  \begin{tabular}{|lccccc|}
    \hline
    Beam energy & Observable &  Particle species & Kinematic cuts & Centrality classes & Ref. \\
      \hline
    5.02~TeV & Yields $\mathrm{d}N/\mathrm{d}\eta$                       & $h^\pm$ &
    $|\eta| < 0.5$ & 0--5, 5--10, 10--20, \ldots, 50--60 & \cite{ALICE:2015juo} \\
    & Yields $\mathrm{d}N/\mathrm{d}y$                       & $p\bar p$ &
    $|y| < 0.5$ & 0--5, 5--10, 10--20, \ldots, 50--60 & \cite{Acharya:2019yoi} \\
     & Mean transverse momentum \meanpt & $\pi^\pm$, $K^\pm$, $p\bar p$ &
    $|y| < 0.5$ & 0--5, 5--10, 10--20, \ldots, 50--60 & \cite{Acharya:2019yoi} \\
     & Two-particle flow cumulants & $h^\pm$ &
    $|\eta| < 0.8$ & 0--5, 5--10, 10--20, \ldots, 50--60 &
    \cite{ Acharya:2020taj} \\
     & $v_2$, $v_3$, $v_4$, $v_5$, [$v_6$, $v_7$]  & & $0.2 <  p_\mathrm{T}< 5.0$ GeV &  & \\
     & Non-linear flow coefficients & $h^\pm$ &
    $|\eta| < 0.8$ & 0--5, 5--10, 10--20, \ldots, 50--60 &
    \cite{Acharya:2020taj} \\
     & $\chi_{4,22}$, $\chi_{5,23}$, [$\chi_{6,222}$, $\chi_{6,33}$]   & & $0.2 < p_\mathrm{T} < 5.0$ GeV &  & \\
     & Normalized Symmetric cumulants & $h^\pm$ &
    $|\eta| < 0.8$ & 0--5, 5--10, 10--20, \ldots, 50--60 &
    \cite{ALICE:2021adw} \\
     & NSC(3,2), NSC(4,2), [NSC(4,3)] & & $0.2 <  p_\mathrm{T} < 5.0$ GeV &  & \\
    & Symmetric Plane correlation & $h^\pm$ &
    $|\eta| < 0.8$ & 0--5, 5--10, 10--20, \ldots, 50--60 &
    \cite{ALICE:2021adw} \\
     & $\rho_{4,22}$, [$\rho_{5,23}$, $\rho_{6,222}$, $\rho_{6,33}$] & & $0.2 <  p_\mathrm{T} < 5.0$ GeV &  & \\
     
        \hline
        \hline
    2.76~TeV & Yields $\mathrm{d}N/\mathrm{d}\eta$                       & $h^\pm$ &
    $|\eta| < 0.5$ & 0--5, 5--10, 10--20, \ldots, 50--60 & \cite{ALICE:2013mez} \\
     & Mean transverse momentum \meanpt & $\pi^\pm$, $K^\pm$, $p\bar p$ &
    $|y| < 0.5$ & 0--5, 5--10, 10--20, \ldots, 50--60 & \cite{ALICE:2013mez} \\
     & Two-particle flow cumulants & $h^\pm$ &
    $|\eta| < 0.8$ & 0--5, 5--10, 10--20, \ldots, 50--60 &
    \cite{ALICE:2011ab,Acharya:2017gsw} \\
     &  $v_2$, $v_3$, $v_4$, [$v_5$] & & $0.2 <  p_\mathrm{T}< 5.0$ GeV &  & \\
     & Non-linear flow coefficients & $h^\pm$ &
    $|\eta| < 0.8$ & 0--5, 5--10, 10--20, \ldots, 50--60 &
    \cite{ALICE:2017fcd} \\
     & $\chi_{4,22}$, $\chi_{5,23}$, [$\chi_{6,222}$, $\chi_{6,33}$]  & & $0.2 < p_\mathrm{T} < 5.0$ GeV &  & \\
     & Normalized Symmetric cumulants & $h^\pm$ &
    $|\eta| < 0.8$ & 0--5, 5--10, 10--20, \ldots, 50--60 &
    \cite{ALICE:2016kpq,Acharya:2017gsw} \\
     & \tiny NSC(3,2), NSC(4,2), [NSC(4,3), NSC(2,3,4), NSC(2,3,5)] & & $0.2 <  p_\mathrm{T} < 5.0$ GeV &  & \\
    & Symmetric Plane correlation & $h^\pm$ &
    $|\eta| < 0.8$ & 0--5, 5--10, 10--20, \ldots, 50--60 &
    \cite{ALICE:2017fcd} \\
     & $\rho_{4,22}$, [$\rho_{5,23}$, $\rho_{6,222}$, $\rho_{6,33}$] & & $0.2 <  p_\mathrm{T} < 5.0$ GeV &  & \\
    \hline
    \hline
200~GeV & Yields $\mathrm{d}N/\mathrm{d}\eta$                       & $h^\pm$ &
    $|\eta| < 0.5$ & 0--5, 5--10, 10--20, \ldots, 50--60 & \cite{PHENIX:2003iij} \\
     & Mean transverse momentum \meanpt & $\pi^\pm$, $K^\pm$ &
    $|y| < 0.5$ & 0--5, 5--10, 10--20, \ldots, 50--60 & \cite{STAR:2008med} \\
     & Two-particle flow cumulants & $h^\pm$ &
    $|\eta| < 1.3,\, 1.0,\, 1.2$ & 0--5, 5--10, 10--20, \ldots, 50--60 &
    \cite{STAR:2003xyj, STAR:2004jwm, STAR:2013qio} \\
    & $v_2$, $v_3$, [$v_4$] & & $0.15 <  p_\mathrm{T}< 2.0$ GeV &  & \\
     & Non-linear flow coefficients & $h^\pm$ &
    $|\eta| < 1.0$ & 0--5, 5--10, 10--20, \ldots, 50--60 &
    \cite{STAR:2020gcl} \\
     & $\chi_{4,22}$, [$\chi_{5,23}$]  & & $0.2 < p_\mathrm{T} < 4.0$ GeV &  & \\
    & Normalized Symmetric cumulants & $h^\pm$ &
    $|\eta| < 1.0$ & 0--5, 5--10, 10--20, \ldots, 50--60 &
    \cite{STAR:2018fpo} \\
     & [NSC(3,2), NSC(4,2)] & & $0.15 <  p_\mathrm{T} < 2.0$ GeV &  & \\
    & Symmetric Plane correlation & $h^\pm$ &
    $|\eta| < 1.0$ & 0--5, 5--10, 10--20, \ldots, 50--60 &
    \cite{STAR:2020gcl} \\
     & $\rho_{4,22}$, [$\rho_{5,23}$] & & $0.2 <  p_\mathrm{T} < 4.0$ GeV &  & \\
    \hline
  \end{tabular}
\end{table*}
The anisotropic pressure-driven expansion of the QGP, referred to as anisotropic flow, can be characterized by a Fourier decomposition of the azimuthal particle distributions as
\begin{equation}
   \frac{\mathrm{d}N}{\mathrm{d}\phi} \propto 1 + 2 \sum_{n=1}^{\infty}v_n \cos\left(n(\phi-\psi_{n})\right), 
   \label{eq:fourier}
\end{equation} 
where $v_n$ and $\Psi_n$ denote the flow amplitudes and symmetry plane angles of the $n$-th harmonic, respectively. Correlations between flow magnitudes of different orders, arising from event-by-event flow fluctuations, are quantified using normalized symmetric cumulants. These are defined as $\mathrm{NSC}(m,n) = (\left<v_{m}^2v_{n}^2\right>-\left<v_{m}^2\right>\left<v_{n}^2\right>)/\left<v_{m}^2\right>\left<v_{n}^2\right>$~\cite{ALICE:2016kpq,ALICE:2021adw}. Furthermore, observables characterising non-linear flow $\chi_{n,mk}$ and correlations between symmetry planes $\rho_{n,mk}$ are used in the analysis. $\chi_{n,mk}$ quantifies the contribution of lower-order harmonic flows to higher-order harmonics (e.g., $\chi_{4,22}$ represents the non-linear contribution of $v_4$ originating from $v_2$; see~\cite{Acharya:2020taj} for details). The symmetry-plane correlations $\rho_{n,mk}$ measure the relationships between different-order symmetry planes; however, they are found to be biased by correlations between different-order flow amplitudes (see discussions in Sec.~\ref{subsec:sensitivity}). These additional observables provide enhanced sensitivity to medium properties and initial conditions, as demonstrated in Refs.~\cite{ALICE:2016kpq,Acharya:2017gsw,Acharya:2020taj,ALICE:2021adw}.

In Ref.~\cite{Parkkila:2021yha}, ALICE measurements in Pb--Pb collisions at \twosevensixnn and \fivenn were used for calibration. These include, in addition to the aforementioned flow observables, the centrality-dependent particle yield $\mathrm{d}N_\mathrm{ch}/\mathrm{d}y$ and mean transverse momentum \meanpt of identified particles as well as the charged hadron yields $N_\mathrm{ch}/\mathrm{d}\eta$. This analysis expands upon the previous study by incorporating an additional collision system: Au--Au at $\snn = 200$~GeV. The inclusion of a collision system with an order of magnitude lower energy than the LHC energies enable the capture of energy-dependent information within the framework. For this new system, same type of observables that were calculated in  Pb--Pb collisions are used for calibration. Table~\ref{tab:observables} provides a comprehensive list of these observables with further details, highlighting the vast energy range now covered in this analysis. The order of utilized observables vary for different collision systems, e.g., $v_n$ are measured with $n=[2-7]$ in \fivenn, $n=[2-5]$ in \twosevensixnn, and $n=[2-4]$ in $\sqrt{s_\mathrm{NN}}=200$~GeV. Here one needs to be cautious as some of the higher-order observables don't behave well in model calculations; their behavior is studied and discussed in Sec.~\ref{subsec:obs_choice} and in Appendix~\ref{sec:appendix_obs}. The orders of utilized observables and the selections are detailed in Tab.~\ref{tab:observables}. 

The observables are extracted with same methods that are applied in the experimental analyses in Refs.~\cite{ALICE:2016kpq,Acharya:2017gsw,Acharya:2020taj,ALICE:2021adw}. To ensure an internally consistent comparison, the centrality classes are chosen to match the centrality classes of the experimental data for which the multiplicity ranges have to be defined. In Ref.~\cite{Parkkila:2021yha}, this was carried out by using the MAP parametrization from~\cite{Bernhard2019} to simulate events and select the resulting minimum bias events based on charged-particle multiplicity $\mathrm{d}N_\mathrm{ch}/\mathrm{\eta}$. For this analysis, the centrality is calibrated separately for each parametrization to reduce multiplicity dependence of the results. To quantify the effect of centrality calibration, this analysis is repeated with the observable configuration from Ref.~\cite{Parkkila:2021yha}. The results are shown and discussed in the Appendix~\ref{sec:appendix_Cent}.

\subsection{The choice of observables}\label{subsec:obs_choice}
The Bayesian framework is optimized to give best fit model parameters given some data. The more data one has, the more improved model constraints one can get --- as long as the model has an ability to describe the data. This can be validated by inspecting the prior distributions of the observables and the emulator accuracy. The first check guarantees that only observables that can be produced reliably by the model are used. The second check ensures that the emulator can interpret the observable's relations to the parameters. These two checks tend to go hand in hand; if the model cannot produce distinct results for different parametrizations, then the emulator will not be able to do so either. Vice versa is not necessarily true as the model calculations don't depend on the emulator.

\begin{figure*}[t]
    \centering
    \includegraphics[width=0.49\textwidth]{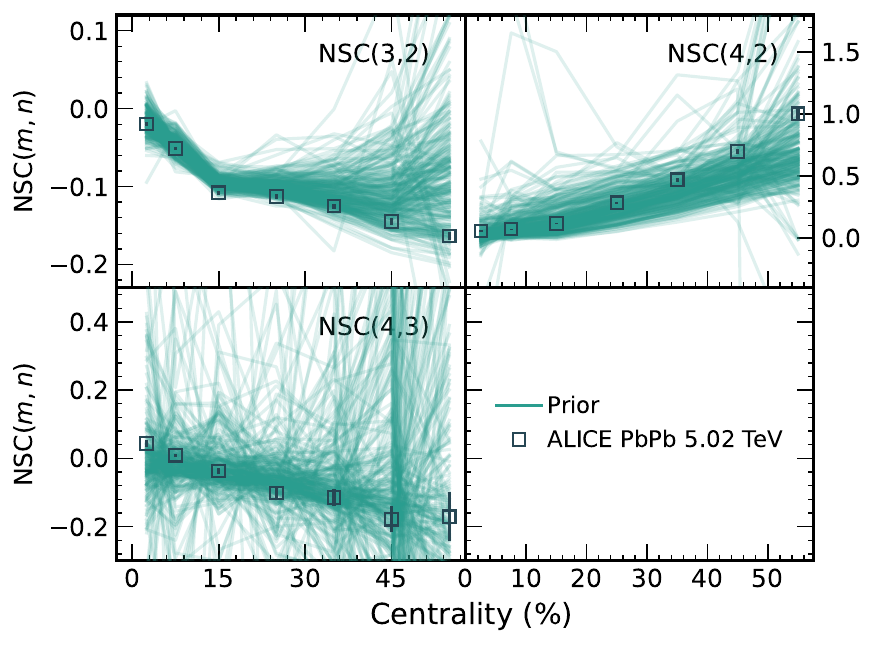}
    \includegraphics[width=0.49\textwidth]{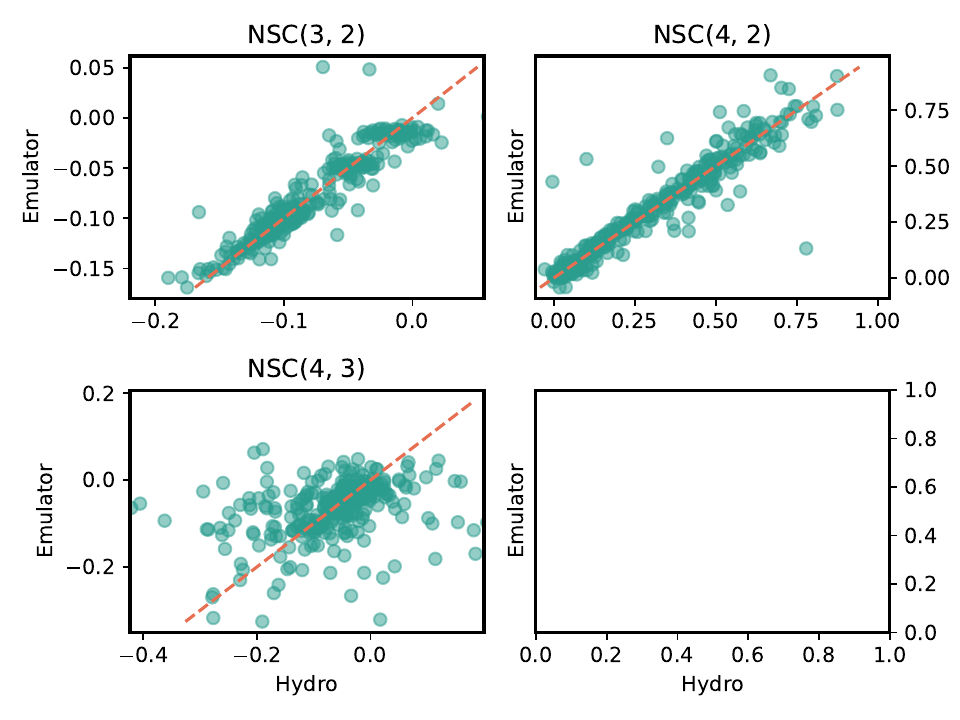}
    \caption{Prior distribution (left) and emulator validation (right) of $\mathrm{NSC}(m,n)$ model calculations in PbPb collisions at $\sqrt{s_\mathrm{NN}}= 5.02$ TeV shown on the right. The orange line on the right is the $y=x$ line.}
    \label{fig:nsc_prior5TeV}
\end{figure*}

As an example, in Fig.~\ref{fig:nsc_prior5TeV} on the left are shown the prior model calculations of NSC in Pb--Pb collisions at $\sqrt{s_\mathrm{NN}}= 5.02$ TeV. In the top panels, experimental results of NSC(3,2) and NSC(4,2) are well within the prior distribution. Moreover, the prior distribution is mostly well behaved. In comparison, the prior distribution of NSC(4,3) is much more spread out with strong peaks around centrality $40-50\%$. Nevertheless, the experimental result of NSC(4,3) is within the prior distribution. To make conclusive decisions one needs to also look at the emulator validation depicted in Fig.~\ref{fig:nsc_prior5TeV} on the right. Similarly to the prior distributions, NSC(3,2) and NSC(4,2) are behaving well as the emulator can make precise predictions. NSC(4,3) on the other hand is not captured by the emulator. Due to the mismatch in both, the prior distribution and the emulator validation, NSC(4,3) is not included in the current analysis. 

The suitability of an observable is quantified with a coefficient of variance of the prior distribution, and a Pearson correlation of the emulator validation. If the coefficient of variance is less than 1.0 then the observable is considered to be well described by the model. Furthermore, if the Pearson correlation between the model calculation and emulator estimation is greater than 0.7 then the emulator evaluation of the observable is considered valid. A detailed table and further discussion on the pragmatic choices of the observables can be found in the Appendix~\ref{sec:appendix_obs}. Additionally, a results comparison between pragmatically choosing observables and forcefully including all observables is studied. The observables that are excluded based on the criteria are shown in parentheses in Tab.~\ref{tab:observables}.

\section{\label{sec:results}Results}
\subsection{\label{subsec:posteriors}Posterior distribution}

\begin{figure}[h!]
    \centering
    \includegraphics[width=0.49\textwidth]{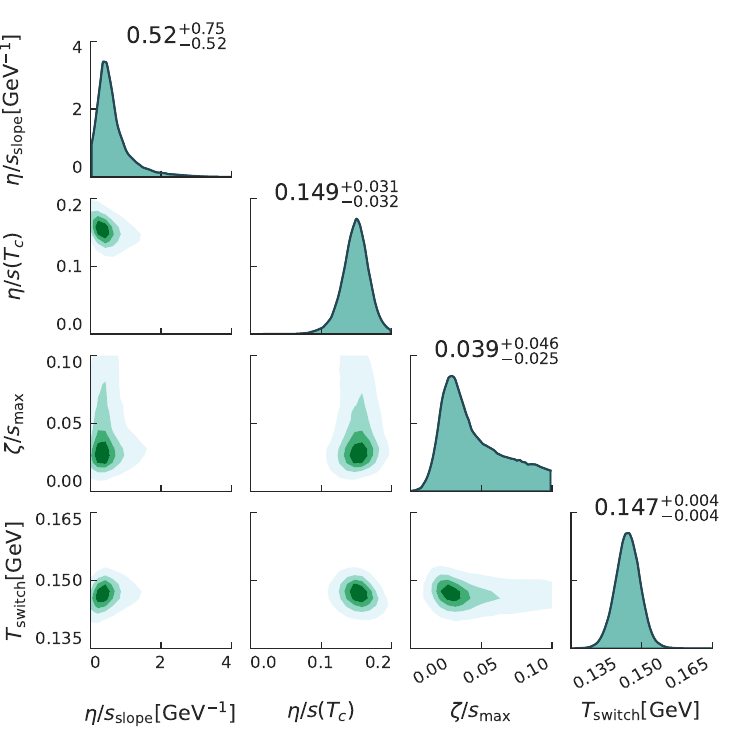}
    \caption{Selected posterior distributions of parameters defining transport coefficients and the switching temperature are shown. The shown distributions are well converged to narrow distributions. The minimum and maximum values of the x- and y-axes correspond to the prior ranges of the respective parameters.}
    \label{fig:posterior_small}
\end{figure}

\begin{figure}[h!]
    \centering
    \includegraphics[width=0.49\textwidth]{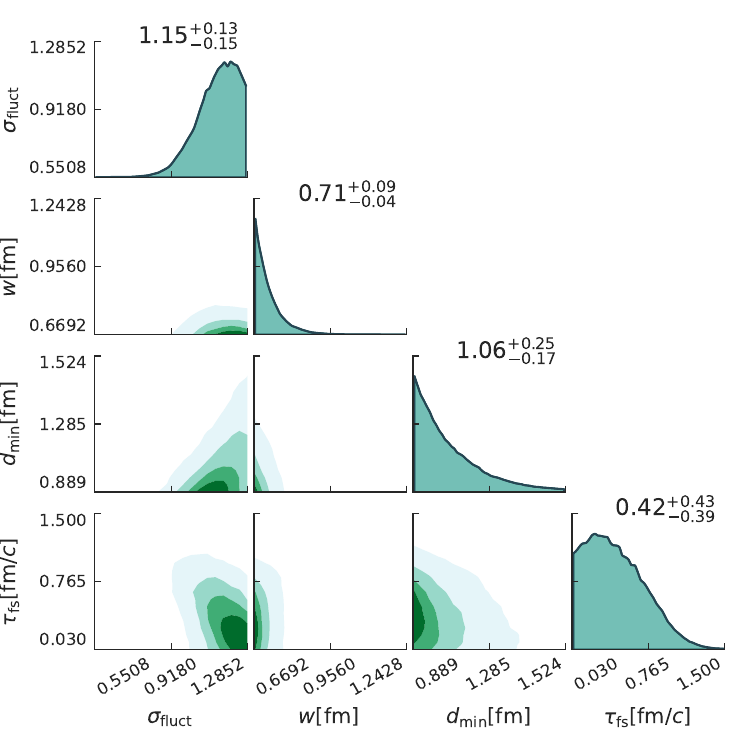}
    \caption{Selected posterior distributions of parameters defining initial condition are shown. The maxima of $w_\mathrm{nucleon}$ and $d_\mathrm{min}$ posteriors are at the boundary of their prior ranges. The minimum and maximum values of the x- and y-axes correspond to the prior ranges of the respective parameters.}
    \label{fig:posterior_ic}
\end{figure}

Figure~\ref{fig:posterior_small} presents the posterior distributions for a set of parameters constituting the transport properties as well as the switching temperature. A narrow, Gaussian-like distributions indicate that the parameters are well constrained. The full range of posterior distributions for all parameters can be found in the appendix in Fig.~\ref{fig:posteriorAvsB}.
%
\begin{table}[htbp]
\centering
\caption{MAP parameter values}
\label{tab:parameter_map}
\begin{tabular}{|l|r|}
\hline
Parameter & Value \\
\hline
Norm(5.02 TeV) & 18.9141 \\
Norm(2.76 TeV) & 14.8912 \\
Norm(200 GeV) & 5.9709 \\
$p$ & 0.0042 \\
$\sigma_k$ & 1.1078 \\
$w_\mathrm{nucleon}$ & 0.6692 \\
$d_{\min}^3$ & 0.7026 \\
$\tau_\mathrm{fs}$ & 0.4576 \\
$T_c$ & 0.1350 \\
$\eta/s(T_c)$ & 0.1578 \\
$(\eta/s)_\mathrm{slope}$ & 0.3426 \\
$(\eta/s)_\mathrm{crv}$ & 0.8476 \\
$(\zeta/s)_\mathrm{peak}$ & 0.1853 \\
$(\zeta/s)_{\max}$ & 0.0746 \\
$(\zeta/s)_\mathrm{width}$ & 0.0186 \\
$T_\mathrm{switch}$ & 0.1469 \\
\hline
\end{tabular}
\end{table}
The optimal model parameters are extracted at the maximum of the respective posterior distributions, denoted MAP (\textit{Maximum A Posteriori}) parameters. This set of parameters with the highest probability of reproducing the experimental data is presented in Tab.~\ref{tab:parameter_map}. Some of the resulting parameters differ from those obtained without the inclusion of the Au--Au collision data or the centrality calibration~\cite{Parkkila:2021yha}. The largest changes are seen in the nucleon width and in the minimum distance between the nucleons, where the new results are at the lower boundary of the corresponding parameter ranges. These posterior distributions are depicted in Fig.~\ref{fig:posterior_ic} along with the $\sigma_\mathrm{fluct}$ and $\tau_\mathrm{fs}$. The implication of the lower values of $w_\mathrm{nucleon}$ and $d_\mathrm{min}$ are further discussed in Sec.~\ref{subsec:issues}. 

\begin{figure}[ht]
    \centering
    \includegraphics[width=0.49\textwidth]{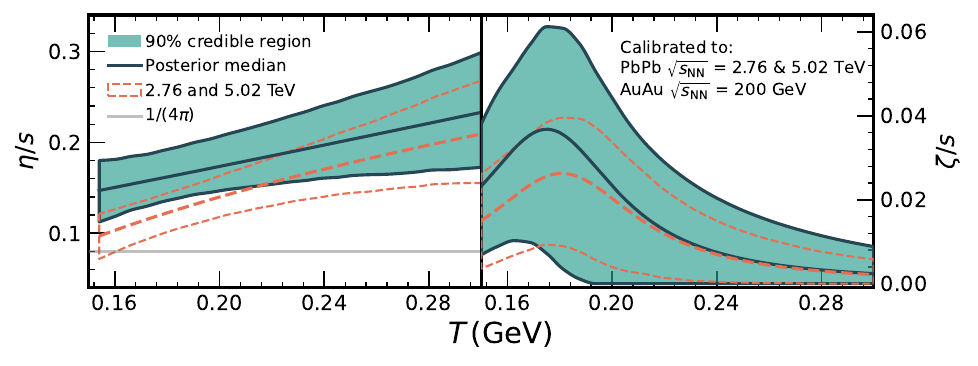}
    \caption{Credibility intervals for specific shear ($\eta/s$) and bulk ($\zeta$) viscosities as functions of temperature are shown in panels on the left and right, respectively. As a comparison, the previous Bayesian analysis (conducted with the two LHC energies $\sqrt{s_\mathrm{NN}}=2.76$ TeV and $\sqrt{s_\mathrm{NN}}=5.02$ TeV) results are shown with dashed lines.}
    \label{fig:params_transport}
\end{figure}

The composed parametrization for the transport properties is presented in Fig.~\ref{fig:params_transport} as the turquoise color band representing the combined 90\% credible interval. Here the general features of the parametrization include a moderate rising temperature dependence of the $\eta/s(T)$ as well as a relatively high $\zeta/s(T)$ peaking at around 180 MeV. There are some discrepancies when compared with the results~\cite{Parkkila:2021yha} obtained without the Au--Au data, shown as the orange dashed outline also representing the credible interval of the same limit. Compared to this previous result, $\eta/s(T)$ and $\zeta/s(T)$ are presenting higher values with larger uncertainties. For $\eta/s(T)$ the uncertainty is increased slightly whereas for $\zeta/s(T)$ the increase of uncertainty is more prominent. The general increase is due to the centrality calibration. Previous results that used only one centrality calibration were more biased favoring parametrizations closer to the calibration parametrization. Nonetheless, the results are well behaved demonstrating the potential for both, the model setup and its underlying framework, to adapt to multiple heavy-ion systems simultaneously.

\subsection{MAP predictions}\label{subsec:MAP_results}
\begin{figure}[ht]
    \centering
    \includegraphics[width=0.49\textwidth]{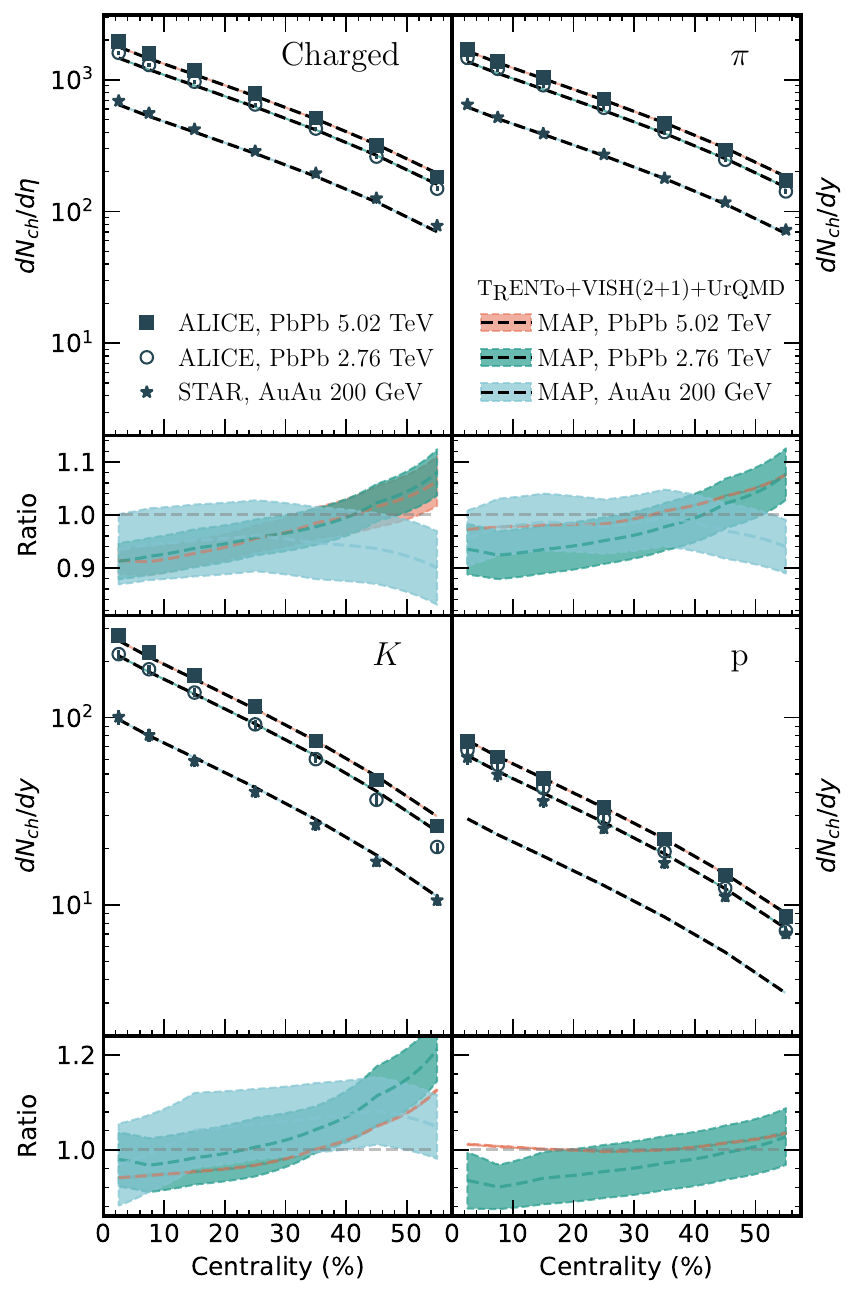}
    \caption{Model predictions with the MAP configuration are shown for the particle yield distributions of charged and identified particles as colored bands. The PbPb \fivenn, PbPb \twosevensixnn and AuAu $\snn = 200 $~GeV data points are presented as squares, open circles and stars, respectively. The ratios between the model prediction and the corresponding data (model/data) are shown in panels below graphs. }
    \label{fig:dndetamap}
\end{figure}

\begin{figure}[ht]
    \centering
    \includegraphics[width=0.49\textwidth]{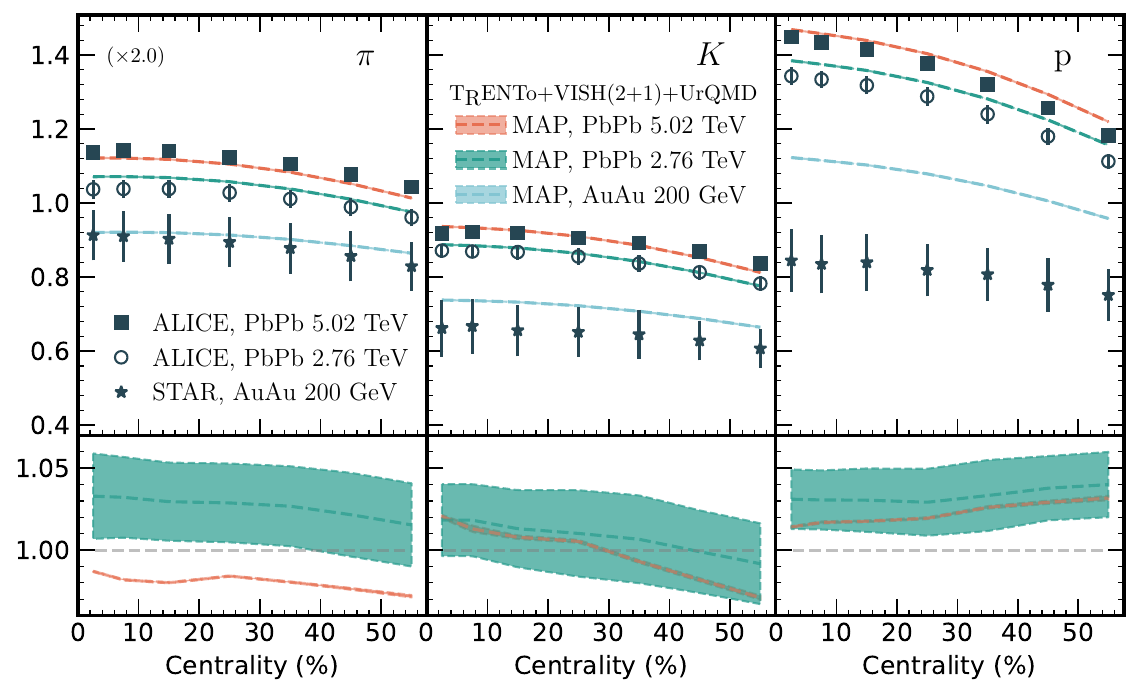}
    \caption{Model predictions with the MAP configuration are shown for the mean transverse momentum of charged and identified particles as colored bands. The PbPb \fivenn, PbPb \twosevensixnn and AuAu $\snn = 200 $~GeV data points are presented as squares, open circles and stars, respectively. The ratios between the model prediction and the corresponding data (model/data) are shown in panels below graphs. }
    \label{fig:meanptmap}
\end{figure}

After the MAP parameters are extracted, the different observables are produced with the respective parametrization in each collision system and compared with the data. In Fig.~\ref{fig:dndetamap}, the comparisons with the data are shown for the particle yields of charged and identified particles in all collision systems. For Pb--Pb systems the yields are underestimated in the central collisions and overestimated in the more peripheral collisions. In Au--Au collisions the predicted yields agree within uncertainties across all centrality intervals for all yields except for proton. The proton yields have been known to vary between measurements from STAR- and PHENIX-experiment which is the reason why proton measurements in Au--Au collisions at $\sqrt{s_\mathrm{NN}}=200$~GeV were not included in the analysis. The comparisons with the data for average transverse momenta of identified particles are shown in Fig.~\ref{fig:meanptmap}. The predictions show good agreement with the data for pion \meanpt in all collision systems. For Pb--Pb collisions the kaon and proton \meanpt are also well captured but for Au--Au collisions they are overestimated. For all identified particle \meanpt the Pb--Pb predictions agree with the data within 5\%. 

\begin{figure}[ht]
    \centering
    \includegraphics[width=0.49\textwidth]{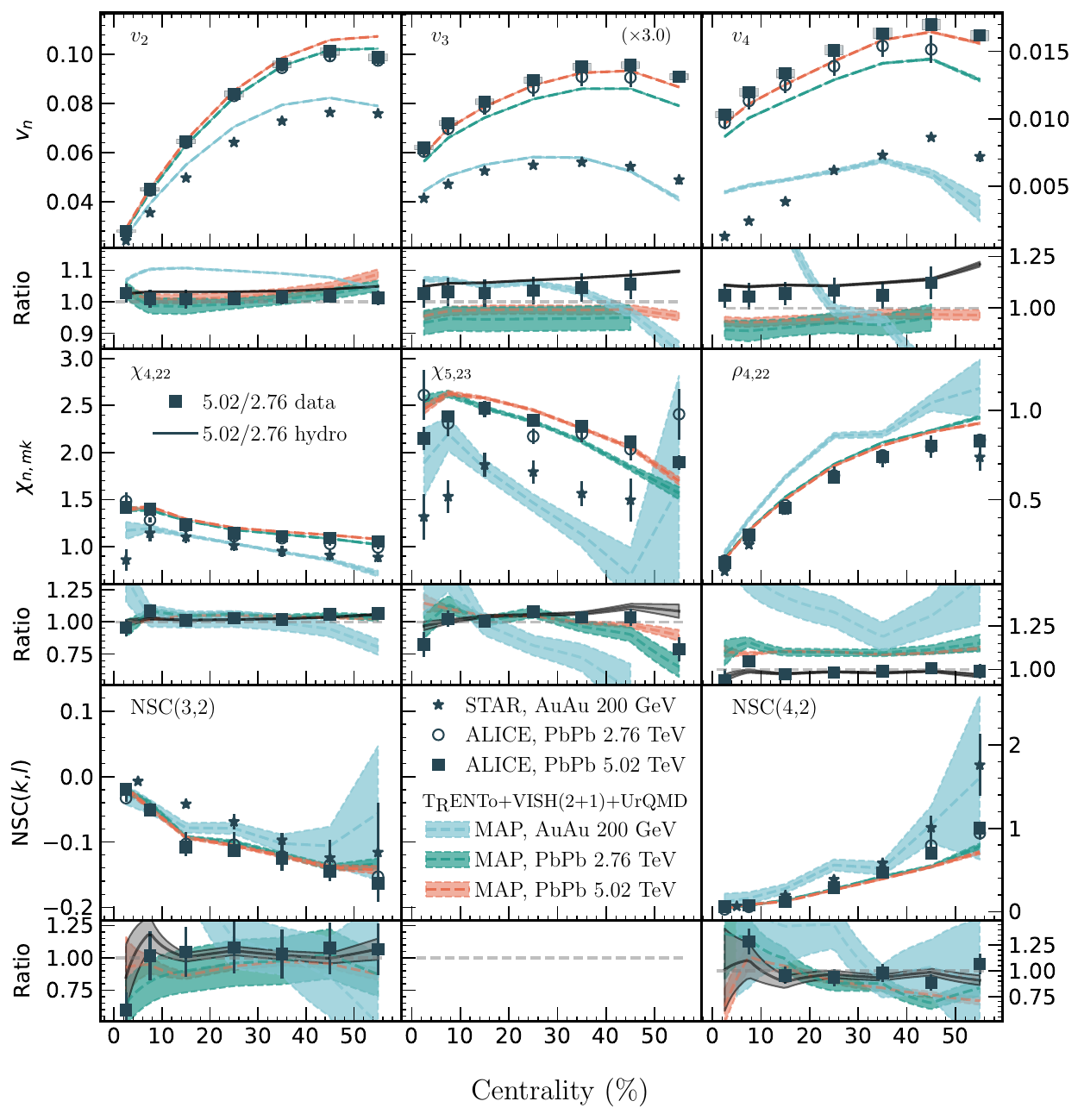}
    \caption{Model predictions with the MAP configuration for the lower order flow observables $v_2$-$v_4$, $\chi_{4,22}$-$\chi_{5,23}$, $\rho_{4,22}$, and NSC(3,2) and NSC(4,2). The Pb--Pb \fivenn data points are presented as black square markers, with a gray rectangle indicating the systematic uncertainty. The Pb--Pb \twosevensixnn and Au--Au $\sqrt{s_\mathrm{NN}} = 200$~GeV data points, denoted as open circles and black stars, respectively, are shown with combined statistical and systematic uncertainty.}
    \label{fig:obsmap}
\end{figure}

Figure~\ref{fig:obsmap} presents the calculations of the two-particle flow coefficients $v_n$ ($n=2,3$, and 4) as a function of centrality in Pb--Pb collisions at 2.76 and 5.02 TeV as well as  Au--Au collisions at 200 GeV. For Pb--Pb the $v_2$ is slightly overestimated in peripheral collisions but is well described in central collisions while the $v_3$ and $v_4$ are underestimated. For the Au--Au collisions, the MAP prediction overestimates the $v_2$ and $v_3$ in most centrality intervals, and is unable to capture the trend of $v_4$. In addition to $v_n$, the figure shows the MAP results of  NSC(2,3), NSC(4,2), $\chi_{4,22}$, $\chi_{5,23}$, and $\rho_{4,22}$. The Pb--Pb MAP predictions for these observables describe the data well with slight over- or underestimation in the most peripheral collisions. However, the MAP predictions for Au--Au collisions are quantitatively disagreeing with the data although the trends are generally captured. Notably interesting is that the MAP results capture the trend of NSC(3,2) and NSC(4,2) although they were not included in the analysis for Au--Au.  

\section{Discussions}\label{sec:discussion}
\subsection{Sensitivity analysis}\label{subsec:sensitivity}
\begin{figure}[h]
    \centering
    \includegraphics[width=0.49\textwidth]{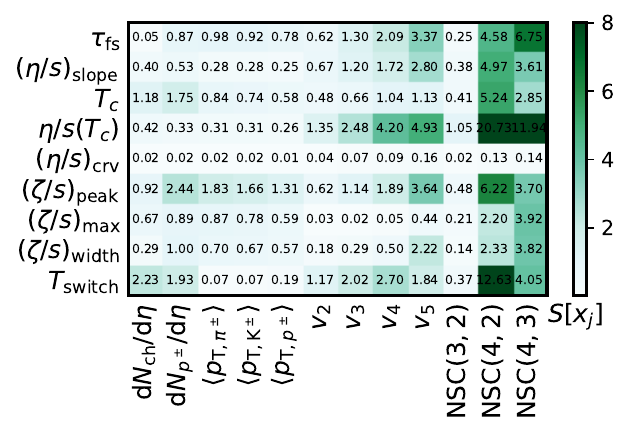}
    \caption{Quantified sensitivity of several observables used in the analysis. The strongest sensitivities are displayed by NSC to switching temperature, $\eta/s$ and $\zeta/s$.}
    \label{fig:sensitivity_obs0}
\end{figure}

\begin{figure}[h]
    \centering
    \includegraphics[width=0.49\textwidth]{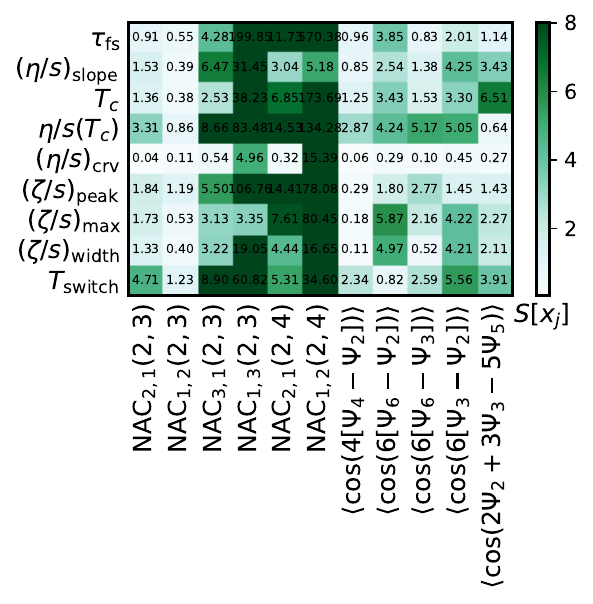}
    \caption{Quantified sensitivity of normalized asymmetric cumulants and unbiased symmetry-plane correlations. Strong sensitivity to the switching temperature, $\eta/s$ and $\zeta/s$ for most observables.}
    \label{fig:sensitivity_obs1}
\end{figure}

The sensitivity of observables to the model parameters is evaluated by following the suggestions from Refs.~\cite{JETSCAPE:2020mzn,Hamby:1994}, and is defined as $S[x_j]=|\hat{O}(\vec{x}')-\hat{O}(\vec{x})|/\delta \hat{O}(\vec{x})$. In the equation, $\hat{O}(\vec{x})$ is the value of an observable at a parameter point $\vec{x}=(x_1,\ldots,x_p)$ and $\vec{x}'$ is a point in the parameter space with a single parameter $x_j$ slightly varied, $\vec{x}'=(x_1,\ldots,(1+\delta)x_j,\ldots,x_p)$. The small change $\delta$ is set to 0.1, but different values of $\delta$ yield similar results. The results for observables used in the analysis are depicted in Fig.~\ref{fig:sensitivity_obs0}, while sensitivity of latest flow observables is shown in Fig.~\ref{fig:sensitivity_obs1}. The results in Fig.~\ref{fig:sensitivity_obs0} are congruent with those in Ref.~\cite{Parkkila:2021yha} with the normalized symmetric cumulants NSC$(m,n)$ being the most sensitive in comparison with the other observables. Furthermore, the sensitivity increases for higher-order harmonics and cumulants. In Refs.~\cite{Alver:2010dn,Teaney:2012ke,Gardim:2020mmy}, the higher harmonics were shown to be sensitive to the modifications of $\eta/s$. The reason for higher sensitivity is that the small variations in the values of $\eta/s$ and $\zeta/s$ will influence the hydrodynamic dissipation effects, which can dissolve the finer structures of the initial energy distribution. Therefore, higher harmonics, which capture these finer structures, will change drastically. Nonetheless, one needs to be cautious with the interpretations, as the emulator precision and model's ability to describe the higher-order observables may have an impact on the sensitivity.

Additionally, sensitivity of latest developments in experimental flow observables, normalized asymmetric cumulants $\mathrm{NAC}_{i,j}(m,n)$~\cite{Bilandzic:2021rgb, ALICE:2023lwx} and non-biased symmetry-plane correlations (SPC)~\cite{Bilandzic:2020csw, ALICE:2023wdn, ALICE:2024fus}, are shown in Fig.~\ref{fig:sensitivity_obs1}. The bias in the currently used symmetry-plane correlations $\rho_{n,mk}$ arises from correlation between different order flow harmonics. Similar to the NSC, the NAC also shows strong sensitivity to most of the transport parameters, averting the bias mentioned earlier. For the higher-order moment NAC, this may be caused by the emulator inaccuracy and model's limitations in describing them. Nevertheless, the lower-order moment NAC will be useful assets in future Bayesian analyses as they show strong sensitivity to $T_\mathrm{switch}$ and $\eta/s(T_c)$. The SPC shows less sensitivity overall when compared to NSC or NAC observables, but has relatively strong sensitivity to $\zeta/s$.      

\subsection{Remaining issues on the model}\label{subsec:issues}
\begin{figure}
    \centering
    \includegraphics[width=0.5\textwidth]{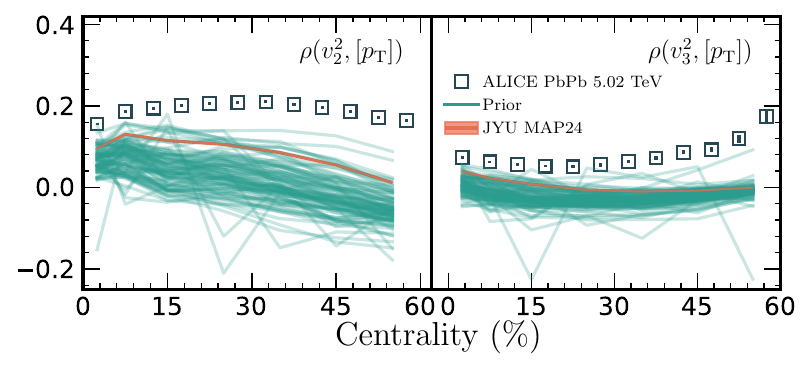}
    \caption{Prior distribution and MAP results of mean transverse momentum and flow coefficient correlation. The data, depicted as open squares, are obtained from Ref.~\cite{ALICE:2021gxt}.}
    \label{fig:pearsonRho}
\end{figure}

The model parametrization obtained in this analysis represents the optimal configurations for the given model and set of  observables. Usage of a certain set of models inevitably causes limitations that are bounded by the individual model's ability to describe reality. The model that is used in this analysis consists of three stages: the initial stage, the hydrodynamic phase and the final hadronic interactions. For the initial stage, {\tt \trento} model, which parametrizes the initial nucleon distributions and computes the initial entropy density with generalized mean, was used. This model is lightweight and flexible due to its many parameters, but has some caveats as well. These limitations arise from correlation measurements between fluctuations of mean transverse momentum and flow coefficient $\rho(v_n^2, [p_\mathrm{T}])$, which have been observed to be sensitive to the initial state~\cite{ALICE:2021gxt}. In Fig.~\ref{fig:pearsonRho} are shown the prior distributions and MAP predictions of $\rho(v_n^2, [p_\mathrm{T}])$ with $n=2,3$ in Pb--Pb collisions at \fivenn. It is evident that the prior distributions don't extend sufficiently to encompass the experimental results. On the other hand, within the prior range, the MAP parametrization favors values closer to the data even though this observable wasn't used in the analysis. The reason for the better agreement could be that the MAP value of $w_\mathrm{nucleon}$ is at lower end of the prior range. This is motivated by a requirement of nucleon width less than 0.7 fm, indicated by the recent studies on hadronic cross-section measurements~\cite{Nijs:2022rme}, which --- like $\rho(v_n^2, [p_\mathrm{T}])$ --- are sensitive to the initial conditions. Therefore, the prior range of $w_\mathrm{nucleon}$ needs to extended in future analysis, although the extension in itself cannot guarantee that the model can describe $\rho(v_n^2, [p_\mathrm{T}])$. An alternative approach would be to adopt another initial state model, such as IP-Glasma~\cite{Schenke:2012wb, Schenke:2012hg} or EKRT~\cite{Niemi:2015qia, Paatelainen:2013eea, Kuha:2024kmq}. These models also have fewer parameters than {\tt \trento} which would reduce the number of parameters that must be constrained. Moreover, this approach is supported by the model comparisons of $\rho(v_n^2, [p_\mathrm{T}])$, which showed that the frameworks using IP-Glasma provide a more accurate description of the data than those using {\tt \trento}~\cite{ALICE:2021gxt}.

The initial stage alone cannot fully account for the model's poor description of higher-order observables. Another reason for their discrepancy may be insufficient statistics in the model calculations. This analysis is calibrated with 3–5 million events per prior parametrization, potentially limiting the precision. Hence, increased computational resources could improve results. Beyond the limited statistics, possible factors contributing to model predictions are the finite grid size of the hydrodynamic simulation and viscous corrections added in Cooper-Frye integrals. The coarse grid may not be able to capture the finer details of the hydrodynamic evolution needed by the higher-order observables. Furthermore, during the Cooper-Frye freeze-out the system is considered to be close equilibrium and the viscous corrections are of linear-order in $\pi^{\mu,\nu}$~\cite{Shen:2014vra, Song:2007ux}. This approximation might be too strict to capture the effects on higher-order observables.

\section{Summary}\label{sec:summary}
In this analysis, for the first time, various flow observables from three different collision systems are utilized to extract optimal model parameters. Previous analyses~\cite{Parkkila:2021tqq, Parkkila:2021yha} have shown that including new observables sensitive to specific shear and bulk viscosity --- primarily reflecting non-linear hydrodynamic response --- improves uncertainty estimates. However, these studies only used measurements from two beam-energy Pb–Pb collisions at the LHC and relied on a single centrality calibration based on the MAP configuration from Ref.~\cite{Bernhard2019}. With the addition of Au--Au $\sqrt{s_\mathrm{NN}}$=200~GeV data and separate centrality calibration, the obtained posterior distributions of the transport coefficients are wider in comparison to the previous results. This is expected as the previous results may have favored parameter values close to those used for the calibration, which caused the narrower posteriors. Nonetheless, the new MAP predictions have better agreement in $v_2$, identified particle yields, and \meanpt compared to the previous results. The remaining observables have similar level of difference with respect to the previous results --- the flow coefficients in general now showing slightly smaller magnitudes. The agreement with data is worse for Au--Au than for Pb--Pb results, although the trends are well captured. 

There is still room for improvement with more statistics for higher-order observables to be included in the analysis and with varied selections of observables. Especially there is a need for the replacement of current symmetry-plane correlations with the unbiased version defined in Ref.~\cite{Bilandzic:2020csw} as well as inclusion of the asymmetric cumulants. Furthermore, the preferred values for the nucleon width, $w$ and the minimum distance between nucleons, $d_\mathrm{min}$ obtained here are at the boundary of the prior ranges calling for wider ranges. Hence, the prior ranges for these parameters need to be relaxed in future.  

To enhance the parameter estimation process, alternative initial state models with fewer parameters, such as IP-Glasma~\cite{Schenke:2012wb, Schenke:2012hg} or EKRT~\cite{Paatelainen:2013eea,Niemi:2015qia,Kuha:2024kmq} are suggested to be used. This approach aims to address limitations in the current model and potentially provide a more accurate description of experimental data. A careful model-dependent Bayesian analysis can reveal discrepancies between model predictions and experimental results, highlighting areas for improvement in both models and experimental precision. Nevertheless, the current work faces significant computational challenges as the evaluation of likelihood is computationally expensive, and calculating the evidence across models is particularly demanding~\cite{Jeffrey:2023stk}. The model evidence, which is crucial for Bayesian model comparison and selection, requires integrating the product of likelihood and prior probability over all possible parameter values. The complexity of these challenges calls for further developments in both theoretical frameworks and computational methods.

\FloatBarrier
\nocite{*}

\bibliography{apssamp}

\providecommand{\noopsort}[1]{}\providecommand{\singleletter}[1]{#1}%
\begin{thebibliography}{99}%
\makeatletter
\providecommand \@ifxundefined [1]{%
 \@ifx{#1\undefined}
}%
\providecommand \@ifnum [1]{%
 \ifnum #1\expandafter \@firstoftwo
 \else \expandafter \@secondoftwo
 \fi
}%
\providecommand \@ifx [1]{%
 \ifx #1\expandafter \@firstoftwo
 \else \expandafter \@secondoftwo
 \fi
}%
\providecommand \natexlab [1]{#1}%
\providecommand \enquote  [1]{``#1''}%
\providecommand \bibnamefont  [1]{#1}%
\providecommand \bibfnamefont [1]{#1}%
\providecommand \citenamefont [1]{#1}%
\providecommand \href@noop [0]{\@secondoftwo}%
\providecommand \href [0]{\begingroup \@sanitize@url \@href}%
\providecommand \@href[1]{\@@startlink{#1}\@@href}%
\providecommand \@@href[1]{\endgroup#1\@@endlink}%
\providecommand \@sanitize@url [0]{\catcode `\\12\catcode `\$12\catcode
  `\&12\catcode `\#12\catcode `\^12\catcode `\_12\catcode `\%12\relax}%
\providecommand \@@startlink[1]{}%
\providecommand \@@endlink[0]{}%
\providecommand \url  [0]{\begingroup\@sanitize@url \@url }%
\providecommand \@url [1]{\endgroup\@href {#1}{\urlprefix }}%
\providecommand \urlprefix  [0]{URL }%
\providecommand \Eprint [0]{\href }%
\providecommand \doibase [0]{http://dx.doi.org/}%
\providecommand \selectlanguage [0]{\@gobble}%
\providecommand \bibinfo  [0]{\@secondoftwo}%
\providecommand \bibfield  [0]{\@secondoftwo}%
\providecommand \translation [1]{[#1]}%
\providecommand \BibitemOpen [0]{}%
\providecommand \bibitemStop [0]{}%
\providecommand \bibitemNoStop [0]{.\EOS\space}%
\providecommand \EOS [0]{\spacefactor3000\relax}%
\providecommand \BibitemShut  [1]{\csname bibitem#1\endcsname}%
\let\auto@bib@innerbib\@empty
\bibitem [{\citenamefont {Borsanyi}\ \emph {et~al.}(2014)\citenamefont
  {Borsanyi}, \citenamefont {Fodor}, \citenamefont {Hoelbling}, \citenamefont
  {Katz}, \citenamefont {Krieg},\ and\ \citenamefont
  {Szabo}}]{Borsanyi:2013bia}%
  \BibitemOpen
  \bibfield  {author} {\bibinfo {author} {\bibfnamefont {S.}~\bibnamefont
  {Borsanyi}}, \bibinfo {author} {\bibfnamefont {Z.}~\bibnamefont {Fodor}},
  \bibinfo {author} {\bibfnamefont {C.}~\bibnamefont {Hoelbling}}, \bibinfo
  {author} {\bibfnamefont {S.~D.}\ \bibnamefont {Katz}}, \bibinfo {author}
  {\bibfnamefont {S.}~\bibnamefont {Krieg}}, \ and\ \bibinfo {author}
  {\bibfnamefont {K.~K.}\ \bibnamefont {Szabo}},\ }\href {\doibase
  10.1016/j.physletb.2014.01.007} {\bibfield  {journal} {\bibinfo  {journal}
  {Phys. Lett. B}\ }\textbf {\bibinfo {volume} {730}},\ \bibinfo {pages} {99}
  (\bibinfo {year} {2014})},\ \Eprint {http://arxiv.org/abs/1309.5258}
  {arXiv:1309.5258 [hep-lat]} \BibitemShut {NoStop}%
\bibitem [{\citenamefont {Bazavov}\ \emph {et~al.}(2014)\citenamefont {Bazavov}
  \emph {et~al.}}]{HotQCD:2014kol}%
  \BibitemOpen
  \bibfield  {author} {\bibinfo {author} {\bibfnamefont {A.}~\bibnamefont
  {Bazavov}} \emph {et~al.} (\bibinfo {collaboration} {HotQCD}),\ }\href
  {\doibase 10.1103/PhysRevD.90.094503} {\bibfield  {journal} {\bibinfo
  {journal} {Phys. Rev. D}\ }\textbf {\bibinfo {volume} {90}},\ \bibinfo
  {pages} {094503} (\bibinfo {year} {2014})},\ \Eprint
  {http://arxiv.org/abs/1407.6387} {arXiv:1407.6387 [hep-lat]} \BibitemShut
  {NoStop}%
\bibitem [{\citenamefont {Braun-Munzinger}\ \emph {et~al.}(2016)\citenamefont
  {Braun-Munzinger}, \citenamefont {Koch}, \citenamefont {Sch\"afer},\ and\
  \citenamefont {Stachel}}]{Braun-Munzinger:2015hba}%
  \BibitemOpen
  \bibfield  {author} {\bibinfo {author} {\bibfnamefont {P.}~\bibnamefont
  {Braun-Munzinger}}, \bibinfo {author} {\bibfnamefont {V.}~\bibnamefont
  {Koch}}, \bibinfo {author} {\bibfnamefont {T.}~\bibnamefont {Sch\"afer}}, \
  and\ \bibinfo {author} {\bibfnamefont {J.}~\bibnamefont {Stachel}},\ }\href
  {\doibase 10.1016/j.physrep.2015.12.003} {\bibfield  {journal} {\bibinfo
  {journal} {Phys. Rept.}\ }\textbf {\bibinfo {volume} {621}},\ \bibinfo
  {pages} {76} (\bibinfo {year} {2016})},\ \Eprint
  {http://arxiv.org/abs/1510.00442} {arXiv:1510.00442 [nucl-th]} \BibitemShut
  {NoStop}%
\bibitem [{\citenamefont {Busza}\ \emph {et~al.}(2018)\citenamefont {Busza},
  \citenamefont {Rajagopal},\ and\ \citenamefont {van~der
  Schee}}]{Busza:2018rrf}%
  \BibitemOpen
  \bibfield  {author} {\bibinfo {author} {\bibfnamefont {W.}~\bibnamefont
  {Busza}}, \bibinfo {author} {\bibfnamefont {K.}~\bibnamefont {Rajagopal}}, \
  and\ \bibinfo {author} {\bibfnamefont {W.}~\bibnamefont {van~der Schee}},\
  }\href {\doibase 10.1146/annurev-nucl-101917-020852} {\bibfield  {journal}
  {\bibinfo  {journal} {Ann. Rev. Nucl. Part. Sci.}\ }\textbf {\bibinfo
  {volume} {68}},\ \bibinfo {pages} {339} (\bibinfo {year} {2018})},\ \Eprint
  {http://arxiv.org/abs/1802.04801} {arXiv:1802.04801 [hep-ph]} \BibitemShut
  {NoStop}%
\bibitem [{\citenamefont {Adams}\ \emph
  {et~al.}(2005{\natexlab{a}})\citenamefont {Adams} \emph
  {et~al.}}]{Adams:2005dq}%
  \BibitemOpen
  \bibfield  {author} {\bibinfo {author} {\bibfnamefont {J.}~\bibnamefont
  {Adams}} \emph {et~al.} (\bibinfo {collaboration} {STAR}),\ }\href {\doibase
  10.1016/j.nuclphysa.2005.03.085} {\bibfield  {journal} {\bibinfo  {journal}
  {Nucl. Phys.}\ }\textbf {\bibinfo {volume} {A757}},\ \bibinfo {pages} {102}
  (\bibinfo {year} {2005}{\natexlab{a}})},\ \Eprint
  {http://arxiv.org/abs/nucl-ex/0501009} {arXiv:nucl-ex/0501009 [nucl-ex]}
  \BibitemShut {NoStop}%
\bibitem [{\citenamefont {Adcox}\ \emph {et~al.}(2005)\citenamefont {Adcox}
  \emph {et~al.}}]{Adcox:2004mh}%
  \BibitemOpen
  \bibfield  {author} {\bibinfo {author} {\bibfnamefont {K.}~\bibnamefont
  {Adcox}} \emph {et~al.} (\bibinfo {collaboration} {PHENIX}),\ }\href
  {\doibase 10.1016/j.nuclphysa.2005.03.086} {\bibfield  {journal} {\bibinfo
  {journal} {Nucl. Phys.}\ }\textbf {\bibinfo {volume} {A757}},\ \bibinfo
  {pages} {184} (\bibinfo {year} {2005})},\ \Eprint
  {http://arxiv.org/abs/nucl-ex/0410003} {arXiv:nucl-ex/0410003 [nucl-ex]}
  \BibitemShut {NoStop}%
\bibitem [{\citenamefont {Arsene}\ \emph {et~al.}(2005)\citenamefont {Arsene}
  \emph {et~al.}}]{Arsene:2004fa}%
  \BibitemOpen
  \bibfield  {author} {\bibinfo {author} {\bibfnamefont {I.}~\bibnamefont
  {Arsene}} \emph {et~al.} (\bibinfo {collaboration} {BRAHMS}),\ }\href
  {\doibase 10.1016/j.nuclphysa.2005.02.130} {\bibfield  {journal} {\bibinfo
  {journal} {Nucl. Phys.}\ }\textbf {\bibinfo {volume} {A757}},\ \bibinfo
  {pages} {1} (\bibinfo {year} {2005})},\ \Eprint
  {http://arxiv.org/abs/nucl-ex/0410020} {arXiv:nucl-ex/0410020 [nucl-ex]}
  \BibitemShut {NoStop}%
\bibitem [{\citenamefont {Back}\ \emph {et~al.}(2005)\citenamefont {Back} \emph
  {et~al.}}]{Back:2004je}%
  \BibitemOpen
  \bibfield  {author} {\bibinfo {author} {\bibfnamefont {B.~B.}\ \bibnamefont
  {Back}} \emph {et~al.} (\bibinfo {collaboration} {PHOBOS}),\ }\href {\doibase
  10.1016/j.nuclphysa.2005.03.084} {\bibfield  {journal} {\bibinfo  {journal}
  {Nucl. Phys.}\ }\textbf {\bibinfo {volume} {A757}},\ \bibinfo {pages} {28}
  (\bibinfo {year} {2005})},\ \Eprint {http://arxiv.org/abs/nucl-ex/0410022}
  {arXiv:nucl-ex/0410022 [nucl-ex]} \BibitemShut {NoStop}%
\bibitem [{\citenamefont {Acharya}\ \emph
  {et~al.}(2024{\natexlab{a}})\citenamefont {Acharya} \emph
  {et~al.}}]{ALICE:2022wpn}%
  \BibitemOpen
  \bibfield  {author} {\bibinfo {author} {\bibfnamefont {S.}~\bibnamefont
  {Acharya}} \emph {et~al.} (\bibinfo {collaboration} {ALICE}),\ }\href
  {\doibase 10.1140/epjc/s10052-024-12935-y} {\bibfield  {journal} {\bibinfo
  {journal} {Eur. Phys. J. C}\ }\textbf {\bibinfo {volume} {84}},\ \bibinfo
  {pages} {813} (\bibinfo {year} {2024}{\natexlab{a}})},\ \Eprint
  {http://arxiv.org/abs/2211.04384} {arXiv:2211.04384 [nucl-ex]} \BibitemShut
  {NoStop}%
\bibitem [{\citenamefont {Akiba}\ \emph {et~al.}(2015)\citenamefont {Akiba}
  \emph {et~al.}}]{Akiba:2015jwa}%
  \BibitemOpen
  \bibfield  {author} {\bibinfo {author} {\bibfnamefont {Y.}~\bibnamefont
  {Akiba}} \emph {et~al.},\ }\href@noop {} {\  (\bibinfo {year} {2015})},\
  \Eprint {http://arxiv.org/abs/1502.02730} {arXiv:1502.02730 [nucl-ex]}
  \BibitemShut {NoStop}%
\bibitem [{\citenamefont {Bernhard}\ \emph {et~al.}(2019)\citenamefont
  {Bernhard}, \citenamefont {Moreland},\ and\ \citenamefont
  {Bass}}]{Bernhard2019}%
  \BibitemOpen
  \bibfield  {author} {\bibinfo {author} {\bibfnamefont {J.~E.}\ \bibnamefont
  {Bernhard}}, \bibinfo {author} {\bibfnamefont {J.~S.}\ \bibnamefont
  {Moreland}}, \ and\ \bibinfo {author} {\bibfnamefont {S.~A.}\ \bibnamefont
  {Bass}},\ }\href {\doibase 10.1038/s41567-019-0611-8} {\bibfield  {journal}
  {\bibinfo  {journal} {Nature Physics}\ } (\bibinfo {year} {2019}),\
  10.1038/s41567-019-0611-8}\BibitemShut {NoStop}%
\bibitem [{\citenamefont {Kovtun}\ \emph {et~al.}(2005)\citenamefont {Kovtun},
  \citenamefont {Son},\ and\ \citenamefont {Starinets}}]{Kovtun:2004de}%
  \BibitemOpen
  \bibfield  {author} {\bibinfo {author} {\bibfnamefont {P.}~\bibnamefont
  {Kovtun}}, \bibinfo {author} {\bibfnamefont {D.~T.}\ \bibnamefont {Son}}, \
  and\ \bibinfo {author} {\bibfnamefont {A.~O.}\ \bibnamefont {Starinets}},\
  }\href {\doibase 10.1103/PhysRevLett.94.111601} {\bibfield  {journal}
  {\bibinfo  {journal} {Phys. Rev. Lett.}\ }\textbf {\bibinfo {volume} {94}},\
  \bibinfo {pages} {111601} (\bibinfo {year} {2005})},\ \Eprint
  {http://arxiv.org/abs/hep-th/0405231} {arXiv:hep-th/0405231 [hep-th]}
  \BibitemShut {NoStop}%
\bibitem [{\citenamefont {Bazavov}\ \emph {et~al.}(2019)\citenamefont
  {Bazavov}, \citenamefont {Karsch}, \citenamefont {Mukherjee},\ and\
  \citenamefont {Petreczky}}]{Bazavov:2019lgz}%
  \BibitemOpen
  \bibfield  {author} {\bibinfo {author} {\bibfnamefont {A.}~\bibnamefont
  {Bazavov}}, \bibinfo {author} {\bibfnamefont {F.}~\bibnamefont {Karsch}},
  \bibinfo {author} {\bibfnamefont {S.}~\bibnamefont {Mukherjee}}, \ and\
  \bibinfo {author} {\bibfnamefont {P.}~\bibnamefont {Petreczky}} (\bibinfo
  {collaboration} {USQCD}),\ }\href {\doibase 10.1140/epja/i2019-12922-0}
  {\bibfield  {journal} {\bibinfo  {journal} {Eur. Phys. J. A}\ }\textbf
  {\bibinfo {volume} {55}},\ \bibinfo {pages} {194} (\bibinfo {year} {2019})},\
  \Eprint {http://arxiv.org/abs/1904.09951} {arXiv:1904.09951 [hep-lat]}
  \BibitemShut {NoStop}%
\bibitem [{\citenamefont {Meyer}(2007)}]{Meyer:2007ic}%
  \BibitemOpen
  \bibfield  {author} {\bibinfo {author} {\bibfnamefont {H.~B.}\ \bibnamefont
  {Meyer}},\ }\href {\doibase 10.1103/PhysRevD.76.101701} {\bibfield  {journal}
  {\bibinfo  {journal} {Phys. Rev. D}\ }\textbf {\bibinfo {volume} {76}},\
  \bibinfo {pages} {101701} (\bibinfo {year} {2007})},\ \Eprint
  {http://arxiv.org/abs/0704.1801} {arXiv:0704.1801 [hep-lat]} \BibitemShut
  {NoStop}%
\bibitem [{\citenamefont {Astrakhantsev}\ \emph {et~al.}(2017)\citenamefont
  {Astrakhantsev}, \citenamefont {Braguta},\ and\ \citenamefont
  {Kotov}}]{Astrakhantsev:2017nrs}%
  \BibitemOpen
  \bibfield  {author} {\bibinfo {author} {\bibfnamefont {N.}~\bibnamefont
  {Astrakhantsev}}, \bibinfo {author} {\bibfnamefont {V.}~\bibnamefont
  {Braguta}}, \ and\ \bibinfo {author} {\bibfnamefont {A.}~\bibnamefont
  {Kotov}},\ }\href {\doibase 10.1007/JHEP04(2017)101} {\bibfield  {journal}
  {\bibinfo  {journal} {JHEP}\ }\textbf {\bibinfo {volume} {04}},\ \bibinfo
  {pages} {101} (\bibinfo {year} {2017})},\ \Eprint
  {http://arxiv.org/abs/1701.02266} {arXiv:1701.02266 [hep-lat]} \BibitemShut
  {NoStop}%
\bibitem [{\citenamefont {Meyer}(2008)}]{Meyer:2007dy}%
  \BibitemOpen
  \bibfield  {author} {\bibinfo {author} {\bibfnamefont {H.~B.}\ \bibnamefont
  {Meyer}},\ }\href {\doibase 10.1103/PhysRevLett.100.162001} {\bibfield
  {journal} {\bibinfo  {journal} {Phys. Rev. Lett.}\ }\textbf {\bibinfo
  {volume} {100}},\ \bibinfo {pages} {162001} (\bibinfo {year} {2008})},\
  \Eprint {http://arxiv.org/abs/0710.3717} {arXiv:0710.3717 [hep-lat]}
  \BibitemShut {NoStop}%
\bibitem [{\citenamefont {Astrakhantsev}\ \emph {et~al.}(2018)\citenamefont
  {Astrakhantsev}, \citenamefont {Braguta},\ and\ \citenamefont
  {Kotov}}]{Astrakhantsev:2018oue}%
  \BibitemOpen
  \bibfield  {author} {\bibinfo {author} {\bibfnamefont {N.~Y.}\ \bibnamefont
  {Astrakhantsev}}, \bibinfo {author} {\bibfnamefont {V.~V.}\ \bibnamefont
  {Braguta}}, \ and\ \bibinfo {author} {\bibfnamefont {A.~Y.}\ \bibnamefont
  {Kotov}},\ }\href {\doibase 10.1103/PhysRevD.98.054515} {\bibfield  {journal}
  {\bibinfo  {journal} {Phys. Rev. D}\ }\textbf {\bibinfo {volume} {98}},\
  \bibinfo {pages} {054515} (\bibinfo {year} {2018})},\ \Eprint
  {http://arxiv.org/abs/1804.02382} {arXiv:1804.02382 [hep-lat]} \BibitemShut
  {NoStop}%
\bibitem [{\citenamefont {Shuryak}(2009)}]{Shuryak:2008eq}%
  \BibitemOpen
  \bibfield  {author} {\bibinfo {author} {\bibfnamefont {E.}~\bibnamefont
  {Shuryak}},\ }\href {\doibase 10.1016/j.ppnp.2008.09.001} {\bibfield
  {journal} {\bibinfo  {journal} {Prog. Part. Nucl. Phys.}\ }\textbf {\bibinfo
  {volume} {62}},\ \bibinfo {pages} {48} (\bibinfo {year} {2009})},\ \Eprint
  {http://arxiv.org/abs/0807.3033} {arXiv:0807.3033 [hep-ph]} \BibitemShut
  {NoStop}%
\bibitem [{\citenamefont {Heinz}\ and\ \citenamefont
  {Snellings}(2013)}]{Heinz:2013th}%
  \BibitemOpen
  \bibfield  {author} {\bibinfo {author} {\bibfnamefont {U.}~\bibnamefont
  {Heinz}}\ and\ \bibinfo {author} {\bibfnamefont {R.}~\bibnamefont
  {Snellings}},\ }\href {\doibase 10.1146/annurev-nucl-102212-170540}
  {\bibfield  {journal} {\bibinfo  {journal} {Ann. Rev. Nucl. Part. Sci.}\
  }\textbf {\bibinfo {volume} {63}},\ \bibinfo {pages} {123} (\bibinfo {year}
  {2013})},\ \Eprint {http://arxiv.org/abs/1301.2826} {arXiv:1301.2826
  [nucl-th]} \BibitemShut {NoStop}%
\bibitem [{\citenamefont {Aamodt}\ \emph
  {et~al.}(2011{\natexlab{a}})\citenamefont {Aamodt} \emph
  {et~al.}}]{Aamodt:2010cz}%
  \BibitemOpen
  \bibfield  {author} {\bibinfo {author} {\bibfnamefont {K.}~\bibnamefont
  {Aamodt}} \emph {et~al.} (\bibinfo {collaboration} {ALICE}),\ }\href
  {\doibase 10.1103/PhysRevLett.106.032301} {\bibfield  {journal} {\bibinfo
  {journal} {Phys. Rev. Lett.}\ }\textbf {\bibinfo {volume} {106}},\ \bibinfo
  {pages} {032301} (\bibinfo {year} {2011}{\natexlab{a}})},\ \Eprint
  {http://arxiv.org/abs/1012.1657} {arXiv:1012.1657 [nucl-ex]} \BibitemShut
  {NoStop}%
\bibitem [{\citenamefont {Abelev}\ \emph
  {et~al.}(2013{\natexlab{a}})\citenamefont {Abelev} \emph
  {et~al.}}]{Abelev:2013vea}%
  \BibitemOpen
  \bibfield  {author} {\bibinfo {author} {\bibfnamefont {B.}~\bibnamefont
  {Abelev}} \emph {et~al.} (\bibinfo {collaboration} {ALICE}),\ }\href
  {\doibase 10.1103/PhysRevC.88.044910} {\bibfield  {journal} {\bibinfo
  {journal} {Phys. Rev. C}\ }\textbf {\bibinfo {volume} {88}},\ \bibinfo
  {pages} {044910} (\bibinfo {year} {2013}{\natexlab{a}})},\ \Eprint
  {http://arxiv.org/abs/1303.0737} {arXiv:1303.0737 [hep-ex]} \BibitemShut
  {NoStop}%
\bibitem [{\citenamefont {Aamodt}\ \emph
  {et~al.}(2011{\natexlab{b}})\citenamefont {Aamodt} \emph
  {et~al.}}]{ALICE:2011ab}%
  \BibitemOpen
  \bibfield  {author} {\bibinfo {author} {\bibfnamefont {K.}~\bibnamefont
  {Aamodt}} \emph {et~al.} (\bibinfo {collaboration} {ALICE}),\ }\href
  {\doibase 10.1103/PhysRevLett.107.032301} {\bibfield  {journal} {\bibinfo
  {journal} {Phys. Rev. Lett.}\ }\textbf {\bibinfo {volume} {107}},\ \bibinfo
  {pages} {032301} (\bibinfo {year} {2011}{\natexlab{b}})},\ \Eprint
  {http://arxiv.org/abs/1105.3865} {arXiv:1105.3865 [nucl-ex]} \BibitemShut
  {NoStop}%
\bibitem [{\citenamefont {Bernhard}\ \emph {et~al.}(2015)\citenamefont
  {Bernhard}, \citenamefont {Marcy}, \citenamefont {Coleman-Smith},
  \citenamefont {Huzurbazar}, \citenamefont {Wolpert},\ and\ \citenamefont
  {Bass}}]{Bernhard:2015hxa}%
  \BibitemOpen
  \bibfield  {author} {\bibinfo {author} {\bibfnamefont {J.~E.}\ \bibnamefont
  {Bernhard}}, \bibinfo {author} {\bibfnamefont {P.~W.}\ \bibnamefont {Marcy}},
  \bibinfo {author} {\bibfnamefont {C.~E.}\ \bibnamefont {Coleman-Smith}},
  \bibinfo {author} {\bibfnamefont {S.}~\bibnamefont {Huzurbazar}}, \bibinfo
  {author} {\bibfnamefont {R.~L.}\ \bibnamefont {Wolpert}}, \ and\ \bibinfo
  {author} {\bibfnamefont {S.~A.}\ \bibnamefont {Bass}},\ }\href {\doibase
  10.1103/PhysRevC.91.054910} {\bibfield  {journal} {\bibinfo  {journal} {Phys.
  Rev. C}\ }\textbf {\bibinfo {volume} {91}},\ \bibinfo {pages} {054910}
  (\bibinfo {year} {2015})},\ \Eprint {http://arxiv.org/abs/1502.00339}
  {arXiv:1502.00339 [nucl-th]} \BibitemShut {NoStop}%
\bibitem [{\citenamefont {Bernhard}\ \emph
  {et~al.}(2016{\natexlab{a}})\citenamefont {Bernhard}, \citenamefont
  {Moreland}, \citenamefont {Bass}, \citenamefont {Liu},\ and\ \citenamefont
  {Heinz}}]{Bernhard:2016bar}%
  \BibitemOpen
  \bibfield  {author} {\bibinfo {author} {\bibfnamefont {J.~E.}\ \bibnamefont
  {Bernhard}}, \bibinfo {author} {\bibfnamefont {J.~S.}\ \bibnamefont
  {Moreland}}, \bibinfo {author} {\bibfnamefont {S.~A.}\ \bibnamefont {Bass}},
  \bibinfo {author} {\bibfnamefont {J.}~\bibnamefont {Liu}}, \ and\ \bibinfo
  {author} {\bibfnamefont {U.}~\bibnamefont {Heinz}},\ }\href {\doibase
  10.1103/PhysRevC.94.024907} {\bibfield  {journal} {\bibinfo  {journal} {Phys.
  Rev. C}\ }\textbf {\bibinfo {volume} {94}},\ \bibinfo {pages} {024907}
  (\bibinfo {year} {2016}{\natexlab{a}})}\BibitemShut {NoStop}%
\bibitem [{\citenamefont {Bernhard}\ \emph
  {et~al.}(2016{\natexlab{b}})\citenamefont {Bernhard}, \citenamefont
  {Moreland}, \citenamefont {Bass}, \citenamefont {Liu},\ and\ \citenamefont
  {Heinz}}]{Bernhard:2016tnd}%
  \BibitemOpen
  \bibfield  {author} {\bibinfo {author} {\bibfnamefont {J.~E.}\ \bibnamefont
  {Bernhard}}, \bibinfo {author} {\bibfnamefont {J.~S.}\ \bibnamefont
  {Moreland}}, \bibinfo {author} {\bibfnamefont {S.~A.}\ \bibnamefont {Bass}},
  \bibinfo {author} {\bibfnamefont {J.}~\bibnamefont {Liu}}, \ and\ \bibinfo
  {author} {\bibfnamefont {U.}~\bibnamefont {Heinz}},\ }\href {\doibase
  10.1103/PhysRevC.94.024907} {\bibfield  {journal} {\bibinfo  {journal} {Phys.
  Rev.}\ }\textbf {\bibinfo {volume} {C94}},\ \bibinfo {pages} {024907}
  (\bibinfo {year} {2016}{\natexlab{b}})},\ \Eprint
  {http://arxiv.org/abs/1605.03954} {arXiv:1605.03954 [nucl-th]} \BibitemShut
  {NoStop}%
\bibitem [{\citenamefont {Bernhard}(2018)}]{Bernhard:2018hnz}%
  \BibitemOpen
  \bibfield  {author} {\bibinfo {author} {\bibfnamefont {J.~E.}\ \bibnamefont
  {Bernhard}},\ }\emph {\bibinfo {title} {{Bayesian parameter estimation for
  relativistic heavy-ion collisions}}},\ \href@noop {} {Ph.D. thesis},\
  \bibinfo  {school} {Duke U.} (\bibinfo {year} {2018}),\ \Eprint
  {http://arxiv.org/abs/1804.06469} {arXiv:1804.06469 [nucl-th]} \BibitemShut
  {NoStop}%
\bibitem [{\citenamefont {Auvinen}\ \emph {et~al.}(2020)\citenamefont
  {Auvinen}, \citenamefont {Eskola}, \citenamefont {Huovinen}, \citenamefont
  {Niemi}, \citenamefont {Paatelainen},\ and\ \citenamefont
  {Petreczky}}]{Auvinen:2020mpc}%
  \BibitemOpen
  \bibfield  {author} {\bibinfo {author} {\bibfnamefont {J.}~\bibnamefont
  {Auvinen}}, \bibinfo {author} {\bibfnamefont {K.~J.}\ \bibnamefont {Eskola}},
  \bibinfo {author} {\bibfnamefont {P.}~\bibnamefont {Huovinen}}, \bibinfo
  {author} {\bibfnamefont {H.}~\bibnamefont {Niemi}}, \bibinfo {author}
  {\bibfnamefont {R.}~\bibnamefont {Paatelainen}}, \ and\ \bibinfo {author}
  {\bibfnamefont {P.}~\bibnamefont {Petreczky}},\ }\href {\doibase
  10.1103/PhysRevC.102.044911} {\bibfield  {journal} {\bibinfo  {journal}
  {Phys. Rev. C}\ }\textbf {\bibinfo {volume} {102}},\ \bibinfo {pages}
  {044911} (\bibinfo {year} {2020})},\ \Eprint
  {http://arxiv.org/abs/2006.12499} {arXiv:2006.12499 [nucl-th]} \BibitemShut
  {NoStop}%
\bibitem [{\citenamefont {Nijs}\ \emph
  {et~al.}(2021{\natexlab{a}})\citenamefont {Nijs}, \citenamefont {van~der
  Schee}, \citenamefont {G\"ursoy},\ and\ \citenamefont
  {Snellings}}]{Nijs:2020ors}%
  \BibitemOpen
  \bibfield  {author} {\bibinfo {author} {\bibfnamefont {G.}~\bibnamefont
  {Nijs}}, \bibinfo {author} {\bibfnamefont {W.}~\bibnamefont {van~der Schee}},
  \bibinfo {author} {\bibfnamefont {U.}~\bibnamefont {G\"ursoy}}, \ and\
  \bibinfo {author} {\bibfnamefont {R.}~\bibnamefont {Snellings}},\ }\href
  {\doibase 10.1103/PhysRevLett.126.202301} {\bibfield  {journal} {\bibinfo
  {journal} {Phys. Rev. Lett.}\ }\textbf {\bibinfo {volume} {126}},\ \bibinfo
  {pages} {202301} (\bibinfo {year} {2021}{\natexlab{a}})},\ \Eprint
  {http://arxiv.org/abs/2010.15130} {arXiv:2010.15130 [nucl-th]} \BibitemShut
  {NoStop}%
\bibitem [{\citenamefont {Nijs}\ \emph
  {et~al.}(2021{\natexlab{b}})\citenamefont {Nijs}, \citenamefont {van~der
  Schee}, \citenamefont {G\"ursoy},\ and\ \citenamefont
  {Snellings}}]{Nijs:2020roc}%
  \BibitemOpen
  \bibfield  {author} {\bibinfo {author} {\bibfnamefont {G.}~\bibnamefont
  {Nijs}}, \bibinfo {author} {\bibfnamefont {W.}~\bibnamefont {van~der Schee}},
  \bibinfo {author} {\bibfnamefont {U.}~\bibnamefont {G\"ursoy}}, \ and\
  \bibinfo {author} {\bibfnamefont {R.}~\bibnamefont {Snellings}},\ }\href
  {\doibase 10.1103/PhysRevC.103.054909} {\bibfield  {journal} {\bibinfo
  {journal} {Phys. Rev. C}\ }\textbf {\bibinfo {volume} {103}},\ \bibinfo
  {pages} {054909} (\bibinfo {year} {2021}{\natexlab{b}})},\ \Eprint
  {http://arxiv.org/abs/2010.15134} {arXiv:2010.15134 [nucl-th]} \BibitemShut
  {NoStop}%
\bibitem [{\citenamefont {Everett}\ \emph {et~al.}(2021)\citenamefont {Everett}
  \emph {et~al.}}]{JETSCAPE:2020mzn}%
  \BibitemOpen
  \bibfield  {author} {\bibinfo {author} {\bibfnamefont {D.}~\bibnamefont
  {Everett}} \emph {et~al.} (\bibinfo {collaboration} {JETSCAPE}),\ }\href
  {\doibase 10.1103/PhysRevC.103.054904} {\bibfield  {journal} {\bibinfo
  {journal} {Phys. Rev. C}\ }\textbf {\bibinfo {volume} {103}},\ \bibinfo
  {pages} {054904} (\bibinfo {year} {2021})},\ \Eprint
  {http://arxiv.org/abs/2011.01430} {arXiv:2011.01430 [hep-ph]} \BibitemShut
  {NoStop}%
\bibitem [{\citenamefont {Parkkila}\ \emph {et~al.}(2021)\citenamefont
  {Parkkila}, \citenamefont {Onnerstad},\ and\ \citenamefont
  {Kim}}]{Parkkila:2021tqq}%
  \BibitemOpen
  \bibfield  {author} {\bibinfo {author} {\bibfnamefont {J.~E.}\ \bibnamefont
  {Parkkila}}, \bibinfo {author} {\bibfnamefont {A.}~\bibnamefont {Onnerstad}},
  \ and\ \bibinfo {author} {\bibfnamefont {D.~J.}\ \bibnamefont {Kim}},\ }\href
  {\doibase 10.1103/PhysRevC.104.054904} {\bibfield  {journal} {\bibinfo
  {journal} {Phys. Rev. C}\ }\textbf {\bibinfo {volume} {104}},\ \bibinfo
  {pages} {054904} (\bibinfo {year} {2021})},\ \Eprint
  {http://arxiv.org/abs/2106.05019} {arXiv:2106.05019 [hep-ph]} \BibitemShut
  {NoStop}%
\bibitem [{\citenamefont {Parkkila}\ \emph {et~al.}(2022)\citenamefont
  {Parkkila}, \citenamefont {Onnerstad}, \citenamefont {Taghavi}, \citenamefont
  {Mordasini}, \citenamefont {Bilandzic}, \citenamefont {Virta},\ and\
  \citenamefont {Kim}}]{Parkkila:2021yha}%
  \BibitemOpen
  \bibfield  {author} {\bibinfo {author} {\bibfnamefont {J.~E.}\ \bibnamefont
  {Parkkila}}, \bibinfo {author} {\bibfnamefont {A.}~\bibnamefont {Onnerstad}},
  \bibinfo {author} {\bibfnamefont {S.~F.}\ \bibnamefont {Taghavi}}, \bibinfo
  {author} {\bibfnamefont {C.}~\bibnamefont {Mordasini}}, \bibinfo {author}
  {\bibfnamefont {A.}~\bibnamefont {Bilandzic}}, \bibinfo {author}
  {\bibfnamefont {M.}~\bibnamefont {Virta}}, \ and\ \bibinfo {author}
  {\bibfnamefont {D.~J.}\ \bibnamefont {Kim}},\ }\href {\doibase
  10.1016/j.physletb.2022.137485} {\bibfield  {journal} {\bibinfo  {journal}
  {Phys. Lett. B}\ }\textbf {\bibinfo {volume} {835}},\ \bibinfo {pages}
  {137485} (\bibinfo {year} {2022})},\ \Eprint
  {http://arxiv.org/abs/2111.08145} {arXiv:2111.08145 [hep-ph]} \BibitemShut
  {NoStop}%
\bibitem [{\citenamefont {Foreman-Mackey}\ \emph {et~al.}(2013)\citenamefont
  {Foreman-Mackey}, \citenamefont {Hogg}, \citenamefont {Lang},\ and\
  \citenamefont {Goodman}}]{foreman2013emcee}%
  \BibitemOpen
  \bibfield  {author} {\bibinfo {author} {\bibfnamefont {D.}~\bibnamefont
  {Foreman-Mackey}}, \bibinfo {author} {\bibfnamefont {D.~W.}\ \bibnamefont
  {Hogg}}, \bibinfo {author} {\bibfnamefont {D.}~\bibnamefont {Lang}}, \ and\
  \bibinfo {author} {\bibfnamefont {J.}~\bibnamefont {Goodman}},\ }\href@noop
  {} {\bibfield  {journal} {\bibinfo  {journal} {Publications of the
  Astronomical Society of the Pacific}\ }\textbf {\bibinfo {volume} {125}},\
  \bibinfo {pages} {306} (\bibinfo {year} {2013})}\BibitemShut {NoStop}%
\bibitem [{\citenamefont {Tang}(1993)}]{TangHypercube}%
  \BibitemOpen
  \bibfield  {author} {\bibinfo {author} {\bibfnamefont {B.}~\bibnamefont
  {Tang}},\ }\href {http://www.jstor.org/stable/2291282} {\bibfield  {journal}
  {\bibinfo  {journal} {Journal of the American Statistical Association}\
  }\textbf {\bibinfo {volume} {88}},\ \bibinfo {pages} {1392} (\bibinfo {year}
  {1993})}\BibitemShut {NoStop}%
\bibitem [{\citenamefont {Morris}\ and\ \citenamefont
  {Mitchell}(1995)}]{MORRIS1995381}%
  \BibitemOpen
  \bibfield  {author} {\bibinfo {author} {\bibfnamefont {M.~D.}\ \bibnamefont
  {Morris}}\ and\ \bibinfo {author} {\bibfnamefont {T.~J.}\ \bibnamefont
  {Mitchell}},\ }\href {\doibase https://doi.org/10.1016/0378-3758(94)00035-T}
  {\bibfield  {journal} {\bibinfo  {journal} {Journal of Statistical Planning
  and Inference}\ }\textbf {\bibinfo {volume} {43}},\ \bibinfo {pages} {381}
  (\bibinfo {year} {1995})}\BibitemShut {NoStop}%
\bibitem [{\citenamefont {Rasmussen}\ and\ \citenamefont
  {Williams}(2005)}]{10.5555/1162254}%
  \BibitemOpen
  \bibfield  {author} {\bibinfo {author} {\bibfnamefont {C.~E.}\ \bibnamefont
  {Rasmussen}}\ and\ \bibinfo {author} {\bibfnamefont {C.~K.~I.}\ \bibnamefont
  {Williams}},\ }\href@noop {} {\emph {\bibinfo {title} {Gaussian Processes for
  Machine Learning (Adaptive Computation and Machine Learning)}}}\ (\bibinfo
  {publisher} {The MIT Press},\ \bibinfo {year} {2005})\BibitemShut {NoStop}%
\bibitem [{\citenamefont {Moreland}\ \emph {et~al.}(2015)\citenamefont
  {Moreland}, \citenamefont {Bernhard},\ and\ \citenamefont
  {Bass}}]{Moreland:2014oya}%
  \BibitemOpen
  \bibfield  {author} {\bibinfo {author} {\bibfnamefont {J.~S.}\ \bibnamefont
  {Moreland}}, \bibinfo {author} {\bibfnamefont {J.~E.}\ \bibnamefont
  {Bernhard}}, \ and\ \bibinfo {author} {\bibfnamefont {S.~A.}\ \bibnamefont
  {Bass}},\ }\href {\doibase 10.1103/PhysRevC.92.011901} {\bibfield  {journal}
  {\bibinfo  {journal} {Phys. Rev.}\ }\textbf {\bibinfo {volume} {C92}},\
  \bibinfo {pages} {011901} (\bibinfo {year} {2015})},\ \Eprint
  {http://arxiv.org/abs/1412.4708} {arXiv:1412.4708 [nucl-th]} \BibitemShut
  {NoStop}%
\bibitem [{\citenamefont {Shen}\ \emph {et~al.}(2016)\citenamefont {Shen},
  \citenamefont {Qiu}, \citenamefont {Song}, \citenamefont {Bernhard},
  \citenamefont {Bass},\ and\ \citenamefont {Heinz}}]{Shen:2014vra}%
  \BibitemOpen
  \bibfield  {author} {\bibinfo {author} {\bibfnamefont {C.}~\bibnamefont
  {Shen}}, \bibinfo {author} {\bibfnamefont {Z.}~\bibnamefont {Qiu}}, \bibinfo
  {author} {\bibfnamefont {H.}~\bibnamefont {Song}}, \bibinfo {author}
  {\bibfnamefont {J.}~\bibnamefont {Bernhard}}, \bibinfo {author}
  {\bibfnamefont {S.}~\bibnamefont {Bass}}, \ and\ \bibinfo {author}
  {\bibfnamefont {U.}~\bibnamefont {Heinz}},\ }\href {\doibase
  10.1016/j.cpc.2015.08.039} {\bibfield  {journal} {\bibinfo  {journal}
  {Comput. Phys. Commun.}\ }\textbf {\bibinfo {volume} {199}},\ \bibinfo
  {pages} {61} (\bibinfo {year} {2016})},\ \Eprint
  {http://arxiv.org/abs/1409.8164} {arXiv:1409.8164 [nucl-th]} \BibitemShut
  {NoStop}%
\bibitem [{\citenamefont {Song}\ and\ \citenamefont
  {Heinz}(2008)}]{Song:2007ux}%
  \BibitemOpen
  \bibfield  {author} {\bibinfo {author} {\bibfnamefont {H.}~\bibnamefont
  {Song}}\ and\ \bibinfo {author} {\bibfnamefont {U.~W.}\ \bibnamefont
  {Heinz}},\ }\href {\doibase 10.1103/PhysRevC.77.064901} {\bibfield  {journal}
  {\bibinfo  {journal} {Phys. Rev. C}\ }\textbf {\bibinfo {volume} {77}},\
  \bibinfo {pages} {064901} (\bibinfo {year} {2008})},\ \Eprint
  {http://arxiv.org/abs/0712.3715} {arXiv:0712.3715 [nucl-th]} \BibitemShut
  {NoStop}%
\bibitem [{\citenamefont {Pratt}\ and\ \citenamefont
  {Torrieri}(2010)}]{Pratt:2010jt}%
  \BibitemOpen
  \bibfield  {author} {\bibinfo {author} {\bibfnamefont {S.}~\bibnamefont
  {Pratt}}\ and\ \bibinfo {author} {\bibfnamefont {G.}~\bibnamefont
  {Torrieri}},\ }\href {\doibase 10.1103/PhysRevC.82.044901} {\bibfield
  {journal} {\bibinfo  {journal} {Phys. Rev. C}\ }\textbf {\bibinfo {volume}
  {82}},\ \bibinfo {pages} {044901} (\bibinfo {year} {2010})},\ \Eprint
  {http://arxiv.org/abs/1003.0413} {arXiv:1003.0413 [nucl-th]} \BibitemShut
  {NoStop}%
\bibitem [{\citenamefont {Bass}\ \emph {et~al.}(1998)\citenamefont {Bass} \emph
  {et~al.}}]{Bass:1998ca}%
  \BibitemOpen
  \bibfield  {author} {\bibinfo {author} {\bibfnamefont {S.~A.}\ \bibnamefont
  {Bass}} \emph {et~al.},\ }\href {\doibase 10.1016/S0146-6410(98)00058-1}
  {\bibfield  {journal} {\bibinfo  {journal} {Prog. Part. Nucl. Phys.}\
  }\textbf {\bibinfo {volume} {41}},\ \bibinfo {pages} {255} (\bibinfo {year}
  {1998})},\ \Eprint {http://arxiv.org/abs/nucl-th/9803035}
  {arXiv:nucl-th/9803035} \BibitemShut {NoStop}%
\bibitem [{\citenamefont {Bleicher}\ \emph {et~al.}(1999)\citenamefont
  {Bleicher} \emph {et~al.}}]{Bleicher:1999xi}%
  \BibitemOpen
  \bibfield  {author} {\bibinfo {author} {\bibfnamefont {M.}~\bibnamefont
  {Bleicher}} \emph {et~al.},\ }\href {\doibase 10.1088/0954-3899/25/9/308}
  {\bibfield  {journal} {\bibinfo  {journal} {J. Phys. G}\ }\textbf {\bibinfo
  {volume} {25}},\ \bibinfo {pages} {1859} (\bibinfo {year} {1999})},\ \Eprint
  {http://arxiv.org/abs/hep-ph/9909407} {arXiv:hep-ph/9909407} \BibitemShut
  {NoStop}%
\bibitem [{\citenamefont {Adam}\ \emph
  {et~al.}(2016{\natexlab{a}})\citenamefont {Adam} \emph
  {et~al.}}]{ALICE:2015juo}%
  \BibitemOpen
  \bibfield  {author} {\bibinfo {author} {\bibfnamefont {J.}~\bibnamefont
  {Adam}} \emph {et~al.} (\bibinfo {collaboration} {ALICE}),\ }\href {\doibase
  10.1103/PhysRevLett.116.222302} {\bibfield  {journal} {\bibinfo  {journal}
  {Phys. Rev. Lett.}\ }\textbf {\bibinfo {volume} {116}},\ \bibinfo {pages}
  {222302} (\bibinfo {year} {2016}{\natexlab{a}})},\ \Eprint
  {http://arxiv.org/abs/1512.06104} {arXiv:1512.06104 [nucl-ex]} \BibitemShut
  {NoStop}%
\bibitem [{\citenamefont {Acharya}\ \emph
  {et~al.}(2020{\natexlab{a}})\citenamefont {Acharya} \emph
  {et~al.}}]{Acharya:2019yoi}%
  \BibitemOpen
  \bibfield  {author} {\bibinfo {author} {\bibfnamefont {S.}~\bibnamefont
  {Acharya}} \emph {et~al.} (\bibinfo {collaboration} {ALICE}),\ }\href
  {\doibase 10.1103/PhysRevC.101.044907} {\bibfield  {journal} {\bibinfo
  {journal} {Phys. Rev. C}\ }\textbf {\bibinfo {volume} {101}},\ \bibinfo
  {pages} {044907} (\bibinfo {year} {2020}{\natexlab{a}})},\ \Eprint
  {http://arxiv.org/abs/1910.07678} {arXiv:1910.07678 [nucl-ex]} \BibitemShut
  {NoStop}%
\bibitem [{\citenamefont {Acharya}\ \emph
  {et~al.}(2020{\natexlab{b}})\citenamefont {Acharya} \emph
  {et~al.}}]{Acharya:2020taj}%
  \BibitemOpen
  \bibfield  {author} {\bibinfo {author} {\bibfnamefont {S.}~\bibnamefont
  {Acharya}} \emph {et~al.} (\bibinfo {collaboration} {ALICE}),\ }\href
  {\doibase 10.1007/JHEP05(2020)085} {\bibfield  {journal} {\bibinfo  {journal}
  {JHEP}\ }\textbf {\bibinfo {volume} {05}},\ \bibinfo {pages} {085} (\bibinfo
  {year} {2020}{\natexlab{b}})},\ \Eprint {http://arxiv.org/abs/2002.00633}
  {arXiv:2002.00633 [nucl-ex]} \BibitemShut {NoStop}%
\bibitem [{\citenamefont {Acharya}\ \emph
  {et~al.}(2021{\natexlab{a}})\citenamefont {Acharya} \emph
  {et~al.}}]{ALICE:2021adw}%
  \BibitemOpen
  \bibfield  {author} {\bibinfo {author} {\bibfnamefont {S.}~\bibnamefont
  {Acharya}} \emph {et~al.} (\bibinfo {collaboration} {ALICE}),\ }\href
  {\doibase 10.1016/j.physletb.2021.136354} {\bibfield  {journal} {\bibinfo
  {journal} {Phys. Lett. B}\ }\textbf {\bibinfo {volume} {818}},\ \bibinfo
  {pages} {136354} (\bibinfo {year} {2021}{\natexlab{a}})},\ \Eprint
  {http://arxiv.org/abs/2102.12180} {arXiv:2102.12180 [nucl-ex]} \BibitemShut
  {NoStop}%
\bibitem [{\citenamefont {Abelev}\ \emph
  {et~al.}(2013{\natexlab{b}})\citenamefont {Abelev} \emph
  {et~al.}}]{ALICE:2013mez}%
  \BibitemOpen
  \bibfield  {author} {\bibinfo {author} {\bibfnamefont {B.}~\bibnamefont
  {Abelev}} \emph {et~al.} (\bibinfo {collaboration} {ALICE}),\ }\href
  {\doibase 10.1103/PhysRevC.88.044910} {\bibfield  {journal} {\bibinfo
  {journal} {Phys. Rev. C}\ }\textbf {\bibinfo {volume} {88}},\ \bibinfo
  {pages} {044910} (\bibinfo {year} {2013}{\natexlab{b}})},\ \Eprint
  {http://arxiv.org/abs/1303.0737} {arXiv:1303.0737 [hep-ex]} \BibitemShut
  {NoStop}%
\bibitem [{\citenamefont {Acharya}\ \emph
  {et~al.}(2018{\natexlab{a}})\citenamefont {Acharya} \emph
  {et~al.}}]{Acharya:2017gsw}%
  \BibitemOpen
  \bibfield  {author} {\bibinfo {author} {\bibfnamefont {S.}~\bibnamefont
  {Acharya}} \emph {et~al.} (\bibinfo {collaboration} {ALICE}),\ }\href
  {\doibase 10.1103/PhysRevC.97.024906} {\bibfield  {journal} {\bibinfo
  {journal} {Phys. Rev.}\ }\textbf {\bibinfo {volume} {C97}},\ \bibinfo {pages}
  {024906} (\bibinfo {year} {2018}{\natexlab{a}})},\ \Eprint
  {http://arxiv.org/abs/1709.01127} {arXiv:1709.01127 [nucl-ex]} \BibitemShut
  {NoStop}%
\bibitem [{\citenamefont {Acharya}\ \emph {et~al.}(2017)\citenamefont {Acharya}
  \emph {et~al.}}]{ALICE:2017fcd}%
  \BibitemOpen
  \bibfield  {author} {\bibinfo {author} {\bibfnamefont {S.}~\bibnamefont
  {Acharya}} \emph {et~al.} (\bibinfo {collaboration} {ALICE}),\ }\href
  {\doibase 10.1016/j.physletb.2017.07.060} {\bibfield  {journal} {\bibinfo
  {journal} {Phys. Lett. B}\ }\textbf {\bibinfo {volume} {773}},\ \bibinfo
  {pages} {68} (\bibinfo {year} {2017})},\ \Eprint
  {http://arxiv.org/abs/1705.04377} {arXiv:1705.04377 [nucl-ex]} \BibitemShut
  {NoStop}%
\bibitem [{\citenamefont {Adam}\ \emph
  {et~al.}(2016{\natexlab{b}})\citenamefont {Adam} \emph
  {et~al.}}]{ALICE:2016kpq}%
  \BibitemOpen
  \bibfield  {author} {\bibinfo {author} {\bibfnamefont {J.}~\bibnamefont
  {Adam}} \emph {et~al.} (\bibinfo {collaboration} {ALICE}),\ }\href {\doibase
  10.1103/PhysRevLett.117.182301} {\bibfield  {journal} {\bibinfo  {journal}
  {Phys. Rev. Lett.}\ }\textbf {\bibinfo {volume} {117}},\ \bibinfo {pages}
  {182301} (\bibinfo {year} {2016}{\natexlab{b}})},\ \Eprint
  {http://arxiv.org/abs/1604.07663} {arXiv:1604.07663 [nucl-ex]} \BibitemShut
  {NoStop}%
\bibitem [{\citenamefont {Adler}\ \emph {et~al.}(2004)\citenamefont {Adler}
  \emph {et~al.}}]{PHENIX:2003iij}%
  \BibitemOpen
  \bibfield  {author} {\bibinfo {author} {\bibfnamefont {S.~S.}\ \bibnamefont
  {Adler}} \emph {et~al.} (\bibinfo {collaboration} {PHENIX}),\ }\href
  {\doibase 10.1103/PhysRevC.69.034909} {\bibfield  {journal} {\bibinfo
  {journal} {Phys. Rev. C}\ }\textbf {\bibinfo {volume} {69}},\ \bibinfo
  {pages} {034909} (\bibinfo {year} {2004})},\ \Eprint
  {http://arxiv.org/abs/nucl-ex/0307022} {arXiv:nucl-ex/0307022} \BibitemShut
  {NoStop}%
\bibitem [{\citenamefont {Abelev}\ \emph {et~al.}(2009)\citenamefont {Abelev}
  \emph {et~al.}}]{STAR:2008med}%
  \BibitemOpen
  \bibfield  {author} {\bibinfo {author} {\bibfnamefont {B.~I.}\ \bibnamefont
  {Abelev}} \emph {et~al.} (\bibinfo {collaboration} {STAR}),\ }\href {\doibase
  10.1103/PhysRevC.79.034909} {\bibfield  {journal} {\bibinfo  {journal} {Phys.
  Rev. C}\ }\textbf {\bibinfo {volume} {79}},\ \bibinfo {pages} {034909}
  (\bibinfo {year} {2009})},\ \Eprint {http://arxiv.org/abs/0808.2041}
  {arXiv:0808.2041 [nucl-ex]} \BibitemShut {NoStop}%
\bibitem [{\citenamefont {Adams}\ \emph {et~al.}(2004)\citenamefont {Adams}
  \emph {et~al.}}]{STAR:2003xyj}%
  \BibitemOpen
  \bibfield  {author} {\bibinfo {author} {\bibfnamefont {J.}~\bibnamefont
  {Adams}} \emph {et~al.} (\bibinfo {collaboration} {STAR}),\ }\href {\doibase
  10.1103/PhysRevLett.127.069901} {\bibfield  {journal} {\bibinfo  {journal}
  {Phys. Rev. Lett.}\ }\textbf {\bibinfo {volume} {92}},\ \bibinfo {pages}
  {062301} (\bibinfo {year} {2004})},\ \bibinfo {note} {[Erratum:
  Phys.Rev.Lett. 127, 069901 (2021)]},\ \Eprint
  {http://arxiv.org/abs/nucl-ex/0310029} {arXiv:nucl-ex/0310029} \BibitemShut
  {NoStop}%
\bibitem [{\citenamefont {Adams}\ \emph
  {et~al.}(2005{\natexlab{b}})\citenamefont {Adams} \emph
  {et~al.}}]{STAR:2004jwm}%
  \BibitemOpen
  \bibfield  {author} {\bibinfo {author} {\bibfnamefont {J.}~\bibnamefont
  {Adams}} \emph {et~al.} (\bibinfo {collaboration} {STAR}),\ }\href {\doibase
  10.1103/PhysRevC.72.014904} {\bibfield  {journal} {\bibinfo  {journal} {Phys.
  Rev. C}\ }\textbf {\bibinfo {volume} {72}},\ \bibinfo {pages} {014904}
  (\bibinfo {year} {2005}{\natexlab{b}})},\ \Eprint
  {http://arxiv.org/abs/nucl-ex/0409033} {arXiv:nucl-ex/0409033} \BibitemShut
  {NoStop}%
\bibitem [{\citenamefont {Adamczyk}\ \emph {et~al.}(2013)\citenamefont
  {Adamczyk} \emph {et~al.}}]{STAR:2013qio}%
  \BibitemOpen
  \bibfield  {author} {\bibinfo {author} {\bibfnamefont {L.}~\bibnamefont
  {Adamczyk}} \emph {et~al.} (\bibinfo {collaboration} {STAR}),\ }\href
  {\doibase 10.1103/PhysRevC.88.014904} {\bibfield  {journal} {\bibinfo
  {journal} {Phys. Rev. C}\ }\textbf {\bibinfo {volume} {88}},\ \bibinfo
  {pages} {014904} (\bibinfo {year} {2013})},\ \Eprint
  {http://arxiv.org/abs/1301.2187} {arXiv:1301.2187 [nucl-ex]} \BibitemShut
  {NoStop}%
\bibitem [{\citenamefont {Adam}\ \emph {et~al.}(2020)\citenamefont {Adam} \emph
  {et~al.}}]{STAR:2020gcl}%
  \BibitemOpen
  \bibfield  {author} {\bibinfo {author} {\bibfnamefont {J.}~\bibnamefont
  {Adam}} \emph {et~al.} (\bibinfo {collaboration} {STAR}),\ }\href {\doibase
  10.1016/j.physletb.2020.135728} {\bibfield  {journal} {\bibinfo  {journal}
  {Phys. Lett. B}\ }\textbf {\bibinfo {volume} {809}},\ \bibinfo {pages}
  {135728} (\bibinfo {year} {2020})},\ \Eprint
  {http://arxiv.org/abs/2006.13537} {arXiv:2006.13537 [nucl-ex]} \BibitemShut
  {NoStop}%
\bibitem [{\citenamefont {Adam}\ \emph {et~al.}(2018)\citenamefont {Adam} \emph
  {et~al.}}]{STAR:2018fpo}%
  \BibitemOpen
  \bibfield  {author} {\bibinfo {author} {\bibfnamefont {J.}~\bibnamefont
  {Adam}} \emph {et~al.} (\bibinfo {collaboration} {STAR}),\ }\href {\doibase
  10.1016/j.physletb.2018.05.076} {\bibfield  {journal} {\bibinfo  {journal}
  {Phys. Lett. B}\ }\textbf {\bibinfo {volume} {783}},\ \bibinfo {pages} {459}
  (\bibinfo {year} {2018})},\ \Eprint {http://arxiv.org/abs/1803.03876}
  {arXiv:1803.03876 [nucl-ex]} \BibitemShut {NoStop}%
\bibitem [{\citenamefont {Hamby}(1994)}]{Hamby:1994}%
  \BibitemOpen
  \bibfield  {author} {\bibinfo {author} {\bibfnamefont {D.}~\bibnamefont
  {Hamby}},\ }\href@noop {} {\bibfield  {journal} {\bibinfo  {journal} {Environ
  Monit Assess}\ }\textbf {\bibinfo {volume} {32}},\ \bibinfo {pages} {135}
  (\bibinfo {year} {1994})}\BibitemShut {NoStop}%
\bibitem [{\citenamefont {Alver}\ \emph {et~al.}(2010)\citenamefont {Alver},
  \citenamefont {Gombeaud}, \citenamefont {Luzum},\ and\ \citenamefont
  {Ollitrault}}]{Alver:2010dn}%
  \BibitemOpen
  \bibfield  {author} {\bibinfo {author} {\bibfnamefont {B.~H.}\ \bibnamefont
  {Alver}}, \bibinfo {author} {\bibfnamefont {C.}~\bibnamefont {Gombeaud}},
  \bibinfo {author} {\bibfnamefont {M.}~\bibnamefont {Luzum}}, \ and\ \bibinfo
  {author} {\bibfnamefont {J.-Y.}\ \bibnamefont {Ollitrault}},\ }\href
  {\doibase 10.1103/PhysRevC.82.034913} {\bibfield  {journal} {\bibinfo
  {journal} {Phys. Rev. C}\ }\textbf {\bibinfo {volume} {82}},\ \bibinfo
  {pages} {034913} (\bibinfo {year} {2010})},\ \Eprint
  {http://arxiv.org/abs/1007.5469} {arXiv:1007.5469 [nucl-th]} \BibitemShut
  {NoStop}%
\bibitem [{\citenamefont {Teaney}\ and\ \citenamefont
  {Yan}(2012)}]{Teaney:2012ke}%
  \BibitemOpen
  \bibfield  {author} {\bibinfo {author} {\bibfnamefont {D.}~\bibnamefont
  {Teaney}}\ and\ \bibinfo {author} {\bibfnamefont {L.}~\bibnamefont {Yan}},\
  }\href {\doibase 10.1103/PhysRevC.86.044908} {\bibfield  {journal} {\bibinfo
  {journal} {Phys. Rev. C}\ }\textbf {\bibinfo {volume} {86}},\ \bibinfo
  {pages} {044908} (\bibinfo {year} {2012})},\ \Eprint
  {http://arxiv.org/abs/1206.1905} {arXiv:1206.1905 [nucl-th]} \BibitemShut
  {NoStop}%
\bibitem [{\citenamefont {Gardim}\ and\ \citenamefont
  {Ollitrault}(2021)}]{Gardim:2020mmy}%
  \BibitemOpen
  \bibfield  {author} {\bibinfo {author} {\bibfnamefont {F.~G.}\ \bibnamefont
  {Gardim}}\ and\ \bibinfo {author} {\bibfnamefont {J.-Y.}\ \bibnamefont
  {Ollitrault}},\ }\href {\doibase 10.1103/PhysRevC.103.044907} {\bibfield
  {journal} {\bibinfo  {journal} {Phys. Rev. C}\ }\textbf {\bibinfo {volume}
  {103}},\ \bibinfo {pages} {044907} (\bibinfo {year} {2021})},\ \Eprint
  {http://arxiv.org/abs/2010.11919} {arXiv:2010.11919 [nucl-th]} \BibitemShut
  {NoStop}%
\bibitem [{\citenamefont {Bilandzic}\ \emph {et~al.}(2021)\citenamefont
  {Bilandzic}, \citenamefont {Lesch}, \citenamefont {Mordasini},\ and\
  \citenamefont {Taghavi}}]{Bilandzic:2021rgb}%
  \BibitemOpen
  \bibfield  {author} {\bibinfo {author} {\bibfnamefont {A.}~\bibnamefont
  {Bilandzic}}, \bibinfo {author} {\bibfnamefont {M.}~\bibnamefont {Lesch}},
  \bibinfo {author} {\bibfnamefont {C.}~\bibnamefont {Mordasini}}, \ and\
  \bibinfo {author} {\bibfnamefont {S.~F.}\ \bibnamefont {Taghavi}},\
  }\href@noop {} {\  (\bibinfo {year} {2021})},\ \Eprint
  {http://arxiv.org/abs/2101.05619} {arXiv:2101.05619 [physics.data-an]}
  \BibitemShut {NoStop}%
\bibitem [{\citenamefont {Acharya}\ \emph
  {et~al.}(2023{\natexlab{a}})\citenamefont {Acharya} \emph
  {et~al.}}]{ALICE:2023lwx}%
  \BibitemOpen
  \bibfield  {author} {\bibinfo {author} {\bibfnamefont {S.}~\bibnamefont
  {Acharya}} \emph {et~al.} (\bibinfo {collaboration} {ALICE}),\ }\href
  {\doibase 10.1103/PhysRevC.108.055203} {\bibfield  {journal} {\bibinfo
  {journal} {Phys. Rev. C}\ }\textbf {\bibinfo {volume} {108}},\ \bibinfo
  {pages} {055203} (\bibinfo {year} {2023}{\natexlab{a}})},\ \Eprint
  {http://arxiv.org/abs/2303.13414} {arXiv:2303.13414 [nucl-ex]} \BibitemShut
  {NoStop}%
\bibitem [{\citenamefont {Bilandzic}\ \emph {et~al.}(2020)\citenamefont
  {Bilandzic}, \citenamefont {Lesch},\ and\ \citenamefont
  {Taghavi}}]{Bilandzic:2020csw}%
  \BibitemOpen
  \bibfield  {author} {\bibinfo {author} {\bibfnamefont {A.}~\bibnamefont
  {Bilandzic}}, \bibinfo {author} {\bibfnamefont {M.}~\bibnamefont {Lesch}}, \
  and\ \bibinfo {author} {\bibfnamefont {S.~F.}\ \bibnamefont {Taghavi}},\
  }\href {\doibase 10.1103/PhysRevC.102.024910} {\bibfield  {journal} {\bibinfo
   {journal} {Phys. Rev. C}\ }\textbf {\bibinfo {volume} {102}},\ \bibinfo
  {pages} {024910} (\bibinfo {year} {2020})},\ \Eprint
  {http://arxiv.org/abs/2004.01066} {arXiv:2004.01066 [nucl-ex]} \BibitemShut
  {NoStop}%
\bibitem [{\citenamefont {Acharya}\ \emph
  {et~al.}(2023{\natexlab{b}})\citenamefont {Acharya} \emph
  {et~al.}}]{ALICE:2023wdn}%
  \BibitemOpen
  \bibfield  {author} {\bibinfo {author} {\bibfnamefont {S.}~\bibnamefont
  {Acharya}} \emph {et~al.} (\bibinfo {collaboration} {ALICE}),\ }\href
  {\doibase 10.1140/epjc/s10052-023-11658-w} {\bibfield  {journal} {\bibinfo
  {journal} {Eur. Phys. J. C}\ }\textbf {\bibinfo {volume} {83}},\ \bibinfo
  {pages} {576} (\bibinfo {year} {2023}{\natexlab{b}})},\ \Eprint
  {http://arxiv.org/abs/2302.01234} {arXiv:2302.01234 [nucl-ex]} \BibitemShut
  {NoStop}%
\bibitem [{\citenamefont {Acharya}\ \emph
  {et~al.}(2024{\natexlab{b}})\citenamefont {Acharya} \emph
  {et~al.}}]{ALICE:2024fus}%
  \BibitemOpen
  \bibfield  {author} {\bibinfo {author} {\bibfnamefont {S.}~\bibnamefont
  {Acharya}} \emph {et~al.} (\bibinfo {collaboration} {ALICE}),\ }\href@noop {}
  {\  (\bibinfo {year} {2024}{\natexlab{b}})},\ \Eprint
  {http://arxiv.org/abs/2409.04238} {arXiv:2409.04238 [nucl-ex]} \BibitemShut
  {NoStop}%
\bibitem [{\citenamefont {Acharya}\ \emph {et~al.}(2022)\citenamefont {Acharya}
  \emph {et~al.}}]{ALICE:2021gxt}%
  \BibitemOpen
  \bibfield  {author} {\bibinfo {author} {\bibfnamefont {S.}~\bibnamefont
  {Acharya}} \emph {et~al.} (\bibinfo {collaboration} {ALICE}),\ }\href
  {\doibase 10.1016/j.physletb.2022.137393} {\bibfield  {journal} {\bibinfo
  {journal} {Phys. Lett. B}\ }\textbf {\bibinfo {volume} {834}},\ \bibinfo
  {pages} {137393} (\bibinfo {year} {2022})},\ \Eprint
  {http://arxiv.org/abs/2111.06106} {arXiv:2111.06106 [nucl-ex]} \BibitemShut
  {NoStop}%
\bibitem [{\citenamefont {Nijs}\ and\ \citenamefont {van~der
  Schee}(2022)}]{Nijs:2022rme}%
  \BibitemOpen
  \bibfield  {author} {\bibinfo {author} {\bibfnamefont {G.}~\bibnamefont
  {Nijs}}\ and\ \bibinfo {author} {\bibfnamefont {W.}~\bibnamefont {van~der
  Schee}},\ }\href {\doibase 10.1103/PhysRevLett.129.232301} {\bibfield
  {journal} {\bibinfo  {journal} {Phys. Rev. Lett.}\ }\textbf {\bibinfo
  {volume} {129}},\ \bibinfo {pages} {232301} (\bibinfo {year} {2022})},\
  \Eprint {http://arxiv.org/abs/2206.13522} {arXiv:2206.13522 [nucl-th]}
  \BibitemShut {NoStop}%
\bibitem [{\citenamefont {Schenke}\ \emph
  {et~al.}(2012{\natexlab{a}})\citenamefont {Schenke}, \citenamefont
  {Tribedy},\ and\ \citenamefont {Venugopalan}}]{Schenke:2012wb}%
  \BibitemOpen
  \bibfield  {author} {\bibinfo {author} {\bibfnamefont {B.}~\bibnamefont
  {Schenke}}, \bibinfo {author} {\bibfnamefont {P.}~\bibnamefont {Tribedy}}, \
  and\ \bibinfo {author} {\bibfnamefont {R.}~\bibnamefont {Venugopalan}},\
  }\href {\doibase 10.1103/PhysRevLett.108.252301} {\bibfield  {journal}
  {\bibinfo  {journal} {Phys. Rev. Lett.}\ }\textbf {\bibinfo {volume} {108}},\
  \bibinfo {pages} {252301} (\bibinfo {year} {2012}{\natexlab{a}})},\ \Eprint
  {http://arxiv.org/abs/1202.6646} {arXiv:1202.6646 [nucl-th]} \BibitemShut
  {NoStop}%
\bibitem [{\citenamefont {Schenke}\ \emph
  {et~al.}(2012{\natexlab{b}})\citenamefont {Schenke}, \citenamefont
  {Tribedy},\ and\ \citenamefont {Venugopalan}}]{Schenke:2012hg}%
  \BibitemOpen
  \bibfield  {author} {\bibinfo {author} {\bibfnamefont {B.}~\bibnamefont
  {Schenke}}, \bibinfo {author} {\bibfnamefont {P.}~\bibnamefont {Tribedy}}, \
  and\ \bibinfo {author} {\bibfnamefont {R.}~\bibnamefont {Venugopalan}},\
  }\href {\doibase 10.1103/PhysRevC.86.034908} {\bibfield  {journal} {\bibinfo
  {journal} {Phys. Rev. C}\ }\textbf {\bibinfo {volume} {86}},\ \bibinfo
  {pages} {034908} (\bibinfo {year} {2012}{\natexlab{b}})},\ \Eprint
  {http://arxiv.org/abs/1206.6805} {arXiv:1206.6805 [hep-ph]} \BibitemShut
  {NoStop}%
\bibitem [{\citenamefont {Niemi}\ \emph {et~al.}(2016)\citenamefont {Niemi},
  \citenamefont {Eskola},\ and\ \citenamefont {Paatelainen}}]{Niemi:2015qia}%
  \BibitemOpen
  \bibfield  {author} {\bibinfo {author} {\bibfnamefont {H.}~\bibnamefont
  {Niemi}}, \bibinfo {author} {\bibfnamefont {K.~J.}\ \bibnamefont {Eskola}}, \
  and\ \bibinfo {author} {\bibfnamefont {R.}~\bibnamefont {Paatelainen}},\
  }\href {\doibase 10.1103/PhysRevC.93.024907} {\bibfield  {journal} {\bibinfo
  {journal} {Phys. Rev. C}\ }\textbf {\bibinfo {volume} {93}},\ \bibinfo
  {pages} {024907} (\bibinfo {year} {2016})},\ \Eprint
  {http://arxiv.org/abs/1505.02677} {arXiv:1505.02677 [hep-ph]} \BibitemShut
  {NoStop}%
\bibitem [{\citenamefont {Paatelainen}\ \emph {et~al.}(2014)\citenamefont
  {Paatelainen}, \citenamefont {Eskola}, \citenamefont {Niemi},\ and\
  \citenamefont {Tuominen}}]{Paatelainen:2013eea}%
  \BibitemOpen
  \bibfield  {author} {\bibinfo {author} {\bibfnamefont {R.}~\bibnamefont
  {Paatelainen}}, \bibinfo {author} {\bibfnamefont {K.~J.}\ \bibnamefont
  {Eskola}}, \bibinfo {author} {\bibfnamefont {H.}~\bibnamefont {Niemi}}, \
  and\ \bibinfo {author} {\bibfnamefont {K.}~\bibnamefont {Tuominen}},\ }\href
  {\doibase 10.1016/j.physletb.2014.02.018} {\bibfield  {journal} {\bibinfo
  {journal} {Phys. Lett. B}\ }\textbf {\bibinfo {volume} {731}},\ \bibinfo
  {pages} {126} (\bibinfo {year} {2014})},\ \Eprint
  {http://arxiv.org/abs/1310.3105} {arXiv:1310.3105 [hep-ph]} \BibitemShut
  {NoStop}%
\bibitem [{\citenamefont {Kuha}\ \emph {et~al.}(2024)\citenamefont {Kuha},
  \citenamefont {Auvinen}, \citenamefont {Eskola}, \citenamefont {Hirvonen},
  \citenamefont {Kanakubo},\ and\ \citenamefont {Niemi}}]{Kuha:2024kmq}%
  \BibitemOpen
  \bibfield  {author} {\bibinfo {author} {\bibfnamefont {M.}~\bibnamefont
  {Kuha}}, \bibinfo {author} {\bibfnamefont {J.}~\bibnamefont {Auvinen}},
  \bibinfo {author} {\bibfnamefont {K.~J.}\ \bibnamefont {Eskola}}, \bibinfo
  {author} {\bibfnamefont {H.}~\bibnamefont {Hirvonen}}, \bibinfo {author}
  {\bibfnamefont {Y.}~\bibnamefont {Kanakubo}}, \ and\ \bibinfo {author}
  {\bibfnamefont {H.}~\bibnamefont {Niemi}},\ }\href@noop {} {\  (\bibinfo
  {year} {2024})},\ \Eprint {http://arxiv.org/abs/2406.17592} {arXiv:2406.17592
  [hep-ph]} \BibitemShut {NoStop}%
\bibitem [{\citenamefont {Jeffrey}\ and\ \citenamefont
  {Wandelt}(2024)}]{Jeffrey:2023stk}%
  \BibitemOpen
  \bibfield  {author} {\bibinfo {author} {\bibfnamefont {N.}~\bibnamefont
  {Jeffrey}}\ and\ \bibinfo {author} {\bibfnamefont {B.~D.}\ \bibnamefont
  {Wandelt}},\ }\href {\doibase 10.1088/2632-2153/ad1a4d} {\bibfield  {journal}
  {\bibinfo  {journal} {Mach. Learn. Sci. Tech.}\ }\textbf {\bibinfo {volume}
  {5}},\ \bibinfo {pages} {015008} (\bibinfo {year} {2024})},\ \Eprint
  {http://arxiv.org/abs/2305.11241} {arXiv:2305.11241 [cs.LG]} \BibitemShut
  {NoStop}%
\bibitem [{\citenamefont {Abelev}\ \emph
  {et~al.}(2013{\natexlab{c}})\citenamefont {Abelev} \emph
  {et~al.}}]{ALICE:2013tla}%
  \BibitemOpen
  \bibfield  {author} {\bibinfo {author} {\bibfnamefont {B.}~\bibnamefont
  {Abelev}} \emph {et~al.} (\bibinfo {collaboration} {ALICE}),\ }\href
  {\doibase 10.1007/JHEP09(2013)049} {\bibfield  {journal} {\bibinfo  {journal}
  {JHEP}\ }\textbf {\bibinfo {volume} {09}},\ \bibinfo {pages} {049} (\bibinfo
  {year} {2013}{\natexlab{c}})},\ \Eprint {http://arxiv.org/abs/1307.1249}
  {arXiv:1307.1249 [nucl-ex]} \BibitemShut {NoStop}%
\bibitem [{\citenamefont {Abelev}\ \emph {et~al.}(2015)\citenamefont {Abelev}
  \emph {et~al.}}]{ALICE:2014mas}%
  \BibitemOpen
  \bibfield  {author} {\bibinfo {author} {\bibfnamefont {B.}~\bibnamefont
  {Abelev}} \emph {et~al.} (\bibinfo {collaboration} {ALICE}),\ }\href
  {\doibase 10.1016/j.physletb.2014.11.028} {\bibfield  {journal} {\bibinfo
  {journal} {Phys. Lett. B}\ }\textbf {\bibinfo {volume} {741}},\ \bibinfo
  {pages} {38} (\bibinfo {year} {2015})},\ \Eprint
  {http://arxiv.org/abs/1406.5463} {arXiv:1406.5463 [nucl-ex]} \BibitemShut
  {NoStop}%
\bibitem [{\citenamefont {Borghini}\ \emph
  {et~al.}(2001{\natexlab{a}})\citenamefont {Borghini}, \citenamefont {Dinh},\
  and\ \citenamefont {Ollitrault}}]{Borghini:2001vi}%
  \BibitemOpen
  \bibfield  {author} {\bibinfo {author} {\bibfnamefont {N.}~\bibnamefont
  {Borghini}}, \bibinfo {author} {\bibfnamefont {P.~M.}\ \bibnamefont {Dinh}},
  \ and\ \bibinfo {author} {\bibfnamefont {J.-Y.}\ \bibnamefont {Ollitrault}},\
  }\href {\doibase 10.1103/PhysRevC.64.054901} {\bibfield  {journal} {\bibinfo
  {journal} {Phys. Rev. C}\ }\textbf {\bibinfo {volume} {64}},\ \bibinfo
  {pages} {054901} (\bibinfo {year} {2001}{\natexlab{a}})},\ \Eprint
  {http://arxiv.org/abs/nucl-th/0105040} {arXiv:nucl-th/0105040} \BibitemShut
  {NoStop}%
\bibitem [{\citenamefont {Taghavi}(2021)}]{Taghavi:2020gcy}%
  \BibitemOpen
  \bibfield  {author} {\bibinfo {author} {\bibfnamefont {S.~F.}\ \bibnamefont
  {Taghavi}},\ }\href {\doibase 10.1140/epjc/s10052-021-09413-0} {\bibfield
  {journal} {\bibinfo  {journal} {Eur. Phys. J. C}\ }\textbf {\bibinfo {volume}
  {81}},\ \bibinfo {pages} {652} (\bibinfo {year} {2021})},\ \Eprint
  {http://arxiv.org/abs/2005.04742} {arXiv:2005.04742 [nucl-th]} \BibitemShut
  {NoStop}%
\bibitem [{\citenamefont {Mordasini}\ \emph {et~al.}(2020)\citenamefont
  {Mordasini}, \citenamefont {Bilandzic}, \citenamefont {Karako\c{c}},\ and\
  \citenamefont {Taghavi}}]{Mordasini:2019hut}%
  \BibitemOpen
  \bibfield  {author} {\bibinfo {author} {\bibfnamefont {C.}~\bibnamefont
  {Mordasini}}, \bibinfo {author} {\bibfnamefont {A.}~\bibnamefont
  {Bilandzic}}, \bibinfo {author} {\bibfnamefont {D.}~\bibnamefont
  {Karako\c{c}}}, \ and\ \bibinfo {author} {\bibfnamefont {S.~F.}\ \bibnamefont
  {Taghavi}},\ }\href {\doibase 10.1103/PhysRevC.102.024907} {\bibfield
  {journal} {\bibinfo  {journal} {Phys. Rev. C}\ }\textbf {\bibinfo {volume}
  {102}},\ \bibinfo {pages} {024907} (\bibinfo {year} {2020})},\ \Eprint
  {http://arxiv.org/abs/1901.06968} {arXiv:1901.06968 [nucl-ex]} \BibitemShut
  {NoStop}%
\bibitem [{\citenamefont {Borghini}\ \emph
  {et~al.}(2001{\natexlab{b}})\citenamefont {Borghini}, \citenamefont {Dinh},\
  and\ \citenamefont {Ollitrault}}]{Borghini:2000sa}%
  \BibitemOpen
  \bibfield  {author} {\bibinfo {author} {\bibfnamefont {N.}~\bibnamefont
  {Borghini}}, \bibinfo {author} {\bibfnamefont {P.~M.}\ \bibnamefont {Dinh}},
  \ and\ \bibinfo {author} {\bibfnamefont {J.-Y.}\ \bibnamefont {Ollitrault}},\
  }\href {\doibase 10.1103/PhysRevC.63.054906} {\bibfield  {journal} {\bibinfo
  {journal} {Phys. Rev. C}\ }\textbf {\bibinfo {volume} {63}},\ \bibinfo
  {pages} {054906} (\bibinfo {year} {2001}{\natexlab{b}})},\ \Eprint
  {http://arxiv.org/abs/nucl-th/0007063} {arXiv:nucl-th/0007063} \BibitemShut
  {NoStop}%
\bibitem [{\citenamefont {Bilandzic}\ \emph {et~al.}(2014)\citenamefont
  {Bilandzic}, \citenamefont {Christensen}, \citenamefont {Gulbrandsen},
  \citenamefont {Hansen},\ and\ \citenamefont {Zhou}}]{Bilandzic:2013kga}%
  \BibitemOpen
  \bibfield  {author} {\bibinfo {author} {\bibfnamefont {A.}~\bibnamefont
  {Bilandzic}}, \bibinfo {author} {\bibfnamefont {C.~H.}\ \bibnamefont
  {Christensen}}, \bibinfo {author} {\bibfnamefont {K.}~\bibnamefont
  {Gulbrandsen}}, \bibinfo {author} {\bibfnamefont {A.}~\bibnamefont {Hansen}},
  \ and\ \bibinfo {author} {\bibfnamefont {Y.}~\bibnamefont {Zhou}},\ }\href
  {\doibase 10.1103/PhysRevC.89.064904} {\bibfield  {journal} {\bibinfo
  {journal} {Phys. Rev. C}\ }\textbf {\bibinfo {volume} {89}},\ \bibinfo
  {pages} {064904} (\bibinfo {year} {2014})},\ \Eprint
  {http://arxiv.org/abs/1312.3572} {arXiv:1312.3572 [nucl-ex]} \BibitemShut
  {NoStop}%
\bibitem [{\citenamefont {Acharya}\ \emph
  {et~al.}(2021{\natexlab{b}})\citenamefont {Acharya} \emph
  {et~al.}}]{ALICE:2021klf}%
  \BibitemOpen
  \bibfield  {author} {\bibinfo {author} {\bibfnamefont {S.}~\bibnamefont
  {Acharya}} \emph {et~al.} (\bibinfo {collaboration} {ALICE}),\ }\href
  {\doibase 10.1103/PhysRevLett.127.092302} {\bibfield  {journal} {\bibinfo
  {journal} {Phys. Rev. Lett.}\ }\textbf {\bibinfo {volume} {127}},\ \bibinfo
  {pages} {092302} (\bibinfo {year} {2021}{\natexlab{b}})},\ \Eprint
  {http://arxiv.org/abs/2101.02579} {arXiv:2101.02579 [nucl-ex]} \BibitemShut
  {NoStop}%
\bibitem [{\citenamefont {Adam}\ \emph
  {et~al.}(2016{\natexlab{c}})\citenamefont {Adam} \emph
  {et~al.}}]{ALICE:2016igk}%
  \BibitemOpen
  \bibfield  {author} {\bibinfo {author} {\bibfnamefont {J.}~\bibnamefont
  {Adam}} \emph {et~al.} (\bibinfo {collaboration} {ALICE}),\ }\href {\doibase
  10.1103/PhysRevC.94.034903} {\bibfield  {journal} {\bibinfo  {journal} {Phys.
  Rev. C}\ }\textbf {\bibinfo {volume} {94}},\ \bibinfo {pages} {034903}
  (\bibinfo {year} {2016}{\natexlab{c}})},\ \Eprint
  {http://arxiv.org/abs/1603.04775} {arXiv:1603.04775 [nucl-ex]} \BibitemShut
  {NoStop}%
\bibitem [{\citenamefont {Aamodt}\ \emph {et~al.}(2010)\citenamefont {Aamodt}
  \emph {et~al.}}]{ALICE:2010suc}%
  \BibitemOpen
  \bibfield  {author} {\bibinfo {author} {\bibfnamefont {K.}~\bibnamefont
  {Aamodt}} \emph {et~al.} (\bibinfo {collaboration} {ALICE}),\ }\href
  {\doibase 10.1103/PhysRevLett.105.252302} {\bibfield  {journal} {\bibinfo
  {journal} {Phys. Rev. Lett.}\ }\textbf {\bibinfo {volume} {105}},\ \bibinfo
  {pages} {252302} (\bibinfo {year} {2010})},\ \Eprint
  {http://arxiv.org/abs/1011.3914} {arXiv:1011.3914 [nucl-ex]} \BibitemShut
  {NoStop}%
\bibitem [{\citenamefont {Di~Francesco}\ \emph {et~al.}(2017)\citenamefont
  {Di~Francesco}, \citenamefont {Guilbaud}, \citenamefont {Luzum},\ and\
  \citenamefont {Ollitrault}}]{DiFrancesco:2016srj}%
  \BibitemOpen
  \bibfield  {author} {\bibinfo {author} {\bibfnamefont {P.}~\bibnamefont
  {Di~Francesco}}, \bibinfo {author} {\bibfnamefont {M.}~\bibnamefont
  {Guilbaud}}, \bibinfo {author} {\bibfnamefont {M.}~\bibnamefont {Luzum}}, \
  and\ \bibinfo {author} {\bibfnamefont {J.-Y.}\ \bibnamefont {Ollitrault}},\
  }\href {\doibase 10.1103/PhysRevC.95.044911} {\bibfield  {journal} {\bibinfo
  {journal} {Phys. Rev. C}\ }\textbf {\bibinfo {volume} {95}},\ \bibinfo
  {pages} {044911} (\bibinfo {year} {2017})},\ \Eprint
  {http://arxiv.org/abs/1612.05634} {arXiv:1612.05634 [nucl-th]} \BibitemShut
  {NoStop}%
\bibitem [{\citenamefont {Jia}(2014)}]{Jia:2014jca}%
  \BibitemOpen
  \bibfield  {author} {\bibinfo {author} {\bibfnamefont {J.}~\bibnamefont
  {Jia}},\ }\href {\doibase 10.1088/0954-3899/41/12/124003} {\bibfield
  {journal} {\bibinfo  {journal} {J. Phys. G}\ }\textbf {\bibinfo {volume}
  {41}},\ \bibinfo {pages} {124003} (\bibinfo {year} {2014})},\ \Eprint
  {http://arxiv.org/abs/1407.6057} {arXiv:1407.6057 [nucl-ex]} \BibitemShut
  {NoStop}%
\bibitem [{\citenamefont {Adam}\ \emph
  {et~al.}(2016{\natexlab{d}})\citenamefont {Adam} \emph
  {et~al.}}]{ALICE:2016ccg}%
  \BibitemOpen
  \bibfield  {author} {\bibinfo {author} {\bibfnamefont {J.}~\bibnamefont
  {Adam}} \emph {et~al.} (\bibinfo {collaboration} {ALICE}),\ }\href {\doibase
  10.1103/PhysRevLett.116.132302} {\bibfield  {journal} {\bibinfo  {journal}
  {Phys. Rev. Lett.}\ }\textbf {\bibinfo {volume} {116}},\ \bibinfo {pages}
  {132302} (\bibinfo {year} {2016}{\natexlab{d}})},\ \Eprint
  {http://arxiv.org/abs/1602.01119} {arXiv:1602.01119 [nucl-ex]} \BibitemShut
  {NoStop}%
\bibitem [{\citenamefont {Abelev}\ \emph {et~al.}(2014)\citenamefont {Abelev}
  \emph {et~al.}}]{ALICE:2014gvd}%
  \BibitemOpen
  \bibfield  {author} {\bibinfo {author} {\bibfnamefont {B.~B.}\ \bibnamefont
  {Abelev}} \emph {et~al.} (\bibinfo {collaboration} {ALICE}),\ }\href
  {\doibase 10.1140/epjc/s10052-014-3077-y} {\bibfield  {journal} {\bibinfo
  {journal} {Eur. Phys. J. C}\ }\textbf {\bibinfo {volume} {74}},\ \bibinfo
  {pages} {3077} (\bibinfo {year} {2014})},\ \Eprint
  {http://arxiv.org/abs/1407.5530} {arXiv:1407.5530 [nucl-ex]} \BibitemShut
  {NoStop}%
\bibitem [{\citenamefont {Aamodt}\ \emph
  {et~al.}(2011{\natexlab{c}})\citenamefont {Aamodt} \emph
  {et~al.}}]{ALICE:2010mlf}%
  \BibitemOpen
  \bibfield  {author} {\bibinfo {author} {\bibfnamefont {K.}~\bibnamefont
  {Aamodt}} \emph {et~al.} (\bibinfo {collaboration} {ALICE}),\ }\href
  {\doibase 10.1103/PhysRevLett.106.032301} {\bibfield  {journal} {\bibinfo
  {journal} {Phys. Rev. Lett.}\ }\textbf {\bibinfo {volume} {106}},\ \bibinfo
  {pages} {032301} (\bibinfo {year} {2011}{\natexlab{c}})},\ \Eprint
  {http://arxiv.org/abs/1012.1657} {arXiv:1012.1657 [nucl-ex]} \BibitemShut
  {NoStop}%
\bibitem [{\citenamefont {Acharya}\ \emph
  {et~al.}(2018{\natexlab{b}})\citenamefont {Acharya} \emph
  {et~al.}}]{ALICE:2017kwu}%
  \BibitemOpen
  \bibfield  {author} {\bibinfo {author} {\bibfnamefont {S.}~\bibnamefont
  {Acharya}} \emph {et~al.} (\bibinfo {collaboration} {ALICE}),\ }\href
  {\doibase 10.1103/PhysRevC.97.024906} {\bibfield  {journal} {\bibinfo
  {journal} {Phys. Rev. C}\ }\textbf {\bibinfo {volume} {97}},\ \bibinfo
  {pages} {024906} (\bibinfo {year} {2018}{\natexlab{b}})},\ \Eprint
  {http://arxiv.org/abs/1709.01127} {arXiv:1709.01127 [nucl-ex]} \BibitemShut
  {NoStop}%
\bibitem [{\citenamefont {Adam}\ \emph {et~al.}(2017)\citenamefont {Adam} \emph
  {et~al.}}]{Adam:2016ddh}%
  \BibitemOpen
  \bibfield  {author} {\bibinfo {author} {\bibfnamefont {J.}~\bibnamefont
  {Adam}} \emph {et~al.} (\bibinfo {collaboration} {ALICE}),\ }\href {\doibase
  10.1016/j.physletb.2017.07.017} {\bibfield  {journal} {\bibinfo  {journal}
  {Phys. Lett. B}\ }\textbf {\bibinfo {volume} {772}},\ \bibinfo {pages} {567}
  (\bibinfo {year} {2017})},\ \Eprint {http://arxiv.org/abs/1612.08966}
  {arXiv:1612.08966 [nucl-ex]} \BibitemShut {NoStop}%
\bibitem [{\citenamefont {Yan}\ and\ \citenamefont
  {Ollitrault}(2015)}]{Yan:2015jma}%
  \BibitemOpen
  \bibfield  {author} {\bibinfo {author} {\bibfnamefont {L.}~\bibnamefont
  {Yan}}\ and\ \bibinfo {author} {\bibfnamefont {J.-Y.}\ \bibnamefont
  {Ollitrault}},\ }\href {\doibase 10.1016/j.physletb.2015.03.040} {\bibfield
  {journal} {\bibinfo  {journal} {Phys. Lett.}\ }\textbf {\bibinfo {volume}
  {B744}},\ \bibinfo {pages} {82} (\bibinfo {year} {2015})},\ \Eprint
  {http://arxiv.org/abs/1502.02502} {arXiv:1502.02502 [nucl-th]} \BibitemShut
  {NoStop}%
\bibitem [{\citenamefont {Luzum}\ and\ \citenamefont
  {Ollitrault}(2010)}]{Luzum:2010ad}%
  \BibitemOpen
  \bibfield  {author} {\bibinfo {author} {\bibfnamefont {M.}~\bibnamefont
  {Luzum}}\ and\ \bibinfo {author} {\bibfnamefont {J.-Y.}\ \bibnamefont
  {Ollitrault}},\ }\href {\doibase 10.1103/PhysRevC.82.014906} {\bibfield
  {journal} {\bibinfo  {journal} {Phys. Rev. C}\ }\textbf {\bibinfo {volume}
  {82}},\ \bibinfo {pages} {014906} (\bibinfo {year} {2010})},\ \Eprint
  {http://arxiv.org/abs/1004.2023} {arXiv:1004.2023 [nucl-th]} \BibitemShut
  {NoStop}%
\bibitem [{\citenamefont {Luzum}\ \emph {et~al.}(2010)\citenamefont {Luzum},
  \citenamefont {Gombeaud},\ and\ \citenamefont {Ollitrault}}]{Luzum:2010ae}%
  \BibitemOpen
  \bibfield  {author} {\bibinfo {author} {\bibfnamefont {M.}~\bibnamefont
  {Luzum}}, \bibinfo {author} {\bibfnamefont {C.}~\bibnamefont {Gombeaud}}, \
  and\ \bibinfo {author} {\bibfnamefont {J.-Y.}\ \bibnamefont {Ollitrault}},\
  }\href {\doibase 10.1103/PhysRevC.81.054910} {\bibfield  {journal} {\bibinfo
  {journal} {Phys. Rev. C}\ }\textbf {\bibinfo {volume} {81}},\ \bibinfo
  {pages} {054910} (\bibinfo {year} {2010})},\ \Eprint
  {http://arxiv.org/abs/1004.2024} {arXiv:1004.2024 [nucl-th]} \BibitemShut
  {NoStop}%
\bibitem [{\citenamefont {Aad}\ \emph {et~al.}(2012)\citenamefont {Aad} \emph
  {et~al.}}]{ATLAS:2011ag}%
  \BibitemOpen
  \bibfield  {author} {\bibinfo {author} {\bibfnamefont {G.}~\bibnamefont
  {Aad}} \emph {et~al.} (\bibinfo {collaboration} {ATLAS}),\ }\href {\doibase
  10.1016/j.physletb.2012.02.045} {\bibfield  {journal} {\bibinfo  {journal}
  {Phys. Lett. B}\ }\textbf {\bibinfo {volume} {710}},\ \bibinfo {pages} {363}
  (\bibinfo {year} {2012})},\ \Eprint {http://arxiv.org/abs/1108.6027}
  {arXiv:1108.6027 [hep-ex]} \BibitemShut {NoStop}%
\bibitem [{\citenamefont {Abelev}\ \emph
  {et~al.}(2013{\natexlab{d}})\citenamefont {Abelev} \emph
  {et~al.}}]{ALICE:2012xs}%
  \BibitemOpen
  \bibfield  {author} {\bibinfo {author} {\bibfnamefont {B.}~\bibnamefont
  {Abelev}} \emph {et~al.} (\bibinfo {collaboration} {ALICE}),\ }\href
  {\doibase 10.1103/PhysRevLett.110.032301} {\bibfield  {journal} {\bibinfo
  {journal} {Phys. Rev. Lett.}\ }\textbf {\bibinfo {volume} {110}},\ \bibinfo
  {pages} {032301} (\bibinfo {year} {2013}{\natexlab{d}})},\ \Eprint
  {http://arxiv.org/abs/1210.3615} {arXiv:1210.3615 [nucl-ex]} \BibitemShut
  {NoStop}%
\bibitem [{\citenamefont {Bazavov}\ \emph {et~al.}(2012)\citenamefont {Bazavov}
  \emph {et~al.}}]{Bazavov:2011nk}%
  \BibitemOpen
  \bibfield  {author} {\bibinfo {author} {\bibfnamefont {A.}~\bibnamefont
  {Bazavov}} \emph {et~al.},\ }\href {\doibase 10.1103/PhysRevD.85.054503}
  {\bibfield  {journal} {\bibinfo  {journal} {Phys. Rev. D}\ }\textbf {\bibinfo
  {volume} {85}},\ \bibinfo {pages} {054503} (\bibinfo {year} {2012})},\
  \Eprint {http://arxiv.org/abs/1111.1710} {arXiv:1111.1710 [hep-lat]}
  \BibitemShut {NoStop}%
\bibitem [{\citenamefont {Bernard}\ \emph {et~al.}(2005)\citenamefont
  {Bernard}, \citenamefont {Burch}, \citenamefont {Gregory}, \citenamefont
  {Toussaint}, \citenamefont {DeTar}, \citenamefont {Osborn}, \citenamefont
  {Gottlieb}, \citenamefont {Heller},\ and\ \citenamefont
  {Sugar}}]{Bernard:2004je}%
  \BibitemOpen
  \bibfield  {author} {\bibinfo {author} {\bibfnamefont {C.}~\bibnamefont
  {Bernard}}, \bibinfo {author} {\bibfnamefont {T.}~\bibnamefont {Burch}},
  \bibinfo {author} {\bibfnamefont {E.~B.}\ \bibnamefont {Gregory}}, \bibinfo
  {author} {\bibfnamefont {D.}~\bibnamefont {Toussaint}}, \bibinfo {author}
  {\bibfnamefont {C.~E.}\ \bibnamefont {DeTar}}, \bibinfo {author}
  {\bibfnamefont {J.}~\bibnamefont {Osborn}}, \bibinfo {author} {\bibfnamefont
  {S.}~\bibnamefont {Gottlieb}}, \bibinfo {author} {\bibfnamefont {U.~M.}\
  \bibnamefont {Heller}}, \ and\ \bibinfo {author} {\bibfnamefont
  {R.}~\bibnamefont {Sugar}} (\bibinfo {collaboration} {MILC}),\ }\href
  {\doibase 10.1103/PhysRevD.71.034504} {\bibfield  {journal} {\bibinfo
  {journal} {Phys. Rev. D}\ }\textbf {\bibinfo {volume} {71}},\ \bibinfo
  {pages} {034504} (\bibinfo {year} {2005})},\ \Eprint
  {http://arxiv.org/abs/hep-lat/0405029} {arXiv:hep-lat/0405029} \BibitemShut
  {NoStop}%
\bibitem [{\citenamefont {Paatelainen}\ \emph {et~al.}(2013)\citenamefont
  {Paatelainen}, \citenamefont {Eskola}, \citenamefont {Holopainen},\ and\
  \citenamefont {Tuominen}}]{Paatelainen:2012at}%
  \BibitemOpen
  \bibfield  {author} {\bibinfo {author} {\bibfnamefont {R.}~\bibnamefont
  {Paatelainen}}, \bibinfo {author} {\bibfnamefont {K.~J.}\ \bibnamefont
  {Eskola}}, \bibinfo {author} {\bibfnamefont {H.}~\bibnamefont {Holopainen}},
  \ and\ \bibinfo {author} {\bibfnamefont {K.}~\bibnamefont {Tuominen}},\
  }\href {\doibase 10.1103/PhysRevC.87.044904} {\bibfield  {journal} {\bibinfo
  {journal} {Phys. Rev. C}\ }\textbf {\bibinfo {volume} {87}},\ \bibinfo
  {pages} {044904} (\bibinfo {year} {2013})},\ \Eprint
  {http://arxiv.org/abs/1211.0461} {arXiv:1211.0461 [hep-ph]} \BibitemShut
  {NoStop}%
\end{thebibliography}%

\appendix
\section{Observable suitability}\label{sec:appendix_obs}
As mentioned in the text suitability of observables used in the analysis must be examined to guarantee reliability and applicability of the results. To do so, the convergence of the prior distribution and the validity of the emulator estimates are studied. The convergence of the prior distribution is quantified by extracting the coefficient of variance (CV) for all observables. The CV, defined as the ratio of the standard deviation to the mean, describes how scattered a distribution is with respect to it's mean. For the emulator validity a separate validation results from the hydrodynamical calculations are obtained. These results are compared to the emulator predictions. The agreement between the emulator and the validations results is quantified with a Pearson correlation between the two. 

\begin{figure*}[t]
    \centering
    \includegraphics[width=0.99\textwidth]{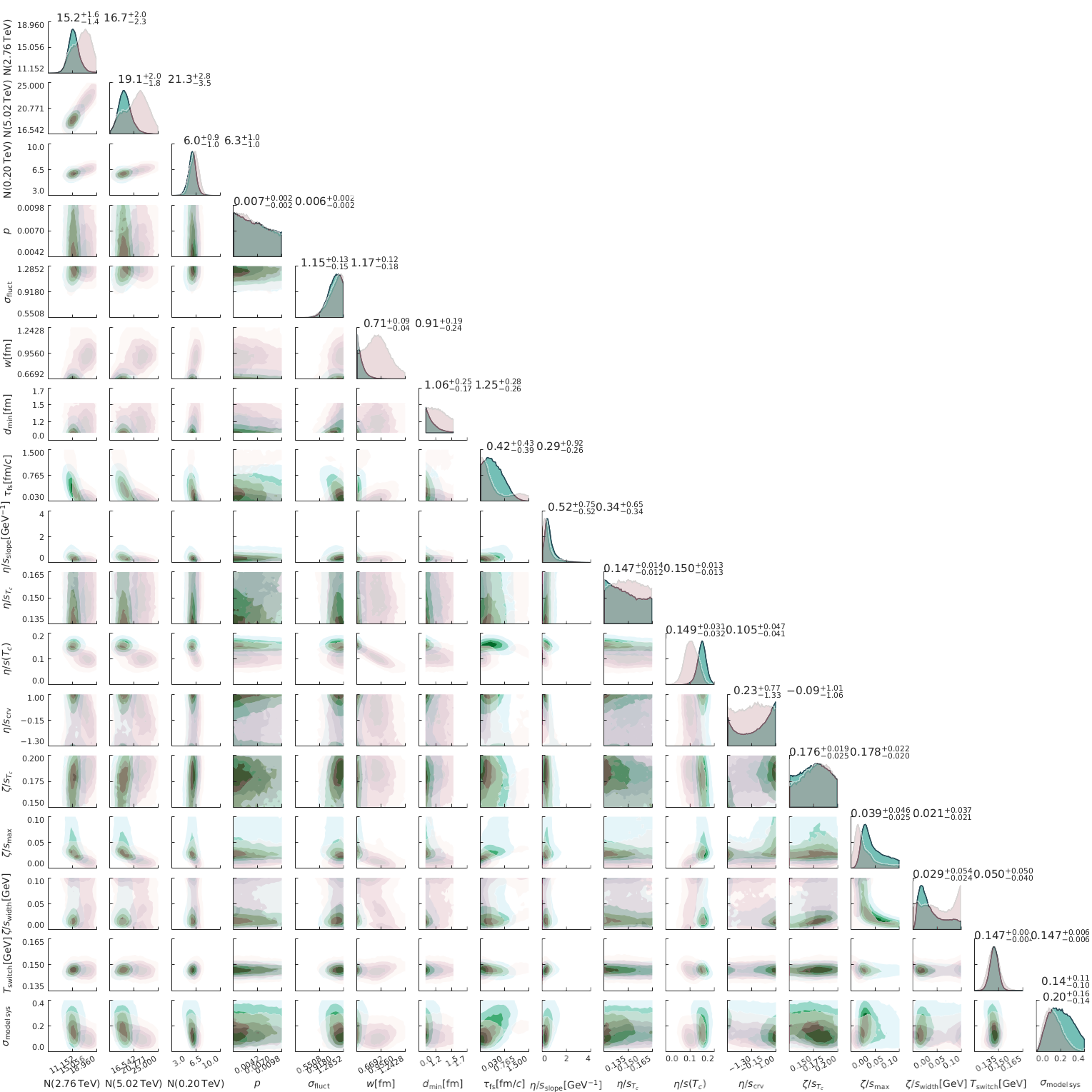}
    \caption{Posterior distribution of model parameters. Posteriors obtained with selected observables are shown in turquoise while the posteriors obtained with all observables are shown in light red. }
    \label{fig:posteriorAvsB}
\end{figure*}

In Tab.~\ref{tab:CVcorr} are shown the CV of the prior distribution, and the Pearson correlation value of the emulator validation for all the observables across the three different collision systems. The observables are considered applicable if the CV is less than 1.0 and the Pearson correlation is greater than 0.7. Values that satisfy these criteria are marked with bold font the Tab.~\ref{tab:CVcorr}.
%
\begin{table*}[tbh!]
  \caption{
    \label{tab:CVcorr}
    Pearson correlation and CV values for all observables. Correlation values that are greater than 0.7 and CV values less than 1.0 are marked with bold font. 
  }
  \begin{tabular}{|m{3cm}|m{6em}|m{6em}|m{6em}|m{6em}|m{6em}|m{6em}|}
    \hline
    \multicolumn{1}{|c}{Observable} & \multicolumn{3}{c}{ Pearson correlation } & \multicolumn{3}{c|}{ CV } \\
      \hline
     & PbPb5020 & PbPb2760 & AuAu200 & PbPb5020 & PbPb2760 & AuAu200 \\
      \hline
    $\mathrm{d}N/\mathrm{d}\eta$ $h^\pm$ & \textbf{1.0000} & \textbf{1.0000} & \textbf{1.0000} & \textbf{0.1288} & \textbf{0.1485} & \textbf{0.2738} \\ \hline
    $\mathrm{d}N/\mathrm{d}y$ $\pi^\pm$ & \textbf{1.0000} & \textbf{1.0000} & \textbf{1.0000} & \textbf{0.1356} & \textbf{0.1546} & \textbf{0.2742} \\ \hline
    $\mathrm{d}N/\mathrm{d}y$ $K^\pm$ & \textbf{1.0000} & \textbf{1.0000} & \textbf{1.0000} & \textbf{0.1303} & \textbf{0.1493} & \textbf{0.2688} \\ \hline
    $\mathrm{d}N/\mathrm{d}y$ $p\bar p$ & \textbf{1.0000} & \textbf{1.0000} & \textit{-1} & \textbf{0.2083} & \textbf{0.2207} & \textit{-1} \\ \hline
    \meanpt $\pi^\pm$ & \textbf{0.9786} & \textbf{0.9907} & \textbf{0.9635} & \textbf{0.1201} & \textbf{0.1215} & \textbf{0.1309} \\ \hline
    \meanpt $K^\pm$ & \textbf{0.9978} & \textbf{0.9800} & \textbf{0.9926} & \textbf{0.1121} & \textbf{0.1153} & \textbf{0.1329} \\ \hline
    \meanpt $p\bar p$  & \textbf{0.9998} & \textbf{0.9997} & \textit{-1} & \textbf{0.0940} & \textbf{0.0976} & \textit{-1} \\ \hline
    $v_2$ & \textbf{0.9996} & \textbf{0.9998} & \textbf{0.9952} & \textbf{0.1444} & \textbf{0.1440} & \textbf{0.1668} \\ \hline
    $v_3$ & \textbf{0.9747} & \textbf{0.9793} & \textbf{0.8582} & \textbf{0.2888} & \textbf{0.2980} & \textbf{0.4025} \\ \hline
    $v_4$ & \textbf{0.9691} & \textbf{0.9801} & \textit{0.6573} & \textbf{0.4279} & \textbf{0.4532} & \textbf{0.6389} \\ \hline
    $v_5$ & \textbf{0.7458} & \textit{0.5938} & \textit{-1} & \textbf{0.6369} & \textbf{0.5933} & \textit{-1} \\ \hline
    $v_6$ & \textit{0.6516} & \textit{-1} & \textit{-1} & \textbf{0.3630} & \textit{-1} & \textit{-1} \\ \hline
    $v_7$ & \textit{0.4660} & \textit{-1} & \textit{-1} & \textbf{0.2975} & \textit{-1} & \textit{-1} \\ \hline
    $\chi_{4,22}$ & \textbf{0.9490} & \textbf{0.9282} & \textbf{0.7246} & \textbf{0.0938} & \textbf{0.1031} & \textbf{0.7131} \\ \hline
    $\chi_{5,23}$ & \textbf{0.8351} & \textbf{0.8861} & \textit{0.2621} & \textbf{0.1892} & \textbf{0.2374} & \textit{7.6973} \\ \hline
    $\chi_{6,222}$ & \textit{0.4865} & \textit{0.2599} & \textit{-1} & \textbf{0.5464} & \textbf{0.8624} & \textit{-1} \\ \hline
    $\chi_{6,33}$ & \textit{0.1022} & \textit{0.0177} & \textit{-1} & \textit{4.2578} & \textit{2.7527} & \textit{-1} \\ \hline
    $\rho_{4,22}$ & \textbf{0.9917} & \textbf{0.9915} & \textbf{0.7193} & \textbf{0.1827} & \textbf{0.2097} & \textbf{0.6983} \\ \hline
    $\rho_{5,23}$ & \textit{0.5157} & \textit{0.5807} & \textit{0.3924} & \textit{1.3046} & \textbf{0.5867} & \textit{8.2865} \\ \hline
    $\rho_{6,222}$ & \textit{0.6791} & \textit{0.6108} & \textit{-1} & \textit{1.8353} & \textit{1.5498} & \textit{-1} \\ \hline
    $\rho_{6,33}$ & \textit{0.3068} & \textit{0.3322} & \textit{-1} & \textit{2.1308} & \textit{1.9215} & \textit{-1} \\ \hline
    NSC(3,2) & \textbf{0.9279} & \textbf{0.9493} & \textit{0.0053} & \textbf{0.4411} & \textbf{0.4946} & \textit{4.7598} \\ \hline
    NSC(4,2)& \textbf{0.9399} & \textbf{0.8095} & \textit{0.2044} & \textbf{0.8541} & \textbf{0.6307} & \textit{21.7216} \\ \hline
    NSC(4,3)& \textit{0.4717} & \textit{0.0133} & \textit{-1} & \textit{11.4635} & \textit{25.7502} & \textit{-1} \\
     
        \hline
  \end{tabular}
\end{table*}
In Fig.~\ref{fig:posteriorAvsB} are shown posterior distributions of the model parameters obtained with selected observables and with all observables. Although the distributions vary parameter to parameter, in general the posteriors obtained with the selected observables are slightly narrower. This is expected as including more observables will increase the uncertainty. There are few notable differences between the results. Firstly, the inclusion of all observables prefers larger normalization values and a shorter free-streaming time. These parameters are correlated though for the larger normalization will increase the particle yields, whereas the shorter free-streaming time decreases the yields. Secondly, the analysis with the selected observables prefers smaller values of the nucleon width, $w$ and minimum distance between the nucleons, $d_\mathrm{min}$. The preferred values are hitting the lower boundaries of the prior ranges for these parameters. Therefore, in future analyses wider parameter ranges are crucial for more precise MAP extractions. Furthermore, according to the latest studies the experimental measurements of nucleus-nucleus cross-sections imply the nucleon width to be less than 0.7 fm~\cite{Nijs:2022rme}. 

\begin{figure*}[t]
    \centering
    \includegraphics[width=0.49\textwidth]{figs/MAP/dndeta_MAP_A.pdf}
    \includegraphics[width=0.49\textwidth]{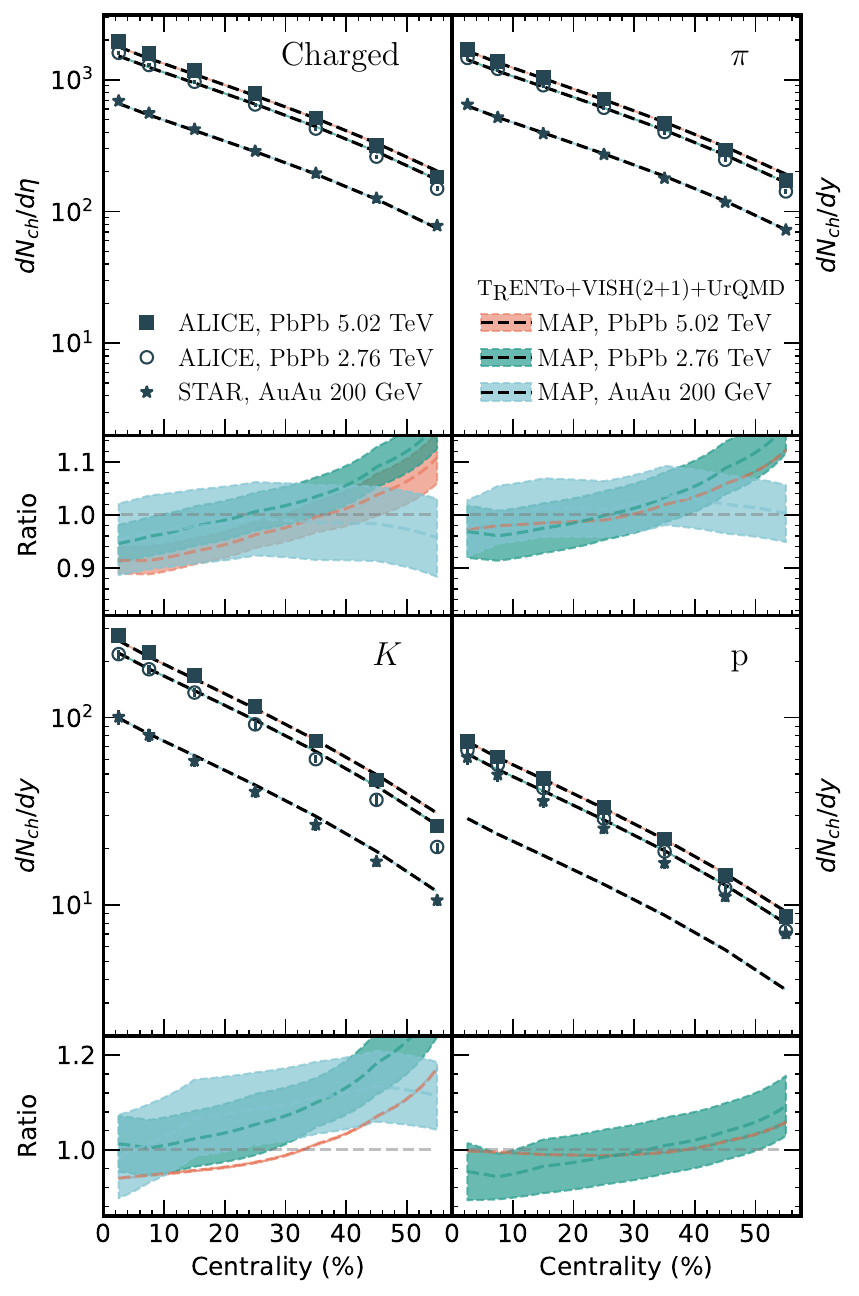}
    \caption{ MAP predictions of charged hadron and identified particle yields obtained by calibrating with selected and all observables depicted on the left and right, respectively. }
    \label{fig:MAPdndetaAvsB}
\end{figure*}

\begin{figure*}[t]
    \centering
    \includegraphics[width=0.49\textwidth]{figs/MAP/meanpT_MAP_A.pdf}
    \includegraphics[width=0.49\textwidth]{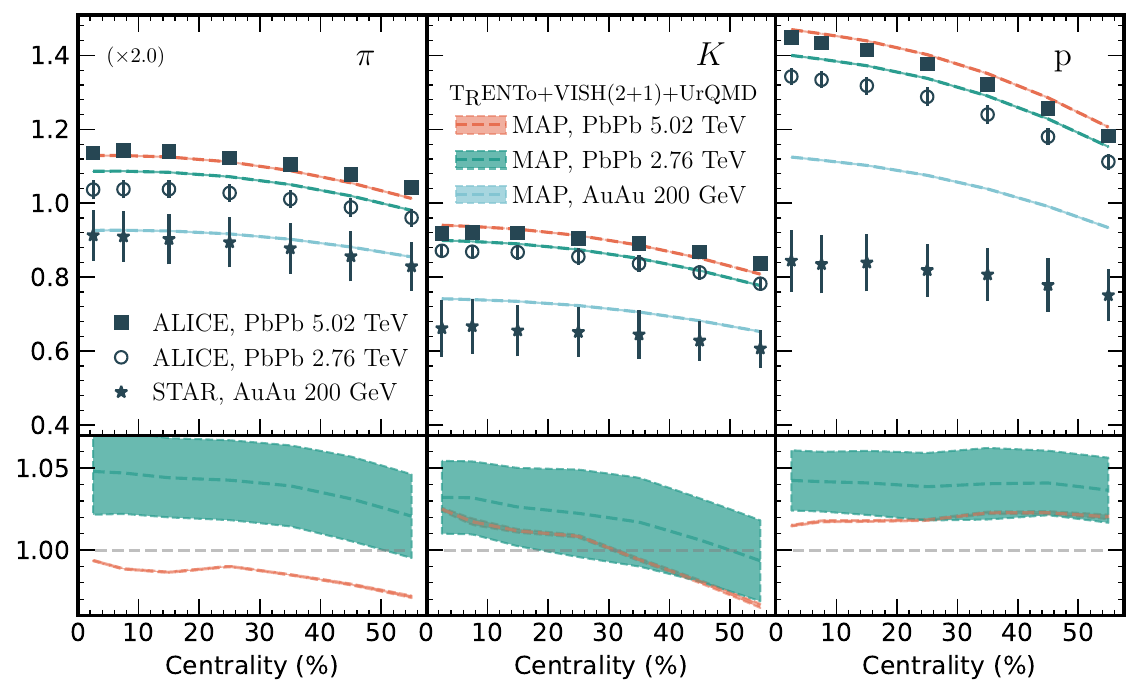}
    \caption{ MAP predictions of transverse momentum of identified particles obtained by calibrating with selected and all observables depicted on the left and right, respectively. }
    \label{fig:MAPmeanptAvsB}
\end{figure*}

\begin{figure*}[t]
    \centering
    \includegraphics[width=0.49\textwidth]{figs/MAP/obs_trio_MAP_A.pdf}
    \includegraphics[width=0.49\textwidth]{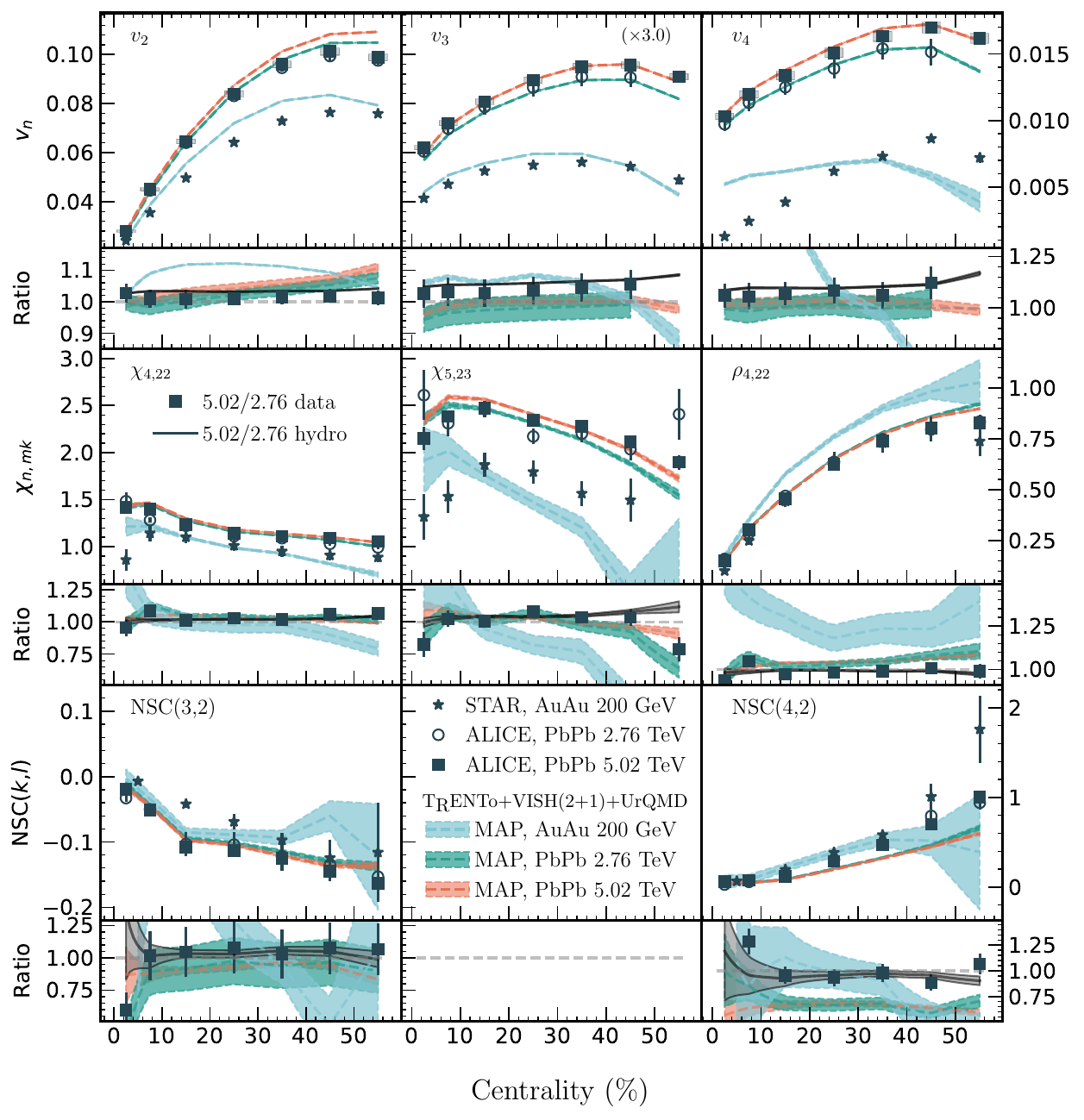}
    \caption{ MAP predictions of lower-order flow observables obtained by calibrating with selected and all observables depicted on the left and right, respectively. }
    \label{fig:MAPflowAvsB}
\end{figure*}

The MAP results two configurations are compared with each other in Figs.~\ref{fig:MAPdndetaAvsB}-\ref{fig:MAPflowAvsB}. The configuration with selected observables shows better agreement with data for charged hadron and identified particle yields, all $\meanpt$, $v_2$ and NSC(4,2). On the other hand, the configuration calibrated with all observables gives slightly better predictions for $v_3$, $v_4$ and $\rho_{4,22}$. The configuration calibrated wit all observables predicts stronger magnitudes for $v_n$ causing the difference in agreements. 

\begin{figure*}
    \centering
    \includegraphics[width=0.49\textwidth]{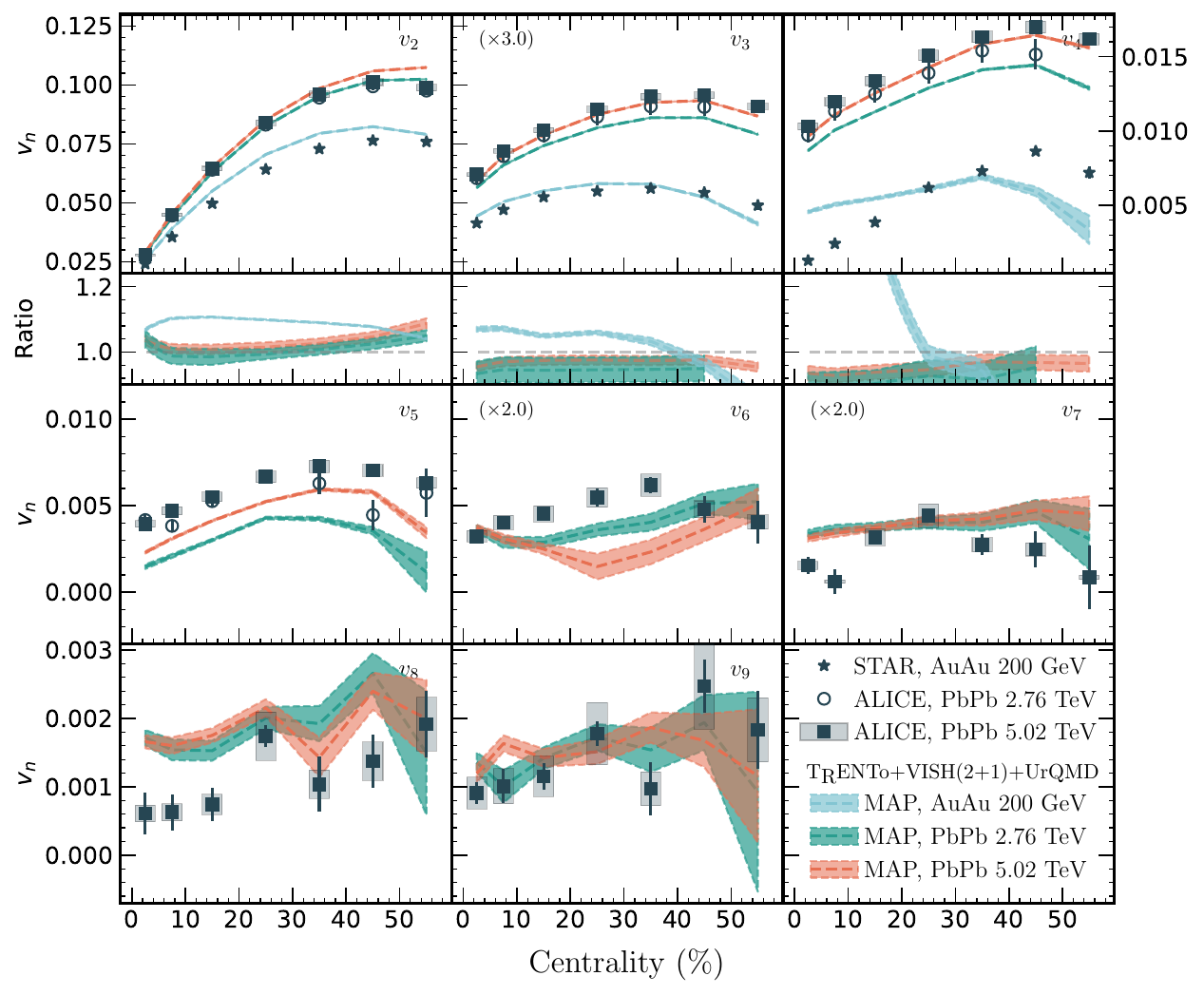}
    \includegraphics[width=0.49\textwidth]{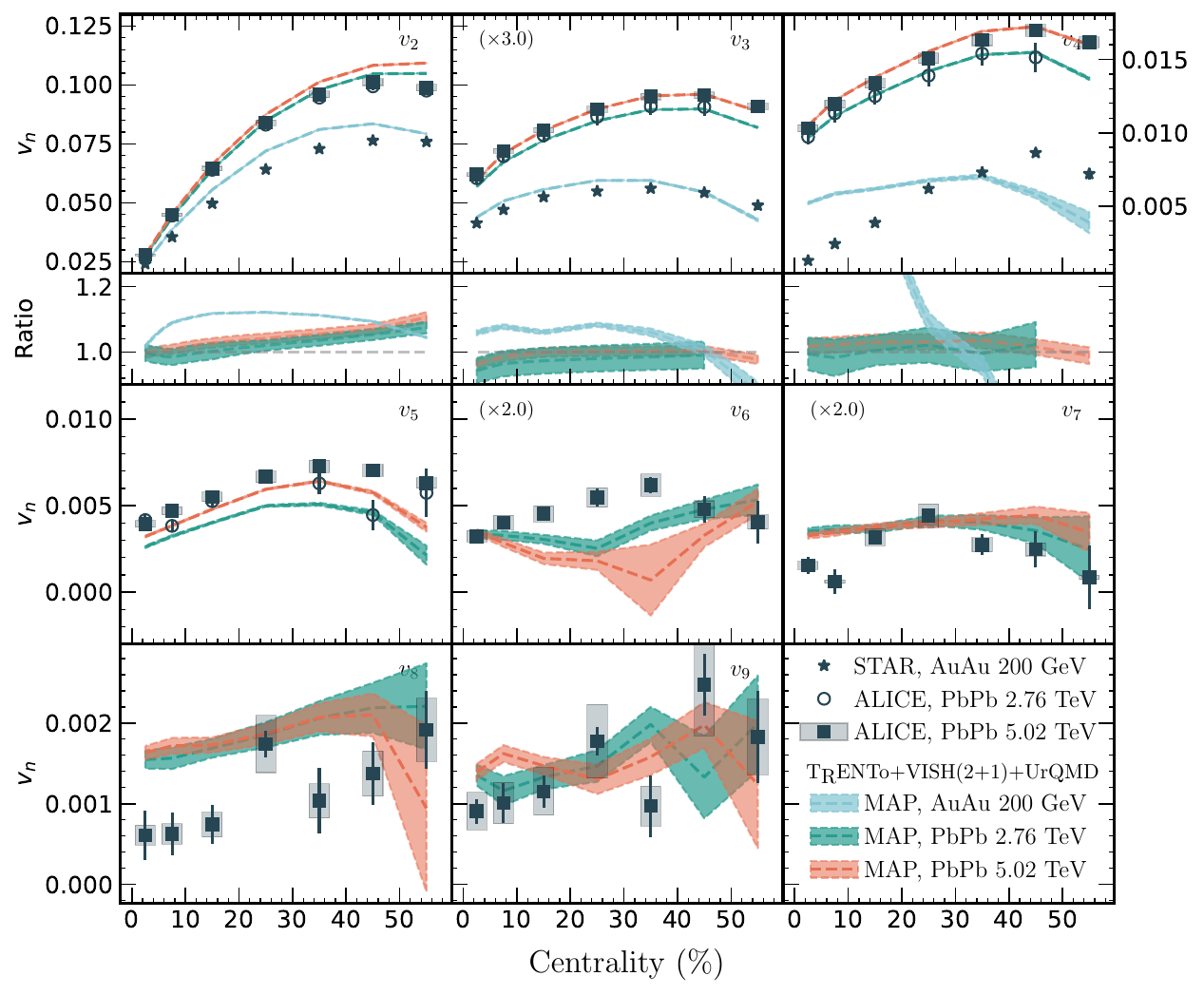}
    \caption{MAP predictions of flow coefficients $v_n$ obtained by calibrating with selected and all observables depicted on the left and right, respectively.}
    \label{fig:vnmapAvsB}
\end{figure*}

\begin{figure*}
    \centering
    \includegraphics[width=0.49\textwidth]{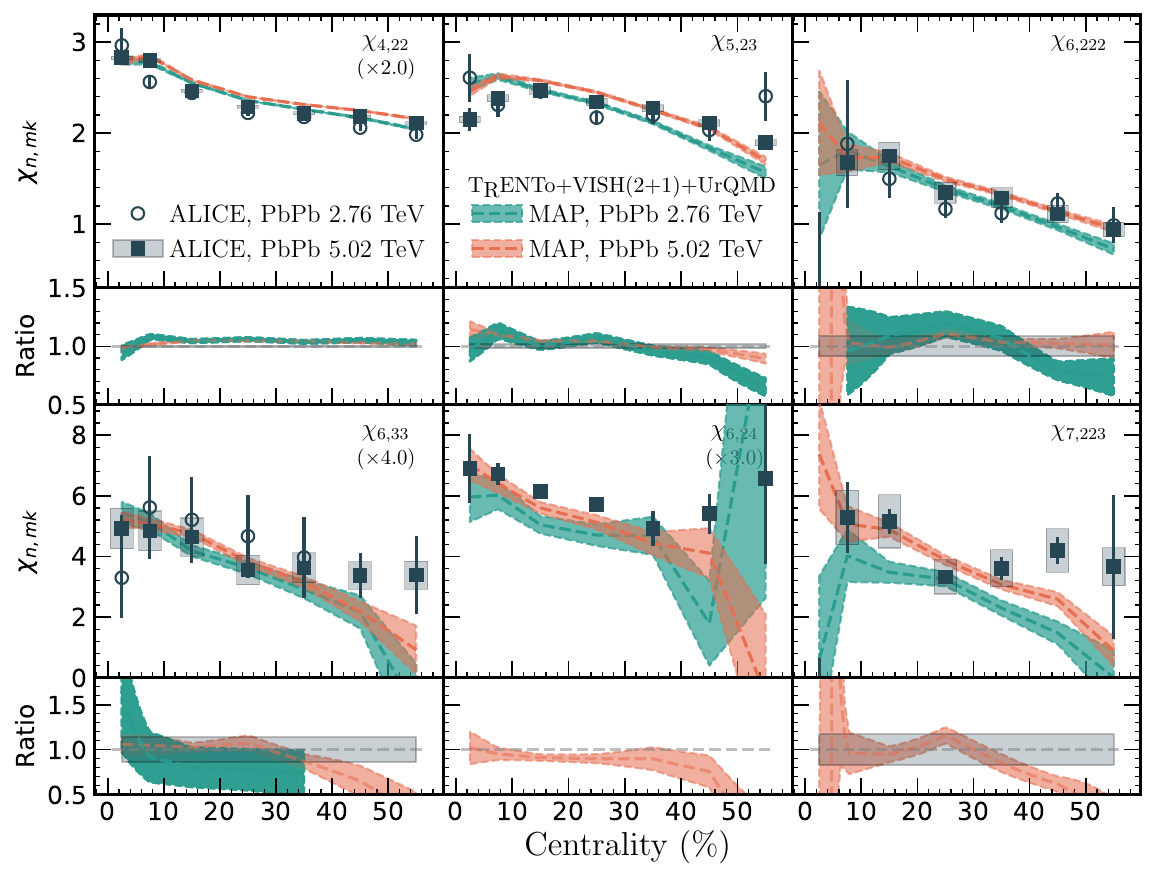}
    \includegraphics[width=0.49\textwidth]{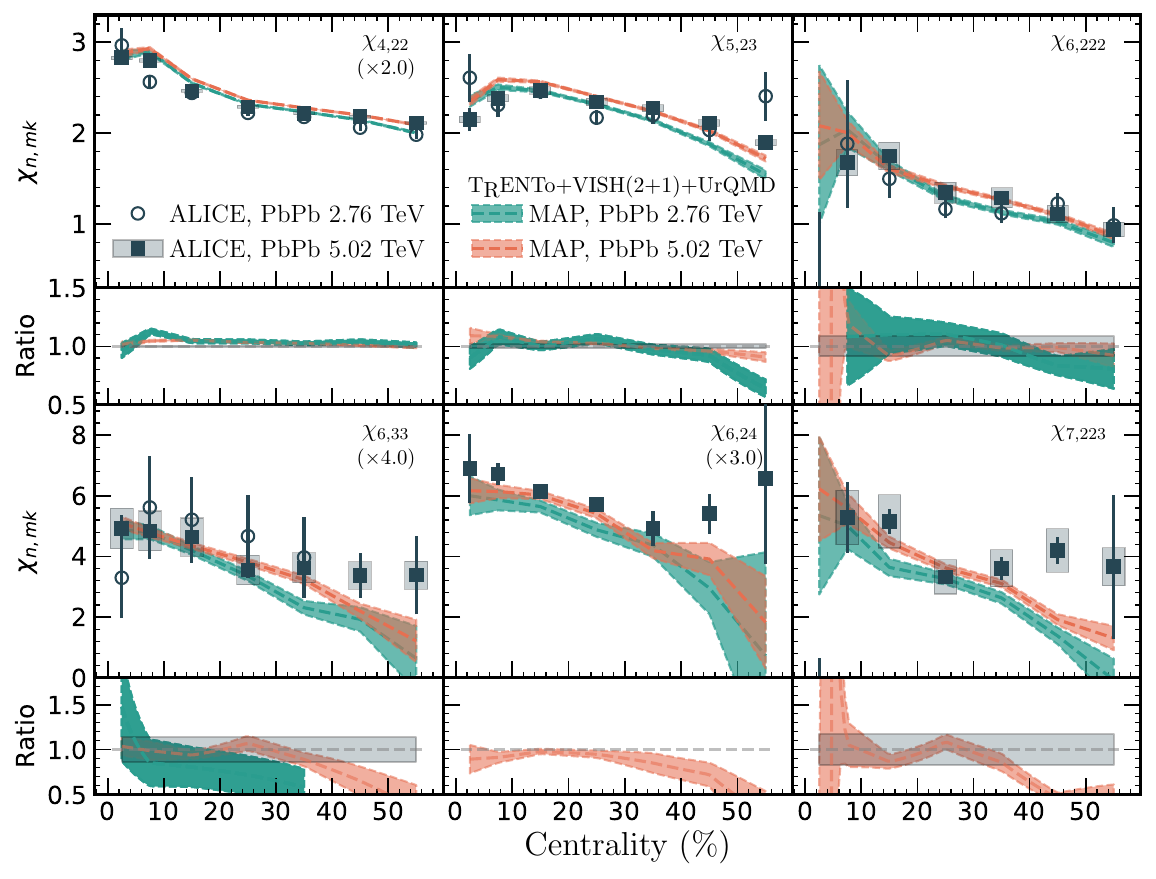}
    \caption{MAP predictions of non-linear flow-mode coefficients $\chi_{n,mk}$ obtained by calibrating with selected and all observables depicted on the left and right, respectively.}
    \label{fig:chimapAvsB}
\end{figure*}

\begin{figure*}
    \centering
    \includegraphics[width=0.49\textwidth]{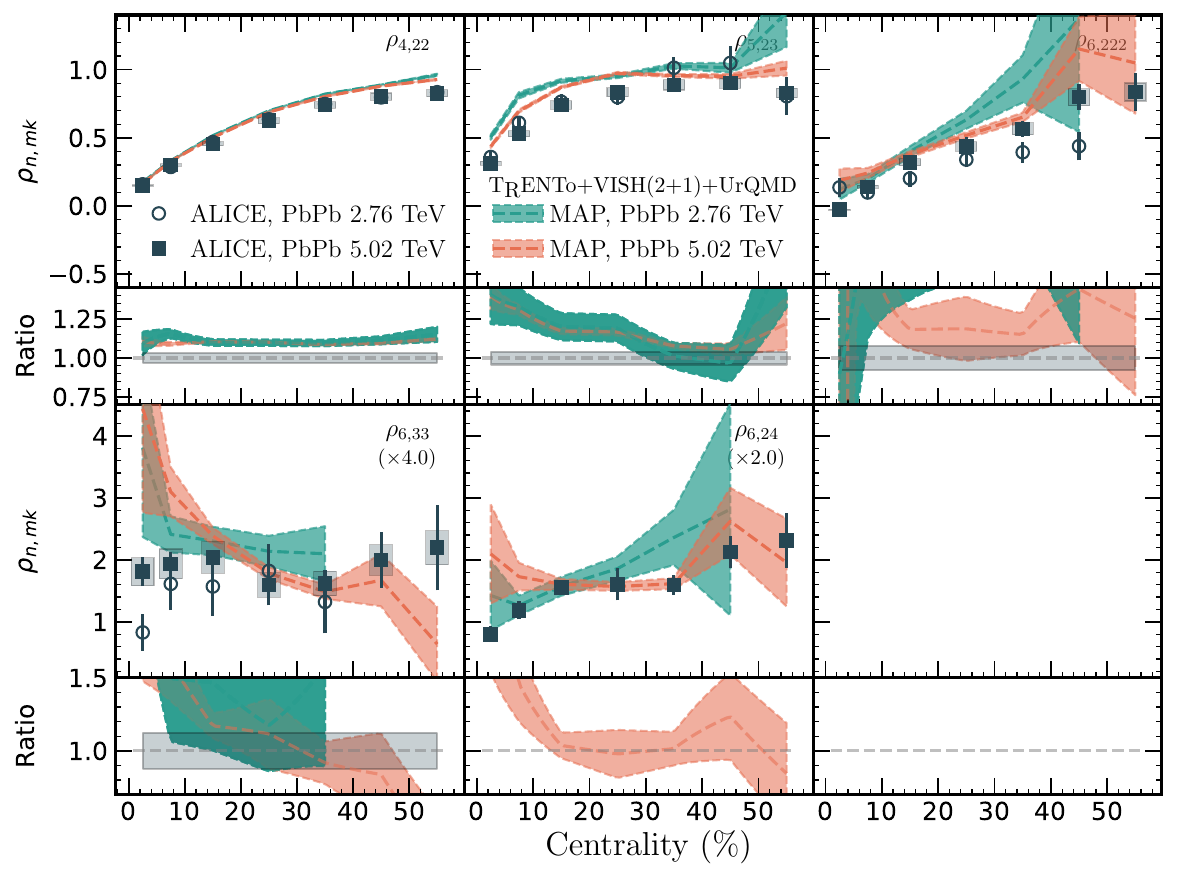}
    \includegraphics[width=0.49\textwidth]{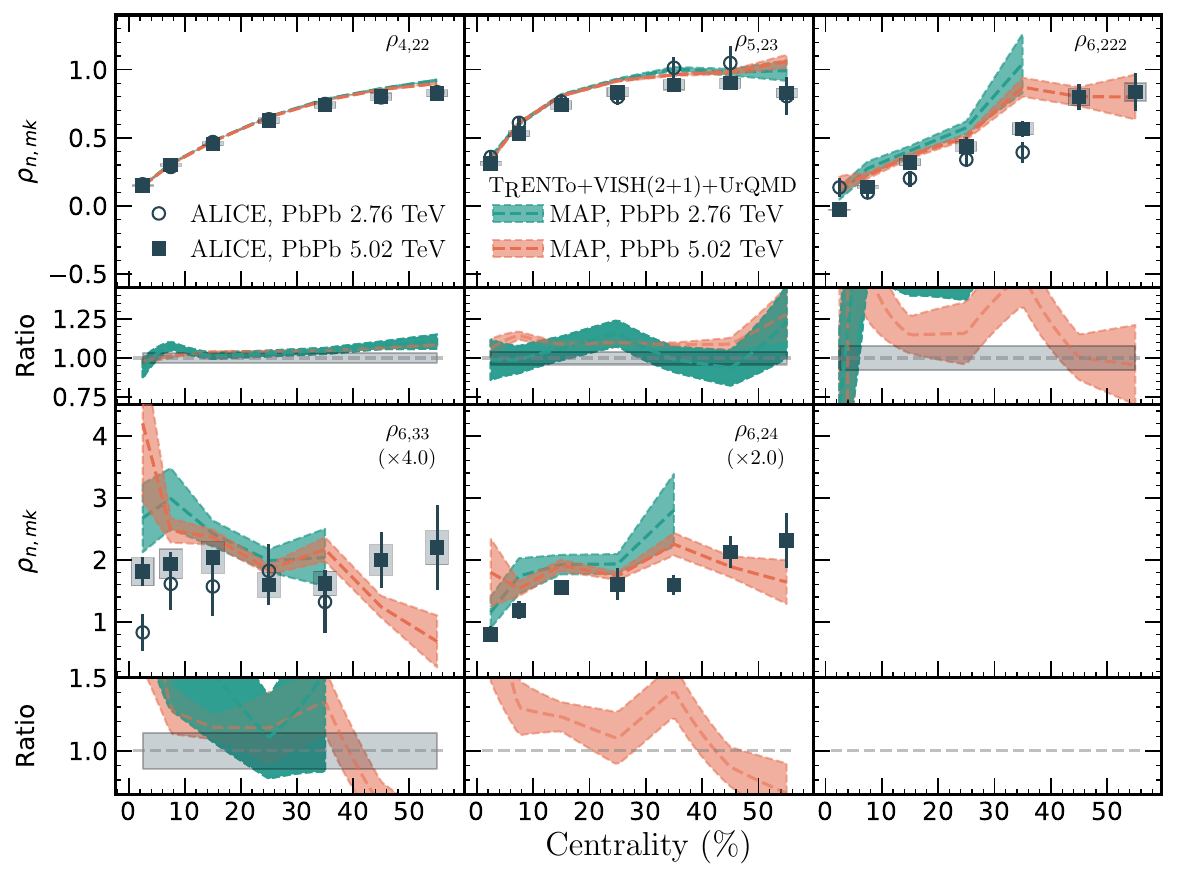}
    \caption{MAP predictions of symmetry-plane correlations $\rho_{n,mk}$ obtained by calibrating with selected and all observables depicted on the left and right, respectively.}
    \label{fig:rhomapAvsB}
\end{figure*}

\begin{figure*}
    \centering
    \includegraphics[width=0.49\textwidth]{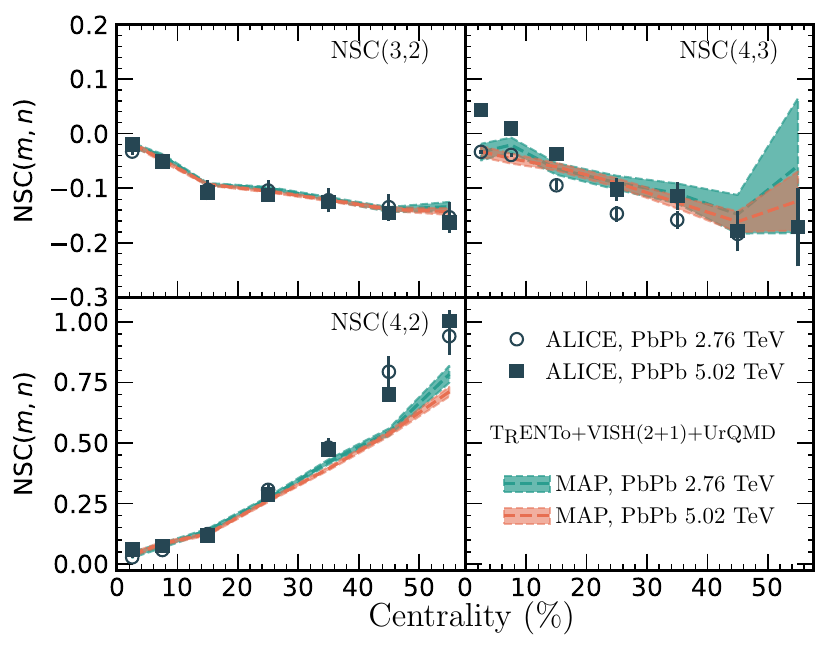}
    \includegraphics[width=0.49\textwidth]{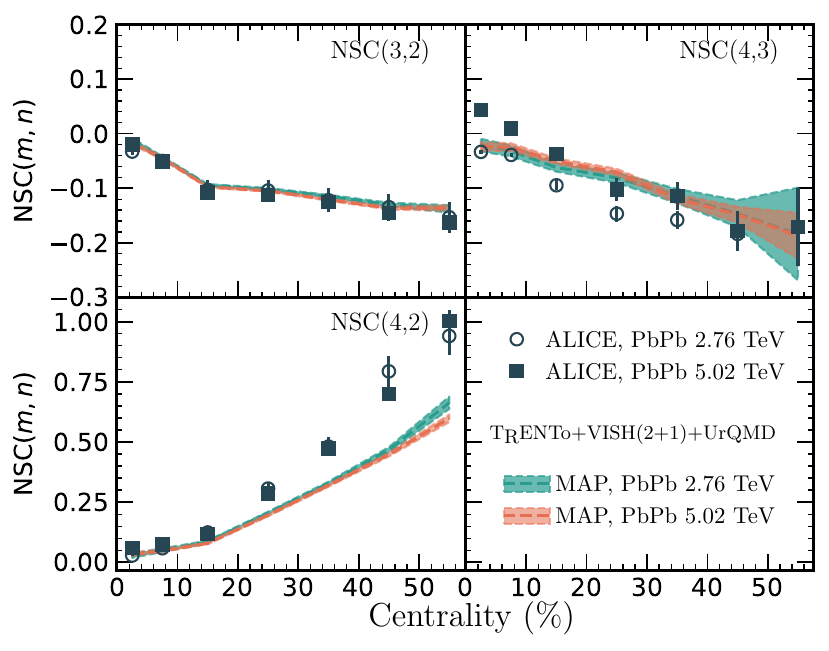}
    \caption{MAP predictions of normalized symmetric cumulants $\mathrm{NSC}(m,n)$ obtained by calibrating with selected and all observables depicted on the left and right, respectively.}
    \label{fig:nscmapAvsB}
\end{figure*}

The comparison between the configuration of the remaining higher-order observables are shown in Figs.~\ref{fig:vnmapAvsB}-\ref{fig:nscmapAvsB}.

\section{Centrality calibration}\label{sec:appendix_Cent}
\begin{figure*}
    \centering
    \includegraphics[width=0.49\textwidth]{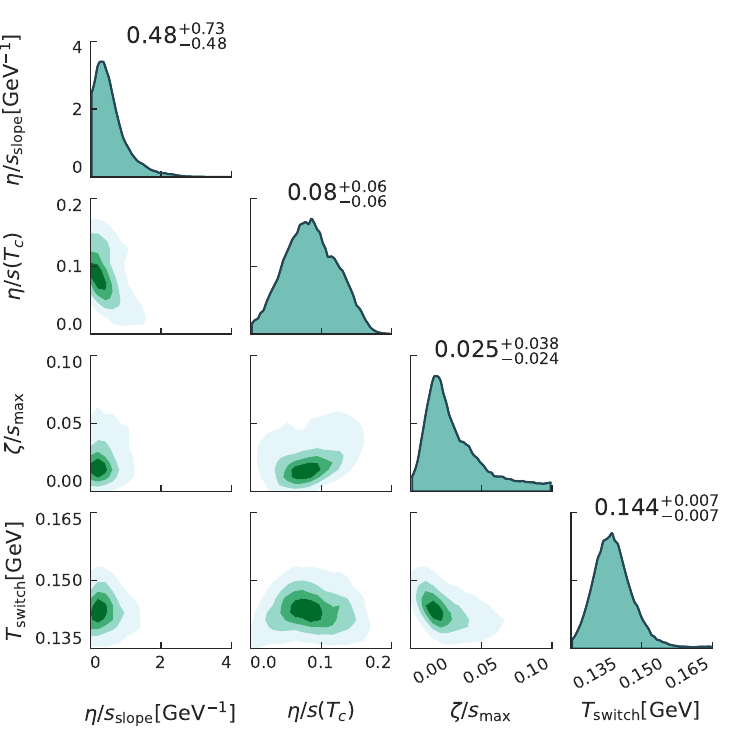}
    \includegraphics[width=0.49\textwidth]{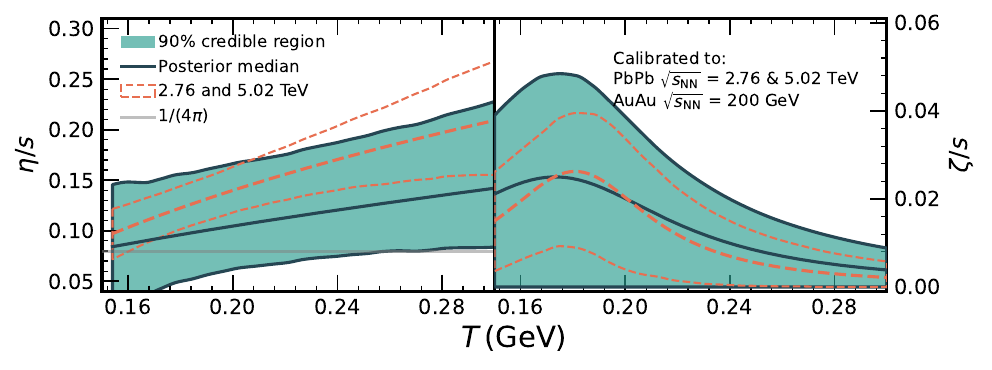}
    \caption{Posterior distribution of transport coefficients obtained with Pb--Pb data only shown on the left. The corresponding credibility region depicted on the right. }
    \label{fig:posteriorC}
\end{figure*}

In previous analyses simulations of the 500 design parametrizations used a singular centrality definition. This was chosen to reduce computational time spent on calibrating the centrality. For this analysis the calibration of centrality was carried out exclusively for all parametrizations. To see the effect of centrality calibration on the transport coefficients the analysis was conducted with the same set of observables as those in Ref.~\cite{Parkkila:2021yha}. The resulting posterior distributions of the transport coefficients and the credibility regions are shown in Fig.~\ref{fig:posteriorC}.
The obtained posteriors deviate from the previous results with increasing uncertainty due to the exclusive centrality calibration.

\end{document}